\documentclass[12pt]{article}

\usepackage{amssymb}
\usepackage[mathscr]{euscript}
\usepackage{graphicx}

\newtheorem{prop}{Proposition}[section]

\newtheorem{defi}{Definition}[section]
\newtheorem{lemm}{Lemma}[section]
\newtheorem{theo}{Theorem}[section]
\newtheorem{coro}{Corollary}[section]

\newcommand{\bbox}{\normalsize {}%
        \nolinebreak \hfill $\blacksquare$ \medbreak \par}

\newcommand{\Ni}{\hbox{ {\vrule height .22cm}{\leaders\hrule\hskip.2cm} }}
\newcommand{\iN}{\hbox{ {\leaders\hrule\hskip.2cm}{\vrule height .22cm} }}

\def\<{\langle} \def\>{\rangle}

\newcommand{\1}{\mathrm{1 \hspace{-0.25em} l}}

\title{Covariant Hamiltonian formalism for the calculus of variations with several variables}
\author{  
Fr\'ed\'eric H\'ELEIN\footnote{helein@cmla.ens-cachan.fr} \\
 CMLA, ENS de Cachan \\
 61,avenue du Pr\'esident Wilson\\
 94235 Cachan Cedex, France 
\\ \\
Joseph KOUNEIHER\footnote{kouneiher@paris7.jussieu.fr}\\
Universit\'e Diderot-Paris 7 \\
   case 7064\\
 75005 Paris, France  \\ 
\& \\
CNRS-URA 2052 (C.E.A.)\\
C.E. Saclay
91191 Gif-sur-Yvette Cedex}

\begin{document}
\maketitle

\begin{abstract}The main purpose in the present paper is to build a Hamiltonian theory
for fields
which is consistent with the principles of relativity. For this we
consider detailed geometric pictures of Lepage theories in the spirit of Dedecker and try
to stress
out the interplay between the Lepage-Dedecker ($LP$) description and the
(more usual) de Donder-Weyl ($dDW$) one. For instance the Legendre transform in the $dDW$ approach is replaced by a
Legendre correspondence in the $LP$ theory, while ignoring singularities whenever the Lagrangian is degenerate. Moreover
we show that there exist two different
definitions of the observable $(n-1)$-forms which allows one to construct
observable functionals by integration\footnote{which correspond to two different points of view: generalizing the law
$\{p,q\} = 1$ or the law
$\frac{dF}{dt} = \{H,F\}$}, oddly enough we prove that these two definitions coincides only in
the $LP$ situation. Finally other contributions concerning this subject and  examples are also
given.
\end{abstract}

\section{Introduction}
\subsection{Presentation}
Multisymplectic formalisms are finite dimensional descriptions of variational problems
with several variables (or field theories for physicists) analogue to the well-known Hamiltonian
theory of point mechanics. For example consider
on the set of maps $u:\Bbb{R}^n\longrightarrow \Bbb{R}$ a Lagrangian action of the type
\[
{\cal L}[u]=\int_{\Bbb{R}^n}L(x,u(x),\nabla u(x))dx^1\cdots dx^n.
\]
Then it is well-known that the maps which are critical points of ${\cal L}$
are characterised by the Euler--Lagrange equation
${\partial \over \partial x^\mu}\left({\partial L\over \partial (\partial _\mu u)}\right)
= {\partial L\over \partial u}$. By analogy with the Hamiltonian theory we can do the change
of variables
$p^\mu:= {\partial L\over \partial (\partial _\mu u)}$ and define the Hamiltonian function
\[
H(x,u,p):= p^\mu{\partial u\over \partial x^\mu} - L(x,u,\nabla u),
\]
where here $\nabla u = \left({\partial u\over \partial x^\mu}\right)$ is a function of $(x,u,p)$ defined
implicitely by $p^\mu:= {\partial L\over \partial (\partial _\mu u)}(x,u,\nabla u)$.
Then the Euler-Lagrange
equation is equivalent to the generalized Hamilton system of equations
\begin{equation}\label{0.h}
\left\{
\begin{array}{ccc}
\displaystyle {\partial u\over \partial x^{\mu}} & =
& \displaystyle {\partial H\over \partial p^{\mu}}(x,u,p)\\
\displaystyle \sum_{\mu}{\partial p^{\mu}\over \partial x^{\mu}} & =
& \displaystyle - {\partial H\over \partial u}(x,u,p).
\end{array}
\right.
\end{equation}
This simple observation is the basis of a theory discovered by T. de Donder \cite{deDonder}
and H. Weyl \cite{Weyl} independently in 1935. This theory can be formulated in a geometric
setting, an analogue of the symplectic geometry, which is governed by the Poincar\'e--Cartan $n$-form
$\theta:=  e \omega + p^{\mu} du\wedge \omega_{\mu}$ (where $\omega:= dx^1\wedge \cdots  \wedge dx^n$ and
$\omega_{\mu}:= \partial _\mu\iN \omega$) and its differential $\Omega:=d\theta$,
often called multisymplectic (or polysymplectic form).\\

\noindent
Although similar to mechanics this theory shows up deep differences. The first one, which
can be noticed by looking at the system (\ref{0.h}), is that there is a disymmetry
between the
``position'' variable $u$ and the ``momentum'' variables $p^\mu$. Since (\ref{0.h}) involves a
divergence of $p^\mu$ one can anticipate that, when formulated in more geometrical terms, $p^\mu$
will be interpreted as the components of a $(n-1)$-form, whereas $u$ as a scalar function. Another difference
is that there exist other theories which are analogues of Hamilton's one: for instance
the first one, constructed by C. Carath\'eodory in 1929 \cite{Caratheodory}.... In fact, as realized by
T. Lepage in 1936 \cite{Lepage}, there are infinitely many theories, due to the fact that one could fix arbitrary
the value of some tensor in the Legendre transform
(see also \cite{Rund}, \cite{GiaquintaHildebrandt}).\\

\noindent
In the present paper, which is a continuation of \cite{HeleinKouneiher},
we expound contributions concerning five important questions in this area:
\begin{enumerate}
\item what are the features and the advantages on the Lepage theories in comparison
with the de Donder--Weyl one ?
\item the (kinematic) observable functionals defined by integration of differential
$(n-1)$-forms on hypersurfaces
\item the (kinematic) observable functionals defined by integration of differential
$(p-1)$-forms for $0\leq p<n$
\item the {\em dynamical} observable functionals defined by integration of forms
\item covariance and agreement with the principles of Relativity.
\end{enumerate}
Let us explain these points in more detail.\\

\begin{figure}[h]\label{fig-pseudo}
\begin{center}
\includegraphics[scale=1]{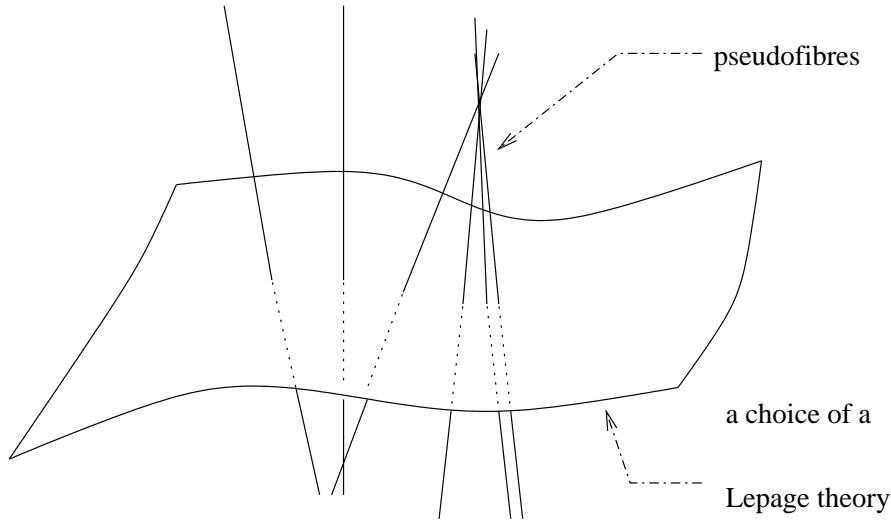}
\caption{Pseudofibers which intersect a submanifold corresponding to the choice of a Lepage theory}
\end{center}
\end{figure}

(i) First, the range of application of the de Donder--Weyl theory is restricted in principle
to variational problems on sections of a bundle ${\cal F}$.
The right framework for it, as expounded in \cite{GIMMSY}
(see also \cite{EcheverriaMunozRoman}), consists in using the affine
first jet bundle $J^1{\cal F}$ and its dual $\left(J^1\right)^*{\cal F}$ as analogues
of the tangent and the cotangent bundles for
mechanics respectively. For non degenerate variational problems the Legendre
transform induces a diffeomorphism between $J^1{\cal F}$ and $\left(J^1\right)^*{\cal F}$.
In constrast the Lepage theories can be applied to more general situations but
involve, in general, many more variables and so are more complicated to deal with,
as noticed in \cite{Kijowski2}. This is probably the reason why
most papers on the subject focus on the de Donder--Weyl theory, e.g.
\cite{Snyatycki}, \cite{GarciaPerezrendon}, \cite{Gawedski}, \cite{GoldschmidtSternberg},
\cite{Kijowski1}, \cite{KijowskiSzczyrba}, \cite{KijowskiTulczyjew},
\cite{deLeonRodrigues}, \cite{Gunther}, \cite{Marsden}, \cite{BinzSnyatyckiFisher},
\cite{Sardanashvily}, \cite{Kanatchikov1}, \cite{GIMMSY}, \cite{CantrijnIbortdeLeon},
\cite{Hrabak1}.
A geometrical framework for building all Lepage theories simultaneously
was first expounded by P. Dedecker in
\cite{Dedecker}: if we view variational problems as being defined on $n$-dimensional
submanifolds embedded in a $(n+k)$-dimensional manifold ${\cal N}$, then what plays the
role of the (projective) tangent bundle to space-time
in mechanics is the Grassmann bundle $Gr^n{\cal N}$
of oriented $n$-dimensional subspaces of tangent spaces to ${\cal N}$. What plays
the role of the cotangent bundle in mechanics is $\Lambda^nT^*{\cal N}$. Note that
$\hbox{dim}Gr^n{\cal N} = n+k+nk$ so that
$\hbox{dim}\Lambda^nT^*{\cal N} = n+k+{(n+k)!\over n!k!}$ is strictly larger than
$\hbox{dim}Gr^n{\cal N} + 1$ unless $n=1$ (classical mechanics) or $k=1$ (submanifolds
are hypersurfaces). This difference between the dimensions reflects
the multiplicity of Lepage theories:
as shown in \cite{Dedecker}, we substitute to the Legendre transform
a Legendre correspondence which to each $n$-subspace $T\in Gr^n_q{\cal N}$
(a ``generalized velocity'') associates an affine subspace of $\Lambda^nT^*_q{\cal N}$
called {\em pseudofibre} by Dedecker. Then two points in the same
pseudofiber do actually represent the same physical (infinitesimal) state, so that the coordinates on
$\Lambda^nT^*{\cal N}$, called {\em momento\"\i des} by Dedecker do not represent
physically observable quantities. In this picture any choice of a Lepage theory corresponds
in selecting a submanifold of $\Lambda^nT^*{\cal N}$,
which --- when the induced Legendre transform is invertible --- intersects
transversally all pseudofibers at one point (see Figure \ref{fig-pseudo}):
so the Legendre correspondence specializes
to a Legendre transform. For instance the de Donder--Weyl theory
can be recovered in this setting by the restriction to some submanifold of $\Lambda^nT^*{\cal N}$
(see Section 2.3).

In \cite{HeleinKouneiher} and in the present paper we consider a geometric pictures of
Lepage theories in the spirit of Dedecker and we try to stress out the interplay
between the Lepage--Dedecker description and the de Donder--Weyl one.
Roughly speaking a comparison between these two points of view shows up some analogy with 
some aspects of the projective geometry,
for which there is no perfect system of coordinates, but basically two: the homogeneous ones,
more symmetric but redundant (analogue to the Dedecker description) and the local ones
(analogue to the choice of a particular Lepage theory like e.g.\,the de Donder--Weyl one).
Note that both points of view are based on the same geometrical framework:
a multisymplectic manifold. Such a structure is a manifold ${\cal M}$
equipped with a closed non degenerate $(n+1)$-form $\Omega$ called the multisymplectic form,
an analogue of the symplectic form (see Section 3.1 for details).
For the de Donder--Weyl theory ${\cal M}$ is $\left(J^1\right)^*{\cal F}$
and for the Lepage--Dedecker theory ${\cal M}$ is $\Lambda^nT^*{\cal N}$. 
In both descriptions solutions of the variational
problem correspond to $n$-dimensional submanifolds $\Gamma$ (analogues of Hamiltonian
trajectories: we call them {\em Hamiltonian $n$-curves}) and are characterized by
the Hamilton equation $X\iN \Omega = (-1)^nd{\cal H}$, where $X$ is a $n$-multivector
tangent to $\Gamma$, ${\cal H}$ is a (Hamiltonian) function defined on ${\cal M}$
and by ``$\iN$'' we mean the interior product.

In section 2 we present a complete derivation of the (Dedecker) Legendre
correspondence and of the generalized Hamilton equations.
We use a method that does not rely on any trivialisation or connection
on the Grassmannian bundle.
A remarkable property, which is illustrated in this paper
through the examples given in section 2.3.2, is that when $n$ and $k$ are greater than 2,
the Legendre correspondence is generically never degenerate. The more spectacular example is
when the Lagrangian density is a constant function --- the most degenerate situation
one can think about --- then the Legendre correspondence is well-defined almost
everywhere except precisely along the de Donder--Weyl submanifold. Such a phenomenon
can be useful when one deals for example with the bosonic string theory with a
skewsymmetric
2-form on the target manifold (a ``$B$-field'', as discussed in \cite{HeleinKouneiher}) or with
the Yang--Mills action in 4 dimensions with a topological term in the Lagrangian: then
the de Donder--Weyl formalism may fail but one can cure this degenerescence by using
another Lepage theory or by working in the full Dedecker setting.

In sections 2, 3 and 4 we explore another aspect of the (Dedecker)
Legendre correspondence: one expects that the resulting Hamiltonian function
on $\Lambda^nT^*{\cal N}$ should satisfy some condition
expressing the ``projective'' invariance along each pseudofiber.
This is indeed the case. We first observe in Section 2.2 that
any smoothly continuous deformation of a Hamiltonian $n$-curve along directions tangent to the
pseudofibers remains a Hamiltonian $n$-curve. Then when we come back to the study
of the geometry of $\Lambda^nT^*{\cal N}$ in section 4 we propose an intrinsic
characterization the subspaces tangent to pseudofibers in Section 4.5. This motivates
the definition given in Paragraph 3.3.5 of the {\em generalized pseudofiber directions}
on any multisymplectic manifold. We also show in Paragraph 3.3.5, that under some hypothese observable functionals on the set of Hamiltonian $n$-curves are invariant under deformation
along the generalized pseudofiber directions.

Lastly another difference between the Lepage--Dedecker point of view and the
de Donder--Weyl one is related to observable forms, the subject of the next paragraph.\\

(ii) The second question concerns the observable functionals defined on the set of all
Hamiltonian $n$-curves $\Gamma$. An important class of such functionals can be constructed
by choosing appropriate $(n-1)$-forms $F$ on the multisymplectic manifold ${\cal M}$ and
a hypersurface $\Sigma$ of ${\cal M}$ which crosses transversally
all Hamiltonian $n$-curves (we shall call {\em slices} such hypersurfaces).
Then $\int_{\Sigma}F:\Gamma \longmapsto \int_{\Sigma\cap \Gamma}F$
is such a functional.
One should however check that such functionals measure physically relevant
quantities. The philosophy adopted here is
inspired from quantum Physics: the formalism should provide us with rules for predicting
the dynamical evolution of an observable. There are two ways to translate this requirement
mathematically: first the ``infinitesimal evolution'' $dF(X)$ of $F$ along a $n$-multivector
$X$ tangent to a Hamiltonian $n$-curve should be completely determined by the value of
$d{\cal H}$ at the point --- this leads to the definition of what we call an {\em
observable $(n-1)$-form} (OF) (see Section 3.3); alternatively, inspired by an
analogy with classical particle mechanics, one can assume that there exists
a tangent vector field
$\xi_F$ such that $\xi_F\iN \Omega + dF = 0$ everywhere --- we call such forms
{\em algebraic observable $(n-1)$-forms} (AOF). We believe that the notion of AOF was
introduced by W. Tulczyjew in 1968 \cite{Tulczyjew-} (see also \cite{Gawedski},
\cite{GoldschmidtSternberg}, \cite{Kijowski1}). To our knowledge the notion of
OF was never considered before; it seems to us however that
it is a more natural definition. It is easy to check that all AOF are actually OF
(see section 3). But the converse is in general not true (see Section 3.3.3), in particular
when we are using the de Donder--Weyl theory. 

It is worth here to insist on the difference of point of view between choosing
OF's or AOF's. The definition of OF is in fact the right notion if we are motivated by the
interplay between the dynamics and observable functionals. It allows us to define a
{\em pseudobracket} $\{{\cal H},F\}$ between the Hamiltonian function and an OF $F$ which leads
to a generalization of the famous equation ${dA\over dt} = \{H,A\}$ of the
Hamiltonian mechanics. This is the relation
\begin{equation}\label{0.dF}
dF_{|\Gamma} = \{{\cal H}, F\}\omega_{|\Gamma},
\end{equation}
where $\Gamma$ is a Hamiltonian $n$-curve and $\omega$ is a given volume $n$-form
on space-time (see Proposition 3.1). In constrast the definition of AOF's
is the right notion if we are motivated
in defining an analogue of the Poisson bracket beween observable $(n-1)$-forms.
This Poisson bracket, for two AOL $F$ and $G$ is given by
$\{F,G\}:= \xi_F\wedge \xi_G\iN \Omega$, a definition reminiscent from classical mechanics.
This allows us to construct a Poisson bracket on functionals by the rule 
$\{\int_\Sigma F,\int_\Sigma G\}: \Gamma \longmapsto \int_{\Sigma\cap \Gamma}\{F,G\}$
(see Section 3.3).

Now here is a subtle difference between the de Donder--Weyl and the Lepage--Dedecker theories: 
OF's and AOF's coincide on $\Lambda^nT^*{\cal N}$. This result is proved in section 4.
But in contrast, on a submanifold of $\Lambda^nT^*{\cal N}$ corresponding
to the choice of a particular Lepage theory, like the de Donder--Weyl one,
there exist OF's which are not AOF's. We call a {\em pataplectic manifold} a
multisymplectic manifold on which OF's and AOF's coincide. So the Lepage--Dedecker theory
is pataplectic but the de Donder--Weyl theory is not so: this plays in favour of
the Lepage--Dedecker formalism, although it could be much more complicated
because of the explosion of the number of
nonphysical variables. However we will also prove (see section 4) that all
OF's in the de Donder--Weyl theory are actually restrictions of AOF's
from the Lepage--Dedecker theory. So one could try to take advantage of both points of view:
working in a multisymplectic manifold, with less variables, and keeping in mind that there is an
extension of this multisymplectic manifold to a pataplectic manifold, where
each OF is the restriction of some AOF. This picture is automatically given for free
by the Legendre correspondence if we start from a variational problem with
a Lagrangian. But more generally, we can consider the following problem:
can we embed any multisymplectic manifold
on which the set of OF's and the set of AOF's differ into another multisymplectic manifold
in such a way that OF's are restrictions of AOF's ?
We shall address this question in a further paper \cite{HeleinKouneiher1.2}:
we prove that such an extension is possible if some hypotheses
are satisfied.

A complementary result ensures that, under suitable hypothese, the observable functionals defined by integration of AOF's are invariant by deformation along pseudo-fibers
(Lemma 3.3 in Section 3.3.5).\\

(iii) Note that it is possible to generalize the notion of observable
$(p-1)$-forms to the case where $0\leq p<n$, as pointed out recently  in \cite{Kanatchikov1}, \cite{Kanatchikov2}.
For example the disymmetry between variables $u$ and $p^\mu$ in system (\ref{0.h})
suggests that, if the $p^\mu$'s are actually the components of the observable $(n-1)$-form
$p^\mu\omega_\mu$, $u$ should be an observable function. Another interesting example
is the Maxwell action, where the gauge potential 1-form $A_\mu dx^\mu$ and the
Faraday $(n-2)$-form $\star dA= \eta^{\mu\lambda}\eta^{\nu\sigma}
(\partial _\mu A_\nu - \partial _\nu A_\mu)\omega_{\lambda\sigma}$
are also ``observable'', as proposed in \cite{Kanatchikov1}. 
Note that again two kinds of approaches for defining such observable forms are possible,
as in the preceding paragraph: either our starting point is to ensure consistency
with the dynamics (this led us in (ii) to the definition of OF's) or we privilegiate the
definition which seems to be the more appropriate for having a notion of Poisson
bracket (this led us in (ii) the definition of AOF's).
If we were follow the second point of view we would be led to the following
definition, in \cite{Kanatchikov1}: a $(p-1)$-form $F$ would be observable (``Hamiltonian'' in
\cite{Kanatchikov1}) if and only if
there exists a $(n-p+1)$-multivector $X_F$ such that
$dF = (-1)^{n-p+1}X_F\iN \Omega$. Note that $X_F$ is far from being unique
in general. This definition has the advantage that --- thanks to a consistent definition of
Lie derivatives of forms with respect to multivectors due to W.M. Tulczyjew
\cite{Tulczyjew} ---
a beautiful notion of graded Poisson bracket between such forms can
be defined, in an intrinsic way (see also \cite{Paufler}, \cite{ForgerRomer}).
These notions were used successfully by S. Hrabak for
constructing a multisymplectic version of the Marsden--Weinstein symplectic reduction 
\cite{Hrabak1} and of the BRST operator \cite{Hrabak2}.
Unfortunately such a definition of observable $(p-1)$-form would not have
nice dynamical properties. For instance if ${\cal M}:= \Lambda^nT^{\star}(\Bbb{R}^n\times \Bbb{R})$
with $\Omega = de\wedge \omega + dp^\mu\wedge d\phi\wedge \omega_\mu$, then
the 0-form $p^1$ would be observable, since
$dp^1 = (-1)^n{\partial \over \partial \phi}\wedge 
{\partial \over \partial x^2}\wedge \cdots \wedge {\partial \over \partial x^n}\iN \Omega$,
but there would be no chance for finding a law for the infinitesimal change
of $p^1$ along a curve inside a Hamiltonian $n$-curve. By that we mean that there would
be no hope for having an analogue of the relation (\ref{0.dF}) (Corollary 3.1).

That is why we have tried to base ourself on the first point of view and to
choose a definition of observable $(p-1)$-forms in order to garantee good
dynamical properties, i.e.\,in the purpose of generalizing
relation (\ref{0.dF}). A first attempt was
in \cite{HeleinKouneiher} for variational problems concerning maps between manifolds.
We propose here another definition working for all Lepagean theories, i.e.\,more
general. Our new definition 
works ``collectively'', requiring to the set of observable $(p-1)$-forms for
$0\leq p<n$ that their differentials
form a sub bundle stable by exterior multiplication and containing differentials of
observable $(n-1)$-forms ({\em copolarisation}, Section 5.1). This definition actually merged
out as the right notion from our efforts to generalize the dynamical relation (\ref{0.dF}).
This is the content of Theorem 5.1.

Once this is done we are left with the question of defining the bracket between an observable
$(p-1)$-form $F$ and an observable $(q-1)$-form $G$. We propose here a (partial)
answer. In Section 5.5. we find necessary conditions on such a bracket
in order to be consistent with the standard bracket used by physicists in quantum
field theory. Recall that this standard bracket is built through an
infinite dimensional Hamiltonian description of fields theory. This allows us to
characterize what should be our correct bracket in two cases: either
$p$ or $q$ is equal to $n$, or $p,q\neq n$ and $p+q=n$. The second situation
arises for example for the Faraday $(n-2)$-form and the gauge potential 1-form
in electromagnetism (see ''Example 8'' in Section 5.5).
However we are  unable to find a general definition: this is left as a
partially open problem.
Regardless, note that  this analysis shows that the right bracket (i.e.\,from the point of
view adopted here) should have a definition which differs from those proposed
in \cite{Kanatchikov1} and also from our
previous definition in \cite{HeleinKouneiher}.\\

(iv) Another question concerns the bracket between observable functionals 
obtained by integration of say $(n-1)$-forms on two {\em different} slices.
This is a crucial question if one is concerned by the relativistic invariance of
a symplectic theory. Indeed the only way to build a relativistic invariant
theory of classical (or quantum) fields is to make sense of functionals
(or observable operators) as defined on the set of solution (each one being a complete
history in space-time), independently of the choice of a time coordinate.
This requires at least that one should be able to define the bracket between
say the observable functionals $\int_\Sigma F$ and $\int_{\widetilde{\Sigma}} G$
even when $\Sigma$ and ${\widetilde{\Sigma}}$ are different (imagine they correspond
to two space-like hypersurfaces). One possibility for that is to assume that one
of the two forms, say $F$ is such that $\int_\Sigma F$ depends uniquely on the
homology class of $\Sigma$. Using Stoke's theorem one checks easily that such
a condition is possible if $\{{\cal H}, F\}= 0$. We call a {\em dynamical} observable
$(n-1)$-form any observable $(n-1)$-form which satisfies such a relation. All that leads
us to the question of finding all such forms.

This problem was investigated in \cite{Kijowski1} and discussed in \cite{GoldschmidtSternberg}
(in collaboration with S. Coleman). It led to an interesting but deceptive
answer: for a linear variational problem (i.e.\,with a linear PDE, or for free fields)
one can find a rich collection of dynamical OF's, roughly speaking in
correspondence with the set of solutions
of the linear PDE. However as soon as the problem becomes nonlinear (so for interacting
fields) the set of dynamical OF's is much more reduced and corresponds to the symmetries
of the problem (so it is in general finite dimensional). We come back here to this question in 
section 6. We are looking at the example of a complex scalar field with one symmetry, so that the 
only dynamical OF's basically correspond to the total charge of the field.
We show there that by a kind of Noether's procedure we can enlarge the set
of dynamical OF's by including all smeared integrals of the current density.
This exemple illustrates the fact that gauge symmetry helps strongly in constructing
dynamical observable functionals. Another possibility in order to enlarge the number
of dynamical functionals is when the nonlinear variational problem can be
approximated by a linear one: this gives rise to observable functionals defined by expansions
\cite{perturb}.\\

(v) The main motivation for multisymplectic formalisms is to build a Hamiltonian theory
which is consistent with the principles of Relativity, i.e.\,being {\em covariant}.
Recall for instance that for all the multisymplectic formalisms which have been proposed
one does not need to use a privilegiate time coordinate. One of our motivations
in this paper was to try to extend this democracy between space and time coordinates
to the coordinates on fiber manifolds (i.e.\,along the fields themselves). This is
quite in the spirit of the Kaluza--Klein theory and its modern avatars: 11-dimensional
supergravity, string theory and M-theory. This concern leads us naturally to replace
de Donder--Weyl by the Dedecker theory.
In particular we do not need in our formalism to split the variables into the
horizontal (i.e.\,corresponding to space-time coordinates) and vertical (i.e.\,non
horizontal) categories. (Such a splitting has several drawbacks, for example it causes
difficulties in order to define the stress-energy tensor.)
Of course, as the reader can imagine many new difficulties appear, if we do not
fix a priori the space-time/fields
splitting,  like for instance, how to define a slice
(see paragraph (ii)), which plays the role of a constant time hypersurface without
referring to a given space-time background ? We propose in Section 3.3.4 a definition
of such a slice which, roughly speaking, requires a slice to be transversal to all
Hamiltonian $n$-curves. Here the idea is that the dynamics only (i.e.\,the Hamiltonian
function) should determine what are the slices. We give in Section 4.1. and 4.2 more
precise characterisations of these slices in the case where the multisymplectic
manifold is $\Lambda^nT^*{\cal N}$. (See also Section 5.4 for the slices of codimension
greater than 1.) In the same spirit the (at first glance unpleasant)
definition of copolarization given in Section 5.1 is our answer
to a similar problematic: how to define forms which --- in a noncovariant
way of thinking --- should be of the type $dx^\mu$, where the $x^\mu$'s are space-time
coordinates, without a space-time background ? Note also that the notion of copolarization
corresponds somehow to the philosophy of general relativity:
the observable quantities again are not measured directly, they are compared each to the other
ones.

In exactly the same spirit we remark that the dynamical law (\ref{0.dF}) can
be expressed in a slightly more general form which is: if $\Gamma$ is a Hamiltonian
$n$-curve then
\begin{equation}\label{0.dFdG}
\{{\cal H}, F\}dG_{|\Gamma} = \{{\cal H}, G\}dF_{|\Gamma},
\end{equation}
for all OF's $F$ and $G$ (Proposition 3.1 and
Theorem 5.1). Mathematically this is not much more difficult than (\ref{0.dF}).
However (\ref{0.dFdG}) is more satifactory from the point of view of relativity:
no volume form $\omega$ is singled out, the dynamics just prescribe how to compare
two observations.

In a future paper \cite{HeleinKouneiher1.1} we investigate gauge theories, addressing the
question of how to formulate a fully covariant multisymplectic for them.
Note that the Lepage--Dedecker theory expounded here does not answer this question completely,
because a connection cannot be seen as a submanifold. We will show there that it is
possible to adapt this theory and that a convenient covariant framework
consists in looking at gauge fields as {\em equivariant} submanifolds
over the principal bundle of the theory, i.e.\,satisfying some suitable
zeroth and first order differential constraints.

\subsection{Notations}

\noindent The Kronecker symbol $\delta^{\mu}_{\nu}$ is equal to 1 if
$\mu = \nu$ and equal to 0 otherwise. We shall also set
\[
\delta^{\mu_1\cdots \mu_p}_{\nu_1\cdots \nu_p}:=
\left| \begin{array}{ccc}
\delta^{\mu_1}_{\nu_1} & \dots  & \delta^{\mu_1}_{\nu_p}\\
\vdots & & \vdots \\
\delta^{\mu_p}_{\nu_1} & \dots & \delta^{\mu_p}_{\nu_p}
\end{array}\right| .
\]
In most examples, $\eta_{\mu\nu}$ is a constant
metric tensor on $\Bbb{R}^n$ (which may be Euclidean or Minkowskian).
The metric on his dual space his $\eta^{\mu\nu}$. Also, $\omega$ will
often denote a volume form on some space-time: in local coordinates
$\omega=dx^1\wedge \cdots \wedge dx^n$ and we will use several times the notation
$\omega_\mu:= {\partial \over \partial x^\mu}\iN \omega$,
$\omega_{\mu\nu}:= {\partial \over \partial x^\mu}\wedge
{\partial \over \partial x^\nu}\iN \omega$, etc. Partial derivatives
${\partial \over \partial x^\mu}$ and ${\partial \over \partial p_{\alpha_1\cdots \alpha_n}}$
will be sometime abbreviated by $\partial _\mu$ and $\partial ^{\alpha_1\cdots \alpha_n}$
respectively.\\

\noindent
When an index or a symbol is omitted in the middle of a
sequence of indices or symbols, we denote this omission by $\widehat{\,}$.
For example $a_{i_1\cdots \widehat{i_p}\cdots i_n}:=
a_{i_1\cdots i_{p-1}i_{p+1}\cdots i_n}$,
$dx^{\alpha_1}\wedge \cdots \wedge \widehat{dx^{\alpha_\mu}}\wedge \cdots \wedge dx^{\alpha_n} :=
dx^{\alpha_1}\wedge \cdots \wedge dx^{\alpha_{\mu-1}}\wedge dx^{\alpha_{\mu+1}}\wedge \cdots \wedge dx^{\alpha_n}$.\\

\noindent If ${\cal N}$ is a manifold
and ${\cal FN}$ a fiber bundle over
${\cal N}$, we denote by $\Gamma({\cal N},{\cal FN})$ the set of smooth sections of
${\cal FN}$. Lastly we use the notations concerning the exterior algebra of multivectors and
differential forms, following W.M.\,Tulczyjew \cite{Tulczyjew}.
If ${\cal N}$ is a differential $N$-dimensional manifold and $0\leq k\leq N$,
$\Lambda^kT{\cal N}$ is the bundle over ${\cal N}$ of $k$-multivectors
($k$-vectors in short)
and $\Lambda^kT^{\star}{\cal N}$ is the bundle of differential forms of degree $k$
($k$-forms in short). Setting $\Lambda T{\cal N}:= \oplus_{k=0}^N\Lambda^kT{\cal N}$
and $\Lambda T^{\star}{\cal N}:= \oplus_{k=0}^N\Lambda^kT^{\star}{\cal N}$, there exists
a unique duality evaluation map between $\Lambda T{\cal N}$ and $\Lambda T^{\star}{\cal N}$
such that for every decomposable $k$-vector field $X$, i.e.\,of the form 
$X=X_1\wedge \cdots \wedge X_k$, and for every $l$-form $\mu$, then
$\langle X,\mu\rangle = \mu(X_1,\cdots ,X_k)$ if $k=l$ and $=0$ otherwise.
Then interior products $\iN $ and $\Ni$ are operations
defined as follows. If $k\leq l$, the product
$\iN :\Gamma({\cal N},\Lambda^kT{\cal N})\times \Gamma({\cal N},\Lambda^lT^{\star}{\cal N})\longrightarrow
\Gamma({\cal N},\Lambda^{l-k}T^{\star}{\cal N})$
is given by
$$\langle Y,X\iN \mu\rangle = \langle X\wedge Y,\mu\rangle ,\quad
\forall (l-k)\hbox{-vector }Y.$$
And if $k\geq l$, the product $\Ni :\Gamma({\cal N},\Lambda^kT{\cal N})\times
\Gamma({\cal N},\Lambda^lT^{\star}{\cal N})\longrightarrow
\Gamma({\cal N},\Lambda^{k-l}T{\cal N})$ is given by
\[
\langle X\Ni \mu,\nu\rangle = \langle X,\mu\wedge \nu\rangle,\quad
\forall (k-l)\hbox{-form }\nu.
\]

\section{Hamiltonian formulation of the calculus of variations}
We recall here how most of the second order variational problems can be restated
as generalized Hamilton equations. Details and computations in coordinates can be
found in \cite{Rund}, \cite{GiaquintaHildebrandt} concerning the first
Section and \cite{Kanatchikov1}, \cite{HeleinKouneiher} for the following.

\subsection{Classical one-dimensional theory}
\subsubsection{The Legendre transform}
Let ${\cal Y}$ be some manifold (the configuration space for a material point),
$T{\cal Y}$ its tangent bundle and $L:T{\cal Y}\longrightarrow \Bbb{R}$ some sufficiently smooth
time-independant Lagrangian density. It leads to the following action, defined on 
a set of ${\cal C}^1$ paths $\gamma:I\longrightarrow {\cal Y}$ (where $I$ is some interval
of $\Bbb{R}$) by
$${\cal L}(\gamma):= \int_I L(\gamma(t),\dot{\gamma}(t))dt.$$
Let $T^{\star}{\cal Y}$ be the cotangent bundle of ${\cal Y}$.
Assuming that the Legendre condition is true, i.e.\,the mapping
$$\begin{array}{cccl}
\phi: & T{\cal Y} & \longrightarrow & T^{\star}{\cal Y}\\
& (y,v) & \longmapsto & \left( y,{\partial L\over \partial v}(y,v)\right)
\end{array}$$
is a diffeomorphism, we can define the Hamiltonian function 
\begin{equation}\label{2.1.H}
H(q,p) = \langle p,V(q,p)\rangle - L(q,V(q,p)),
\end{equation}
where $(y,p)\longmapsto (y, V(y,p))$ is the inverse map of $\phi$. Then the
Euler-Lagrange equation of motion
${d\over dt}
\left( {\partial L\over \partial v^i}(\gamma(t),\dot{\gamma}(t))\right) =
{\partial L\over \partial y^i}(\gamma(t),\dot{\gamma}(t))$ is transformed into the
Hamilton equations system 
\begin{equation}\label{2.1.hamilton}
{dq^i\over dt}(t) = {\partial H\over \partial p_i}(q(t),p(t)),
\quad {dp_i\over dt}(t) = -{\partial H\over \partial q^i}(q(t),p(t))
\end{equation}
through the substitution $(q(t),p(t)):= \phi(\gamma(t),\dot{\gamma}(t))$.

\subsubsection{Poincar\'e-Cartan form and symplectic form}
Hamilton equations (\ref{2.1.hamilton}) can be rewritten in a more
geometrical way using the symplectic form
on $T^{\star}{\cal Y}$. For that purpose we define the Poincar\'e-Cartan form $\theta$ on
$T^{\star}{\cal Y}$ to be the 1-form
$$\theta:= \sum_{i=1}^np_idq^i,$$
where $(q^i,p_i)$ must be thought here as the standard coordinates functions. Then
the canonical symplectic 2-form on $T^{\star}{\cal Y}$ is just $\Omega:= d\theta$.
We let the Hamiltonian vector field associated to $H$ to be the unique vector
field $\xi_H$ such that $\Omega(\xi_H,V) = -dH(V)$, $\forall V\in T_{(q,p)}(T^{\star}{\cal Y})$.
An equivalent statement is to write $\xi_H\iN \Omega = -dH$.
Then Equation (\ref{2.1.hamilton}) is just equivalent to ${d\over dt}(q,p) = \xi_H(q,p)$,
i.e. $t\longmapsto (q(t),p(t))$ parametrizes an integral curve of $\xi_H$.

\subsubsection{Time dependant Legendre transform}
In case where $L$ is not time independant, we have to consider a time dependant
Legendre mapping
$$\begin{array}{cccl}
\phi_t: & T{\cal Y} & \longrightarrow & T^{\star}{\cal Y}\\
& (y,v) & \longmapsto & \left( y,{\partial L\over \partial v}(t,y,v)\right) .
\end{array}$$
Assuming that for all $t$, $\phi_t$ is still invertible and denoting
$(y,p)\longmapsto (y, V(t,y,p))$ its inverse map, we can define a time dependant
Hamiltonian function on $I\times T^{\star}{\cal Y}$:
$H(t,q,p):= \langle p,V(t,q,p)\rangle - L(t,q,V(t,q,p))$. Then, by the substitution
$(q(t),p(t)):= \phi_t(\gamma(t),\dot{\gamma}(t))$, the Euler-Lagrange equation
for $\gamma$ is equivalent to the Hamilton system of equations 
\begin{equation}\label{2.1.timedepe}
{dq^i\over dt}(t) = {\partial H\over \partial p_i}(t,q(t),p(t)),
\quad {dp_i\over dt}(t) = -{\partial H\over \partial q^i}(t,q(t),p(t)).
\end{equation}
If however we take care of relativistic principles, we may treat time and space variables
on the same footing by the following construction. We consider
$\widetilde{\cal Y}:= I\times {\cal Y}$. To each path $t\longmapsto \gamma(t)$ with values in
${\cal Y}$ it corresponds a path $t\longmapsto (t,\gamma(t))$ with values in
$\widetilde{\cal Y}$, which can be lift to a path
$t\longmapsto (t,\gamma(t),{dt\over dt},{d\gamma\over dt}(t)))$ with values in
$T\widetilde{\cal Y}$. Similarly we could try to lift a solution $m:t\longmapsto (q(t),p(t))$
of (\ref{2.1.timedepe}) into a Hamiltonian curve $\widetilde{m}$ in $T^{\star}\widetilde{\cal Y}$. This is
not straightforward because, since $v^0={dt\over dt}=1$, the analogue of the Legendre mapping
$\widetilde{\phi}(t,y,v^0,v) =
(t,y,{\partial L\over \partial v^0}(t,y,v),{\partial L\over \partial v}(t,y,v))$ is not defined.\\

\noindent By constrast its inverse $T^{\star}\widetilde{\cal Y}\longrightarrow T\widetilde{\cal Y}$ still makes sense,
i.e.\,the map $\widetilde{\psi}(t,y,p_0,p_i) = (t,y,1,V(t,y,p_i))$, where $V(t,y,p_i)$ is defined above.
So we can define a Hamiltonian function
${\cal H}: T^{\star}\widetilde{\cal Y}\longrightarrow \Bbb{R}$ in a natural way by
${\cal H}(t,y,p_0,p_i) := p_0 + \sum_ip_iV^i(t,y,p_i) - L(t,y,1,V(t,y,p_i))$. We obtain
$${\cal H}(q^0,q^i,p_0,p_i) = p_0 + H(q^0,q^i,p_i),$$
where $H$ is the same function as the one defined above.\\

\noindent 
Let us come back to the problem of lifting $m:t\longmapsto (q(t),p(t))$.
A first step is to set $\widetilde{m}(t)= (q^0(t),q^i(t),p_0(t),p_i(t)):=
(t,\gamma(t),e(t),{\partial L\over \partial v}(t,\gamma(t),\dot{\gamma}(t)))$, where
$e(t)$ is an arbitrary function, i.e.\,we do not attribute a physical meaning to $p_0$.
Anyway this is enough to describe the dynamical equations. Indeed
we remark that $\gamma$ is a solution of the Euler-Lagrange equation if and only if
${dq^0\over dt}=\xi^0$, ${dq^i\over dt}=\xi^i$ and ${dp_i\over dt}=-\xi_i$ for $i\neq 0$,
where 
\begin{equation}\label{2.1.temps}
\xi^0(q^0,q^i,p_0,p_i) = {\partial {\cal H}\over \partial p_0}(q^0,q^i,p_0,p_i)\;\hbox{(a)},
\quad \xi_0(q^0,q^i,p_0,p_i) =
-{\partial {\cal H}\over \partial q^0}(q^0,q^i,p_0,p_i)\;\hbox{(b)}
\end{equation}
and, for $i\neq 0$,
\begin{equation}\label{2.1.espace}
\xi^i(q^0,q^i,p_0,p_i) = {\partial {\cal H}\over \partial p_i}(q^0,q^i,p_0,p_i),
\quad \xi_i(q^0,q^i,p_0,p_i) = -{\partial {\cal H}\over \partial q^i}(q^0,q^i,p_0,p_i).
\end{equation}
Let us check that: equation (a) in (\ref{2.1.temps}) gives ${dq^0\over dt}=1$, which implies $q^0(t) = t$,
up to an additive constant. Equations in (\ref{2.1.espace}) then give respectively
${dq^i\over dt}(t) = {\partial H\over \partial p_i}(t,q(t),p(t))$ and
${dp_i\over dt}(t) = -{\partial H\over \partial q^i}(t,q(t),p(t))$, which are equivalent to
the Euler-Lagrange equation. This was more or less the approach which was used
in \cite{HeleinKouneiher}.\\

\noindent 
We still have the freedom to require further $p_0$ to be a solution of ${dp_0\over dt}=-\xi_0$,
i.e.
$${dp_0\over dt}(t) = -{\partial H\over \partial q^0}(t,q^i(t),p_i(t)).$$
It then implies that $p_0(t)+ H(t,q(t),p(t))$ is a conserved quantity. So
$p_0(t) = E_0-H(t,q(t),p(t))$, where $E_0$ is some constant. This is why one may
see the energy as canonically conjugate to the time. Note that here what is
time and what is energy depend on the split of $\widetilde{\cal Y}$ into space and
time. This second point of view is more satisfactory because it enforces the
relativistic invariance properties.

\subsubsection{Including the time in a geometric picture}
Again if $L$ is time dependant a geometrical formulation is possible, by using
the Poincar\'e-Cartan form $\widetilde{\theta}:= p_0dq^0 + \sum_{i=1}^np_idq^i$ and the symplectic
form $\widetilde{\Omega}:= d\widetilde{\theta}$ on $T^{\star}(I\times {\cal Y})$. The Hamiltonian
vector field $\xi_{{\cal H}}$ defined by (\ref{2.1.temps}) and (\ref{2.1.espace})
is characterised by 
\begin{equation}\label{2.1.hamiltongeo}
\xi_{{\cal H}}\iN \widetilde{\Omega} = -d{\cal H}.
\end{equation}
A solution of the variational problem can thus be pictured geometrically as a
{\em Hamiltonian curve}
$\widetilde{\Gamma}$ embedded in $T^{\star}\widetilde{\cal Y}$ such that (i) $dq^0_{|\widetilde{\Gamma}}$
does not vanish and (ii) for any $\widetilde{m}\in \widetilde{\Gamma}$, $\xi_{{\cal H}}(\widetilde{m})$
is tangent to $\widetilde{\Gamma}$ at $\widetilde{m}$. \footnote{Recall
that other curves do actually project down to the same solution since, as we have seen in
preceeding paragraph, we can choose $p_0(t)$ to be any arbitrary function $e(t)$ instead of
being equal to $E_0-H(t,q(t),p(t))$. }

\subsection{Variational problems with several variables}
\subsubsection{Lagrangian formulation}
We now generalize the above theory to variational problems with several variables
and for Lagrangian depending on first partial derivatives. The category of
Lagrangian variational problems we start with is quite general and is described as
follows.\\

\noindent We consider $n,k\in \Bbb{N}^*$ and a smooth manifold ${\cal N}$ of dimension
$n+k$; ${\cal N}$ will be equipped with a closed nowhere vanishing ``space-time volume''
$n$-form $\omega$. We define
\begin{itemize}
\item the Grassmannian bundle $Gr^n{\cal N}$, it is the fiber bundle over
${\cal N}$ whose fiber over $q\in{\cal N}$ is $Gr^n_q{\cal N}$,
the set of all oriented $n$-dimensional vector subspaces of $T_q{\cal N}$.
\item the subbundle $Gr^\omega{\cal N}:= \{(q,T)\in Gr^n{\cal N}/
\omega_{q|T}>0\}$.
\item the set ${\cal G}^\omega$, it is the set of all oriented $n$-dimensional submanifolds
$G\subset {\cal N}$, such that $\forall q\in G$,
$T_qG\in Gr^\omega_q{\cal N}$ (i.e.\,the restriction of $\omega$ on
$G$ is positive everywhere).
\end{itemize}
Lastly we consider any Lagrangian density $L$, i.e.\, a smooth function
$L:Gr^\omega{\cal N}\longmapsto \Bbb{R}$. Then the Lagrangian of any $G\in {\cal G}^\omega$
is the integral
\begin{equation}\label{2.2.1.action}
{\cal L}[G]:= \int_GL\left(q,T_qG\right)\omega
\end{equation}
(we also denote, for all $K\subset {\cal N}$, ${\cal L}_K[G]:= \int_{G\cap K}L\left(q,T_qG\right)\omega$).
We are interested in submanifolds $G\in {\cal G}^\omega$ which are critical points of ${\cal L}$
(by that we mean that, for any compact $K\subset {\cal N}$, $G\cap K$ is a critical
point of ${\cal L}_K$ with respect to variations with support in $K$).\\

\noindent It will be useful to represent
$Gr^n{\cal N}$ differentely, by means of $n$-vectors. For any $q\in {\cal N}$, we
define $D^n_q{\cal N}$ to be the set of decomposable
$n$-vectors\footnote{another notation for this set would be $D\Lambda^nT_q{\cal N}$, for it
reminds that it is a subset of $\Lambda^nT_q{\cal N}$, but we have choosen to lighten
the notation.}, i.e.\,elements
$z\in \Lambda^nT_q{\cal N}$ such that there exists $n$ vectors $z_1$,...,$z_n\in 
T_q{\cal N}$ satisfying $z = z_1\wedge \cdots \wedge z_n$. Then $D^n{\cal N}$ 
is the fiber bundle whose fiber at each $q\in {\cal N}$ is
$D^n_q{\cal N}$. Moreover the map
$$\begin{array}{ccc}
D^n_q{\cal N} & \longrightarrow & Gr^n_q{\cal N}\\
z_1\wedge \cdots \wedge z_n & \longmapsto & T(z_1,\cdots ,z_n),
\end{array}$$
where $T(z_1,\cdots ,z_n)$ is the vector space spanned and oriented by $(z_1,\cdots ,z_n)$,
induces a diffeomorphism between $\left(D^n_q{\cal N}\setminus \{0\}\right)/\Bbb{R}^*_+$
and $Gr^n_q{\cal N}$. If we set also
$D^\omega_q{\cal N}:=\{(q,z)\in D^n_q{\cal N}/ \omega_q(z)=1\}$, the same map
² us also to identify $Gr^\omega_q{\cal N}$ with $D^\omega_q{\cal N}$.\\

\noindent This framework includes a large variety of situations as illustrated below.\\

\noindent {\bf Example 1} --- {\em Classical point mechanics --- The motion of a point moving
in a manifold ${\cal Y}$ can be represented by its graph
$G\subset {\cal N}:=\Bbb{R}\times {\cal Y}$. If $\pi:{\cal N}\longrightarrow \Bbb{R}$
is the canonical projection and $t$ is the time coordinate on $\Bbb{R}$, then $\omega:=
\pi^*dt$.}\\
\noindent {\bf Example 2} --- {\em Maps between manifolds --- We consider
maps $u:{\cal X}\longrightarrow {\cal Y}$, where ${\cal X}$ and ${\cal Y}$ are manifolds
of dimension $n$ and $k$ respectively and ${\cal X}$ is equipped with some nonvanishing volume form $\omega$.
A first order Lagrangian density can represented as a function
$l: T{\cal Y}\otimes _{{\cal X}\times {\cal Y}}T^{\star}{\cal X}\longmapsto \Bbb{R}$,
where $T{\cal Y}\otimes _{{\cal X}\times {\cal Y}}T^{\star}{\cal X}:=
\{ (x,y,v)/(x,y)\in {\cal X}\times {\cal Y},v\in T_y{\cal Y}\otimes T_x^*{\cal X}\}$.
(We use here a notation which exploits
the canonical identification of $T_y{\cal Y}\otimes T_x^*{\cal X}$ with the set of
linear mappings from $T_x{\cal X}$ to $T_y{\cal Y}$). The action of a map $u$ is
$$\ell [u]:= \int_{\cal X}l(x,u(x),du(x))\omega.$$
In local coordinates $x^{\mu}$ such that $\omega=dx^1\wedge \cdots \wedge dx^n$,
critical points of $\ell$
satisfy the Euler-Lagrange equation
$\sum_{\mu=1}^n{\partial \over \partial x^{\mu}}\left(
{\partial l \over \partial v^i_{\mu}}(x,u(x),du(x))\right) =
{\partial l \over \partial y^i}(x,u(x),du(x))$, $\forall i=1,\cdots ,k$.\\
Then we set ${\cal N}:={\cal X}\times {\cal Y}$ and denoting by $\pi:{\cal N}\longrightarrow
{\cal X}$ the canonical projection, we use the volume form $\omega\simeq \pi^*\omega$.
Any map $u$ can be represented by its graph
$G_u:=\{(x,u(x))/\,x\in {\cal X}\}\in {\cal G}^\omega$, (and conversely if
$G\in {\cal G}^\omega$ then the condition $\omega_{|G}>0$ forces $G$ to
be the graph of some map). For all $(x,y)\in{\cal N}$ we also have a diffeomorphism
$$\begin{array}{ccc}
T_y{\cal Y}\otimes T_x^*{\cal X} & \longrightarrow & Gr^\omega_{(x,y)}{\cal N}
\simeq D^\omega_{(x,y)}{\cal N}\\
v & \longmapsto & T(v),
\end{array}$$
where $T(v)$ is the graph of the linear map $v:T_x{\cal X}\longrightarrow T_y{\cal Y}$.
Then if we set $L(x,y,T(v)):= l(x,y,v)$, the action defined by (\ref{2.2.1.action})
coincides with $\ell$.}\\
\noindent {\bf Example 3} --- {\em Sections of a fiber bundle --- This is a particular
case of our setting, where ${\cal N}$ is the total space of a fiber bundle with base
manifold ${\cal X}$. The set ${\cal G}^\omega$ is then just the set of smooth sections.}\\
\noindent {\bf Example 4} --- {\em Gauge theories --- Since a connection is not a
section of a bundle, a variational problem on connections such as the Yang--Mills equations
cannot be described as a problem on embeddings of submanifolds directly. If we want to reduce
ourself to such a situation two methods are possible: either we work on a local trivialisation
of the bundle over ${\cal X}$, a situation achieved by choosing a flat connexion
$\nabla^0$ and working in the coordinates provided by sections parallel for $\nabla^0$:
then any other connection $\nabla$ can be identified with $A:=\nabla - \nabla^0$, a 1-form
on ${\cal X}$ with coefficients in a Lie algebra, i.e.\,a section
of a bundle (see Section 2.4 below). Or we consider an equivariant
lift of the connexion on the corresponding principal bundle. In this case the connexion
can be represented globally and in a covariant way by a 1-form
with coefficients in the Lie algebra on the principal bundle satisfying some equivariance
constraints. We shall compare both points of view in \cite{HeleinKouneiher1.1}.}

\subsubsection{The Legendre correspondence}
\noindent Now we consider the manifold $\Lambda^nT^*{\cal N}$ and the
projection mapping $\Pi:\Lambda^nT^*{\cal N}\longrightarrow {\cal N}$. We shall
denote by $p$ an $n$-form in the fiber $\Lambda^nT^*_q{\cal N}$. There is a
canonical $n$-form $\theta$ called the {\em Poincar\'e--Cartan} form defined on
$\Lambda^nT^*{\cal N}$ as follows: $\forall (q,p)\in \Lambda^nT^*{\cal N}$,
$\forall X_1,\cdots ,X_n\in T_{(q,p)}\left(\Lambda^nT^*{\cal N}\right)$,

\[
\theta_{(q,p)}(X_1,\cdots ,X_n):= p\left( \Pi^*X_1,\cdots ,\Pi^*X_n\right)
 = \langle \Pi^*X_1\wedge \cdots \wedge \Pi^*n,p\rangle ,
\]
where $\Pi_*X_\mu := d\Pi_{(q,p)}(X_\mu)$.
If we use local coordinates $\left(q^\alpha\right)_{1\leq \alpha\leq n+k}$ on
${\cal N}$, then a basis of $\Lambda^nT^*_q{\cal N}$ is the family
$\left( dq^{\alpha_1}\wedge \cdots \wedge dq^{\alpha_n}\right)_{1\leq \alpha_1<
\cdots <\alpha_n \leq n+k}$ and we denote by $p_{\alpha_1\cdots \alpha_n}$ the coordinates
on $\Lambda^nT^*_q{\cal N}$ in this basis. Then $\theta$ writes
\begin{equation}\label{2.2.2.theta}
\theta:= \sum_{1\leq \alpha_1<\cdots <\alpha_n\leq n+k}
p_{\alpha_1\cdots \alpha_n}dq^{\alpha_1}\wedge \cdots \wedge dq^{\alpha_n}.
\end{equation}
Its differential is the {\em multisymplectic} form $\Omega:= d\theta$ and
will play the role of generalized symplectic form.\\

\noindent In order to build the analogue of the Legendre transform we consider the fiber bundle
$Gr^\omega{\cal N}\times _{\cal N}\Lambda^nT^*{\cal N}:=\{(q,z,p)/q\in {\cal N},
z\in Gr^\omega_q{\cal N}\simeq D^\omega_q{\cal N},
p\in \Lambda^nT^*_q{\cal N}\}$ and we denote by
$\widehat{\Pi}:Gr^\omega{\cal N}\times _{\cal N}\Lambda^nT^*{\cal N}
\longrightarrow {\cal N}$ the canonical projection:
\[
\begin{array}{cccrcrc}
& & Gr^\omega{\cal N}\times _{\cal N}\Lambda^nT^*{\cal N} &
\longrightarrow & \Lambda^nT^*{\cal N} & \supset & {\cal M}\\
& & \downarrow & \widehat{\Pi} \searrow & \downarrow \Pi &
\swarrow \Pi & \\
Gr^n{\cal N} & \supset & Gr^\omega{\cal N}\simeq D^\omega{\cal N} &
\longrightarrow & {\cal N} & &.
\end{array}
\]
\noindent
We define on $Gr^\omega{\cal N}\times _{\cal N}\Lambda^nT^*{\cal N}$ the function
\[
W(q,z,p):= \langle z,p\rangle -L(q,z).
\]
Note that for each $(q,z,p)$ there a vertical subspace $V_{(q,z,p)}\subset
T_{(q,z,p)} ( Gr^\omega{\cal N}\times _{\cal N}\Lambda^nT^*{\cal N})$,
which is canonically defined as the kernel of
\[
d\widehat{\Pi}_{(q,z,p)}:T_{(q,z,p)}
\left( Gr^\omega{\cal N}\times _{\cal N}\Lambda^nT^*{\cal N}\right)
\longrightarrow T_q{\cal N}.
\]
Moreover it makes sense to split $V_{(q,z,p)}\simeq \{0\}\times T_zD^\omega_q{\cal N}
\times T_p\Lambda^nT^*_q{\cal N}\simeq T_zD^\omega_q{\cal N}
\oplus T_p\Lambda^nT^*_q{\cal N}$. Hence, for any function $F$ defined on
$Gr^\omega{\cal N}\times _{\cal N}\Lambda^nT^*{\cal N}$, we can define 
the restrictions of the differential $dF_{(q,z,p)}$ on both factors,
i.e.\,$T_zD^\omega_q{\cal N}$ and $T_p\Lambda^nT^*_q{\cal N}$, which
will be denoted respectively by $\partial F/\partial z(q,z,p)$
and $\partial F/\partial p(q,z,p)$. [However in order to make sense
of ``$\partial F/\partial q(q,z,p)$'' we would need to define a ``horizontal''
subspace of $T_{(q,z,p)} \left( Gr^\omega{\cal N}\times _{\cal N}\Lambda^nT^*{\cal N}\right)$,
which could be obtained for instance by using a connection on the bundle
$Gr^\omega{\cal N}\times _{\cal N}\Lambda^nT^*{\cal N}\longrightarrow {\cal N}$.]\\

\noindent
Instead of a Legendre transform we shall rather use a Legendre correspondence:
we write
\begin{equation}\label{2.2.2.corr}
(q,z) \longleftrightarrow (q,p)\quad \hbox{if and only if}
\quad {\partial W\over \partial z}(q,z,p) = 0.
\end{equation}

\noindent Let us try to picture geometrically the situation (see figure \ref{fig-tdn}):
\begin{figure}[h]\label{fig-tdn}
\begin{center}
\includegraphics[scale=1]{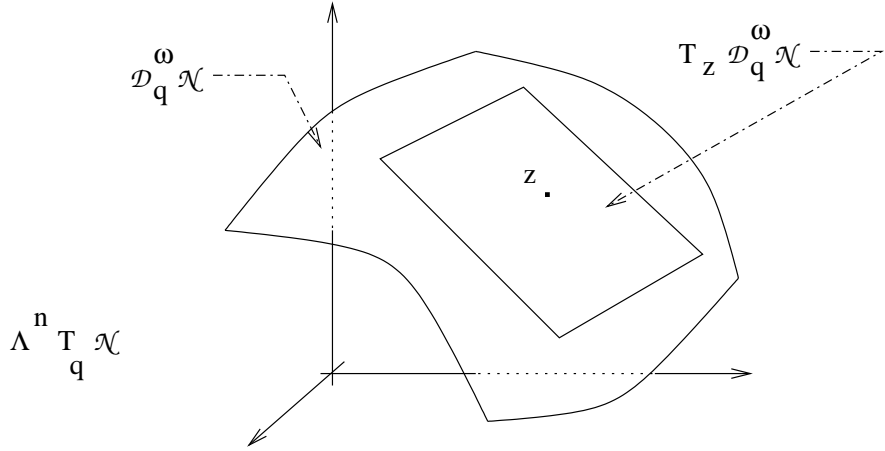}
\caption{$T_zD_q^\omega{\cal N}$ is a
vector subspace of $\Lambda^nT_q{\cal N}$}
\end{center}
\end{figure}
$D_q^\omega{\cal N}$
is a smooth submanifold of dimension $nk$ of the vector space $\Lambda^nT_q{\cal N}$,
which is of dimension ${(n+k)!\over n!k!}$; $T_zD_q^\omega{\cal N}$ is thus a
vector subspace of $\Lambda^nT_q{\cal N}$. And ${\partial L\over \partial z}(q,z)$
or ${\partial W\over \partial z}(q,z,p)$
can be understood as linear forms on $T_zD_q^\omega{\cal N}$ whereas
$p\in \Lambda^nT^*_q{\cal N}$ as a linear form on $\Lambda^nT_q{\cal N}$.\\

\noindent The meaning of the right hand side of (\ref{2.2.2.corr}) is thus that
the restriction of $p$ at $T_zD_q^\omega{\cal N}$ coincides with
${\partial L\over \partial z}(q,z,p)$:
\begin{equation}\label{2.2.2.p=dl}
p_{|T_zD_q^\omega{\cal N}} = {\partial L\over \partial z}(q,z).
\end{equation}
Given $(q,z)\in Gr^\omega{\cal N}$ we define the {\em enlarged pseudofiber} to be:
\[
P_q(z):=\{p\in \Lambda^nT^*_q{\cal N}/{\partial W\over \partial z}(q,z,p) = 0\}.
\]
In other words, $p\in P_q(z)$ if it is a solution of (\ref{2.2.2.p=dl}).
Obviously $P_q(z)$ is not empty; moreover given some $p_0\in P_q(z)$,
\begin{equation}\label{2.2.2.p-p}
p_1\in P_q(z),\ \Longleftrightarrow \ p_1-p_0\in
\left( T_zD_q^\omega{\cal N}\right)^{\perp}:=
\{p\in \Lambda^nT^*_q{\cal N}/\forall \zeta\in T_zD_q^\omega{\cal N},p(\zeta)=0\}.
\end{equation}
So $P_q(z)$ is an affine subspace of $\Lambda^nT^*_q{\cal N}$ of dimension
${(n+k)!\over n!k!} - nk$. Note that in case where $n=1$ (the classical mechanics
of point) then dim\,$P_q(z)=1$: this is due to the fact that we are still free to fix
arbitrarily the momentum component dual to the time (i.e.\,the energy).\\

\noindent
We now define
\[
{\cal P}_q:= \bigcup_{z\in D^\omega_q{\cal N}}P_q(z) \subset \Lambda^nT^*_q{\cal N},
\quad \forall q\in {\cal N}
\]
and we denote by ${\cal P}:=\cup_{q\in {\cal N}}{\cal P}_q$
the associated bundle over ${\cal N}$. We also
let, for all $(q,p)\in \Lambda^nT^*{\cal N}$,
\[
Z_q(p):=\{ z\in Gr^\omega_q{\cal N}/ p\in P_q(z)\}.
\]
It is clear that $Z_q(p)\neq \emptyset \Longleftrightarrow p\in {\cal P}_q$....
Now in order to go further we need to choose some submanifold ${\cal M}_q\subset {\cal P}_q$,
its dimension is not fixed {\em a priori}.\\

\noindent {\bf Legendre Correspondence Hypothesis} ---
{\em We assume that there exists a subbundle manifold ${\cal M}\subset {\cal P}\subset
\Lambda^nT^*{\cal N}$ over ${\cal N}$ of dimension $M:=\dim {\cal M}$
such that,
\begin{itemize}
\item for all $q\in {\cal N}$ the fiber ${\cal M}_q$ is a smooth submanifold,
possibly with boundary, of dimension $1\leq M-n+k\leq {(n+k)!\over n!k!}$
\item for any $(q,p)\in {\cal M}$, $Z_q(p)$ is a non empty smooth connected 
submanifold of $Gr^\omega_q{\cal N}$
\item if $z_0\in Z_q(p)$, then we have $Z_q(p)=\{z\in D^\omega_q{\cal N}/
\forall \dot{p}\in T_p{\cal M}_q, \langle z-z_0,\dot{p}\rangle = 0\}$.
\end{itemize}
}
\noindent 
{\em Remark} --- In the case where $M={(n+k)!\over n!k!}+n+k$, then ${\cal M}_q$
is an open subset of $\Lambda^nT^*_q{\cal N}$ and so $T_p{\cal M}_q\simeq
\Lambda^nT^*_q{\cal N}$. Hence the last assumption of the Legendre Correspondence
Hypothesis means that $Z_q(p)$ is reduced to a point. In general this condition
will imply that the inverse correspondence can be rebuild by using the Hamiltonian
function (see Lemma \ref{2.2.2.dh} below).

\begin{lemm}\label{2.2.2.h=cste}
Assume that the Legendre correspondence hypothesis is true. Then for all $(q,p)\in {\cal M}$,
the restriction of $W$ to $\{z\} \times Z_q(p)\times \{p\}$ is constant.
\end{lemm}
{\em Proof} --- Since $Z_q(p)$ is smooth and connected, it suffices to prove that
$W$ is constant along any smooth path inside $\{(q,z,p)/q,p\hbox{ fixed },z\in Z_q(p)\}$.
Let $s\longmapsto z(s)$ be a smooth path with values into $Z_q(p)$, then
\[
{d \over d s}\left( W(q,z(s),p)\right)  = 
{\partial W\over \partial z}(q,z(s),p)
\left( {dz\over ds}\right) = 0,
\]
because of (\ref{2.2.2.corr}). \bbox

\noindent 
A straightforward consequence of Lemma \ref{2.2.2.h=cste} is that we can define the
{\bf Hamiltonian function} ${\cal H}:{\cal M}\longrightarrow \Bbb{R}$ by
\[
{\cal H}(q,p):= W(q,z,p),\quad \hbox{where }z\in Z_q(p).
\]

\noindent
In the following, for all $(q,p)\in {\cal M}$ and for all $z\in D^n_q{\cal N}$
we denote by
\[
\begin{array}{cccc}
z_{|T_p{\cal M}_q}: & T_p{\cal M}_q & \longrightarrow & \Bbb{R}\\
 & \dot{p} & \longmapsto & \langle z,\dot{p}\rangle
\end{array}
\]
the linear map induced by $z$ on $T_p{\cal M}_q$. Then:
\begin{lemm}\label{2.2.2.dh}
Assume that the Legendre Correspondence Hypothesis is true. Then\\
(i) $\forall (q,p)\in {\cal M}$ and $\forall z\in Z_q(p)$,
\begin{equation}\label{2.2.2.hl}
{\partial {\cal H}\over \partial p}(q,p) = z_{|T_p{\cal M}_q}.
\end{equation}
As a corollary of the above formula, $z_{|T_p{\cal M}_q}$ does not depend on the
choice of $z\in Z_q(p)$.\\
(ii) Conversely if $(q,p)\in {\cal M}$ and $z\in D^\omega_q{\cal N}$ satisfy condition
(\ref{2.2.2.hl}), then $z\in Z_q(p)$ or equivalentely $p\in P_q(z)$.
\end{lemm}
{\em Remark} --- The advised Reader may expect to have also the relation
``${\partial {\cal H}\over \partial q}(q,p) = -  {\partial L\over \partial q}(q,z)$''. But
as remarked above the meaning of ${\partial {\cal H}\over \partial q}$ and
${\partial L\over \partial q}$ is not clearly defined, because we did not introduce
a connection on the bundle $Gr^\omega{\cal N}\times _{\cal N}\Lambda^nT^*{\cal N}$.
This does not matter and we shall make the economy of this relation later !\\ 

\noindent 
{\em Proof} --- Let $(q,p)\in {\cal M}$ and $(0,\dot{p})\in T_{(q,p)}{\cal M}$,
where $\dot{p}\in T_p{\cal M}_q$.
In order to compute $d{\cal H}_{(q,p)}(0,\dot{p})$, we consider a smooth
path $s\longmapsto (q,p(s))$ with values into ${\cal M}_q$
whose derivative at $s=0$ coincides with
$(0,\dot{p})$. We can further lift this path into another one
$s\longmapsto (q,z(s),p(s))$ with values into
$Gr^\omega{\cal N}\times _{\cal N}\Lambda^nT^*{\cal N}$, in such a way that
$z(s)\in Z_{q}(p(s))$, $\forall s$. Then using (\ref{2.2.2.p=dl}) we obtain
\[
\begin{array}{ccl}
\displaystyle {d\over ds}\left({\cal H}(q,p(s))\right)_{|s=0} & = &
\displaystyle {d\over ds}\left( \langle z(s),p(s)\rangle - L(q,p(s))\; \right)_{|s=0}\\
 & = & \displaystyle 
\langle \dot{z},p\rangle + \langle z,\dot{p}\rangle -
{\partial L\over \partial z}(q,z)(\dot{z}) = 
\langle z,\dot{p}\rangle ,
\end{array}
\]
from which (\ref{2.2.2.hl}) follows. This proves (i).\\
The proof of (ii) uses the Legendre Correspondence Hypothesis:
consider $z,z_0\in D^n_q{\cal N}$ and assume that $z_0\in Z_q(p)$ and that
$z$ satisfies (\ref{2.2.2.hl}).
Then by applying the conclusion (i) of the Lemma to $z_0$ we deduce
that $\partial {\cal H}/\partial p(q,p) = z_{0|T_p{\cal M}_q}$ and thus
$(z-z_0)_{|T_p{\cal M}_q}=0$. Hence by the Legendre Correspondence Hypothesis
we deduce that $z\in Z_q(p)$.
\bbox

\noindent A further property is that, given $(q,z)\in D^\omega{\cal N}$, it is possible to
find a $p\in P_q(z)$ and to choose the value of ${\cal H}(q,p)$ simultaneously. This
property will be useful in the following in order to simplify the Hamilton equations.
For that purpose we define, for all $h\in \Bbb{R}$, the {\em pseudofiber}:
\[
P^h_q(z):= \{ p\in P_q(z)/ {\cal H}(q,p)=h\}.
\]
We then have:
\begin{lemm}\label{2.2.2.dernierlemme}
For all $(q,z)\in Gr^\omega{\cal N}$ the pseudofiber $P^h_q(z)$ is a affine subspace of $\Lambda^nT^*_q{\cal N}$ parallel
to $\left( T_zD^n_q{\cal N}\right)^\perp$.
Hence $\hbox{dim }P^h_q(z)=\hbox{dim }P_q(z)-1 = {(n+k)!\over n!k!}-nk-1$.
\end{lemm}
{\em Proof} --- We first remark that, $\forall q\in {\cal N}$
and $\forall z\in D^\omega_q{\cal N}$, $\omega_q$
belongs to $\left( T_zD^\omega_q{\cal N}\right)^\perp$, because of the definition of $D^\omega_q{\cal N}$.
So $\forall \lambda \in \Bbb{R}$, $\forall p\in P_q(z)$, we deduce from (\ref{2.2.2.p-p}) that
$p+\lambda \omega_q\in P_q(z)$ and thus
\[
\begin{array}{ccl}
{\cal H}(q,p+\lambda \omega_q) & = & \langle z,p+\lambda \omega_q\rangle - L(q,z)\\
& = & {\cal H}(q,p) +\lambda \langle z,\omega_q\rangle = {\cal H}(q,p) +\lambda.
\end{array}
\]
Hence we deduce that $\forall h\in \Bbb{R}$, $\forall p\in P_q(z)$,
$\exists !\lambda\in \Bbb{R}$ such that
\[
{\cal H}(q,p+\lambda\omega_q) = h,
\]
so that $P^h_q(z)$ is non empty. Moreover
if $p_0\in P^h_q(z)$ then $p_1\in P^h_q(z)$ if and only if $p_1-p_0\in
\left( T_zD^\omega_q{\cal N}\right)^\perp \cap z^\perp$. In order to conclude 
observe that
$\left( T_zD^\omega_q{\cal N}\right)^\perp \cap z^\perp =
\left( T_zD^n_q{\cal N}\right)^\perp$. \bbox

\subsubsection{Critical points}
We now look at critical points of the Lagrangian functional using the above
framework. Instead of the usual approach using jet bundles and contact structure,
we shall derive Hamilton equations directly, without writing the
Euler--Lagrange equation.\\

\noindent First we extend the form $\omega$ on ${\cal M}$ by setting
$\omega\simeq \Pi^*\omega$, where $\Pi:{\cal M}\longrightarrow {\cal N}$ is the
bundle projection, and we define $\widehat{\cal G}^\omega$ to be the set of oriented
$n$-dimensional submanifolds $\Gamma$ of ${\cal M}$, such that $\omega_{|\Gamma}>0$
everywhere. A consequence of this inequality is that the restriction of
the projection $\Pi$ to any $\Gamma\in \widehat{\cal G}^\omega$ is an
embedding into ${\cal N}$: we denote by $\Pi(\Gamma)$ its image. It is clear
that $\Pi(\Gamma)\in {\cal G}^\omega$. Then we can
view $\Gamma$ as (the graph of) a section $q\longmapsto p(q)$
of the pull-back of the bundle ${\cal M}\longrightarrow {\cal N}$ by the
inclusion $\Pi(\Gamma)\subset {\cal N}$.\\

\noindent
Second, we define the subclass $\mathfrak{p}\widehat{\cal G}^\omega\subset \widehat{\cal G}^\omega$
as the set of $\Gamma\in \widehat{\cal G}^\omega$ such that,
$\forall (q,p)\in \Gamma$, $p\in P_q(T_q\Pi(\Gamma))$ (a contact condition).... [As we will see later
it can be viewed as the subset of $\Gamma\in \widehat{\cal G}^\omega$ which satisfy half
of the Hamilton equations.] And
given some $G\in {\cal G}^\omega$,
we denote by $\mathfrak{p}\widehat{G}\subset \mathfrak{p}\widehat{\cal G}^\omega$ the family
of submanifolds $\Gamma\in \mathfrak{p}\widehat{\cal G}^\omega$ such that $\Pi(\Gamma)=G$
and we say that $\mathfrak{p}\widehat{G}$ is the set of {\em Legendre lifts} of $G$.
We hence have $\mathfrak{p}\widehat{\cal G}^\omega=\cup_{G\in {\cal G}^\omega}\mathfrak{p}\widehat{G}$.\\

\noindent
Lastly, we define the functional on $\widehat{\cal G}^\omega$
\[
{\cal I}[\Gamma]:= \int_\Gamma \theta - {\cal H}\omega.
\]
{\bf Properties of the restriction of ${\cal I}$ to $\mathfrak{p}\widehat{\cal G}^\omega$}\\
First we claim that 
\begin{equation}\label{2.2.3.i=l}
{\cal I}[\Gamma] = {\cal L}[G],\quad 
\forall G\in {\cal G}^\omega, \forall \Gamma \in \mathfrak{p}\widehat{G}.
\end{equation}
This follows from
\[
\begin{array}{ccl}
\displaystyle \int_\Gamma \theta - {\cal H}\omega & = &
\displaystyle \int_G\langle z_G,p(q)\rangle \omega
- {\cal H}(q,p(q))\omega\\
 & = & \displaystyle \int_G \left( \langle z_G,p(q)\rangle 
- \langle z_G,p(q)\rangle + L(q,z_G)\right) \omega = \int_GL(q,z_G)\omega,
\end{array}
\]
where $G\longrightarrow {\cal M}:q\longmapsto (q,p(q))$ is the parametrization of
$\Gamma$ and where $z_G$ is the unique $n$-vector in $D^\omega_q{\cal N}$ (for $q\in G$)
which spans $T_qG$.\\
Second let us exploit relation (\ref{2.2.3.i=l}) to compute the first variation of
${\cal I}$ at any submanifold $\Gamma\in \mathfrak{p}\widehat{G}$, i.e.\,a Legendre lift of
$G\in {\cal G}^\omega$. We let $\xi\in \Gamma({\cal N},T{\cal N})$ be
a smooth vector field with compact support and $G_s$, for $s\in \Bbb{R}$, be
the image of $G$ by the flow diffeomorphism $e^{s\xi}$. For small values of $s$,
$G_s$ is still in ${\cal G}^\omega$ and for all $q_s:=e^{s\xi}(q)\in G_s$
we shall denote by $z_s$ the unique $n$-vector in $D^\omega_{q_s}{\cal N}$ which
spans $T_{q_s}G_s$. Then we choose a smooth section $(s,q_s)\longmapsto p(q)_s$
in such a way that $p(q)_s\in P_{q_s}(z_s)$. This builds a family of Legendre lifts
$\Gamma_s=\{(q_s,p(q)_s)\}$. We can now use relation (\ref{2.2.3.i=l}):
${\cal I}[\Gamma_s]= {\cal L}[G_s]$ and derivate it with respect to $s$.
Denoting by $\widehat{\xi}\in T_{(q,p(q))}{\cal M}$ the vector $d(q_s,p(q)_s)/ds_{|s=0}$, we obtain
\begin{equation}\label{2.2.3.xi}
\delta{\cal I}[\Gamma](\widehat{\xi}) = {d\over ds}{\cal I}[\Gamma_s]_{|s=0} = 
{d\over ds}{\cal L}[G_s]_{|s=0} = \delta{\cal L}[G](\xi).
\end{equation}
{\bf Variations of ${\cal I}$ along $T_p{\cal M}_q$}\\
On the other hand for all $\Gamma\in \widehat{\cal G}^\omega$ and for all
{\em vertical} tangent vector field along $\Gamma$ $\zeta$, i.e.\,such that
$d\Pi_{(q,p)}(\zeta) = 0$ or such that $\zeta\in T_p{\cal M}_q\subset T_{(q,p)}{\cal M}$,
we have
\begin{equation}\label{2.2.3.zeta}
\delta{\cal I}[\Gamma](\zeta) = \int_\Gamma \left( \langle z_{\Pi(\Gamma)},\zeta\rangle
- {\partial {\cal H}\over \partial p}(q,p)(\zeta)\right) \omega,
\end{equation}
where $z_{\Pi(\Gamma)}$ is the unique $n$-vector in $D^\omega_q{\cal N}$ (for $q\in G(\Gamma)$)
which spans $T_q\Pi(\Gamma)$. Note that in the special case where
$\Gamma\in \mathfrak{p}\widehat{\cal G}^\omega$,
we have $z_{\Pi(\Gamma)}\in Z_q(p)$, so we deduce from (\ref{2.2.2.hl}) and (\ref{2.2.3.zeta})
that $\delta{\cal I}[\Gamma](\zeta) = 0$. And the converse is true. So $\mathfrak{p}\widehat{\cal G}^\omega$
can be characterized by requiring that condition (\ref{2.2.3.zeta}) is true for all vertical
vector fields $\zeta$.\\

\noindent {\bf Conclusion}\\
\noindent The key point is now that any vector field along $\Gamma$ can be written
$\widehat{\xi} + \zeta$, where $\widehat{\xi}$ and $\zeta$ are as above. And
for any $G\in {\cal G}^\omega$ and for all $\Gamma\in \mathfrak{p}\widehat{G}$,
the first variation of ${\cal I}$ at $\Gamma$ with respect to a vector field
$\widehat{\xi} + \zeta$, where locally $\widehat{\xi}$ lifts $\xi\in T_q{\cal N}$
and $\zeta \in  T_p{\cal M}_q$, satisfies
\begin{equation}\label{2.2.3.firstv}
\delta{\cal I}[\Gamma](\widehat{\xi} + \zeta) = \delta{\cal L}[G](\xi).
\end{equation}
We deduce the following.
\begin{theo}\label{2.2.3.equivalence}
(i) For any $G\in {\cal G}^\omega$ and for all Legendre lift $\Gamma\in \mathfrak{p}\widehat{G}$,
$G$ is a critical point of ${\cal L}$ if and only if $\Gamma$ is a critical
point of ${\cal I}$.\\
(ii) Moreover for all $\Gamma\in \widehat{\cal G}^\omega$, if $\Gamma$ is a
critical point of ${\cal I}$ then $\Gamma$ is a Legendre lift,
i.e.\,$\Gamma\in \mathfrak{p}\widehat{\Pi(\Gamma)}$
and $\Pi(\Gamma)$ is a critical point of ${\cal L}$.
\end{theo}
{\em Proof} --- (i) is a straightforward consequence of (\ref{2.2.3.firstv}). Let us
prove (ii): if $\Gamma\in \widehat{\cal G}^\omega$ is a
critical point of ${\cal I}$, then in particular for all vertical tangent vector
field $\zeta\in T_p{\cal M}_q$, $\delta{\cal I}[\Gamma](\zeta)=0$ and
by (\ref{2.2.3.zeta}) this implies $(z_{\Pi(\Gamma)})_{|T_p^*{\cal M}_q} =
(\partial {\cal H}/ \partial p)(q,p)$. Then by applying Lemma \ref{2.2.2.dh}--(ii)
we deduce that $z_{\Pi(\Gamma)}\in Z_q(p)$. Hence $\Gamma$ is a Legendre lift.
Lastly we use the conclusion of the part (i) of the Theorem to conclude
that $G(\Gamma)$ is a critical point of ${\cal L}$. \bbox

\begin{coro}\label{2.2.3.coro1}
Let $\Gamma\in \widehat{\cal G}^\omega$ be a critical point of ${\cal I}$ and let a
smooth section $\pi:\Gamma\longrightarrow \Lambda^nT^*{\cal N}$ satisfy:
$\forall (q,p)\in \Gamma$,
$\pi(q,p)\simeq \pi(q)\in \left( T_zD^\omega_q{\cal N}\right) ^\perp$ (where $z\in Z_q(p)$).
Then $\tilde{\Gamma}:= \{ (q,p+\pi(q))/(q,p)\in \Gamma \}$ is another
critical point of ${\cal I}$.
\end{coro}
{\em Proof} --- By using Theorem \ref{2.2.3.equivalence}--(ii) we deduce that $\Gamma$
has the form $\Gamma=\{(q,p)/q\in \Pi(\Gamma),p\in P_q(z_{\Pi(\Gamma)})\}$ and thus
$\tilde{\Gamma}=\{(q,p+\pi(q))/q\in \Pi(\Gamma),p\in P_q(z_{\Pi(\Gamma)})\}$.  This implies,
by using (\ref{2.2.2.p-p}), that $\tilde{\Gamma}\in \mathfrak{p}\widehat{\Pi(\Gamma)}$;
then $\tilde{\Gamma}$ is also a critical point of ${\cal I}$ because of
Theorem \ref{2.2.3.equivalence}--(i). \bbox

\noindent
Note that, for any constant $h\in \Bbb{R}$, by choosing $\Pi(q)=\left( h - {\cal H}(q,p)\right) \omega_q$
(see the proof of Lemma \ref{2.2.2.dernierlemme}) in the above Corollary we deform any critical
point $\Gamma$ of ${\cal I}$ $\Gamma\in \widehat{\cal G}^\omega$ into a critical point $\tilde{\Gamma}$ of ${\cal I}$
contained in ${\cal M}^h:=\{m\in {\cal M}/{\cal H}(m)=h\}$.

\begin{defi}
An {\bf Hamiltonian $n$-curve} is a critical point $\Gamma$ of ${\cal I}$ such that
there exists a constant $h\in \Bbb{R}$ such that $\Gamma\subset {\cal M}^h$....
\end{defi}

\subsubsection{Hamilton equations}
\noindent
We now end this section by looking at the equation satisfied by critical points
of ${\cal I}$. Let $\Gamma\in \widehat{\cal G}^\omega$ and $\xi\in\Gamma({\cal M},
T{\cal M})$ be a smooth vector field with compact support. (Here ${\cal X}$ is an
$n$-dimensional manifold diffeomorphic to $\Gamma$.) We let $e^{s\xi}$
be the flow mapping of $\xi$ and $\Gamma_s$ be the image of $\Gamma$ by $e^{s\xi}$.
We denote by
\[
\begin{array}{cccl}
\sigma : & (0,1)\times {\cal X} & \longrightarrow & {\cal M}\\
 & (s,x) & \longmapsto & \sigma(s,x)
\end{array}
\]
a map such that if $\gamma_s:x\longmapsto \sigma(s,x)$, then $\gamma=\gamma_0$
is a parametrization of $\Gamma$, $\gamma_s$ is a parametrization of $\Gamma_s$
and ${\partial \over \partial s}\left(\sigma (s,x)\right) = \xi\left(\sigma (s,x)\right)$.
Then
\[
\begin{array}{ccl}
\displaystyle 
{\cal I}[\Gamma_s]- {\cal I}[\Gamma] & = &
\displaystyle \int_{\cal X}\gamma_s^*(\theta -{\cal H}\omega)
- \gamma^*(\theta -{\cal H}\omega)\\
 & = & \displaystyle \int_{\partial \left( (0,s)\times {\cal X}\right)}
\sigma^*(\theta -{\cal H}\omega)
= \int_{(0,s)\times {\cal X}}d\left(\sigma^*(\theta -{\cal H}\omega) \right)\\
& = & \displaystyle \int_{(0,s)\times {\cal X}}\sigma^*(\Omega -d{\cal H}\wedge \omega) ).
\end{array}
\]
Thus
\[
\begin{array}{ccl}
\displaystyle 
\lim_{s\rightarrow 0}{{\cal I}[\Gamma_s]- {\cal I}[\Gamma]\over s} & = &
\displaystyle \lim_{s\rightarrow 0}{1\over s}\int_{(0,s)\times {\cal X}}
\sigma^*(\Omega -d{\cal H}\wedge \omega)\\
& = & \displaystyle \int_{\cal X}{\partial \over \partial s}\iN
\sigma^*(\Omega -d{\cal H}\wedge \omega)
= \int_{\cal X}\gamma^*(\xi\iN (\Omega -d{\cal H}\wedge \omega) )\\
& = & \displaystyle \int_\Gamma \xi\iN (\Omega -d{\cal H}\wedge \omega) .
\end{array}
\]
We hence conclude that $\Gamma$ is a critical point of ${\cal I}$ if and only if
$\forall m\in \Gamma$, $\forall \xi\in T_m{\cal M}$, $\forall X\in \Lambda^nT_m\Gamma$,
\[
\xi\iN (\Omega -d{\cal H}\wedge \omega)(X) = 0
\quad \Longleftrightarrow \quad
X\iN (\Omega -d{\cal H}\wedge \omega)(\xi) = 0.
\]
We thus deduce the following.
\begin{theo}\label{2.2.4.theo}
A submanifold $\Gamma\in \widehat{\cal G}^\omega$ is a critical point of ${\cal I}$
if and only if
\begin{equation}\label{2.2.4.hamilton-}
\forall m\in \Gamma, \forall X\in \Lambda^nT_m\Gamma,
\quad X\iN (\Omega -d{\cal H}\wedge \omega) = 0.
\end{equation}
Moreover, if there exists some $h\in \Bbb{R}$ such that $\Gamma\subset {\cal M}^h$
(i.e.\,$\Gamma$ is a Hamiltonian $n$-curve) then
\begin{equation}\label{2.2.4.hamilton}
\forall m\in \Gamma, \exists ! X\in \Lambda^nT_m\Gamma,
\quad X\iN \Omega = (-1)^nd{\cal H}.
\end{equation}
\end{theo}
Recall that,
because of Lemma \ref{2.2.2.dernierlemme} and Corollary \ref{2.2.3.coro1}, it is always
possible to deform a Hamiltonian $n$-curve $\Gamma\longmapsto \tilde{\Gamma}$
in such a way that ${\cal H}$ be constant on $\tilde{\Gamma}$ and
$\Pi(\Gamma) = \Pi(\tilde{\Gamma})$.\\
{\em Proof} --- We just need to check (\ref{2.2.4.hamilton}). Let $\Gamma\subset {\cal M}^h$.
Since $d{\cal H}_{|\Gamma}=0$, $\forall X\in \Lambda^nT_m\Gamma$,
$X\iN d{\cal H}\wedge \omega = (-1)^n\langle X,\omega\rangle d{\cal H}$. So by choosing
the unique $X$ such that $\langle X,\omega\rangle =1$, we obtain 
$X\iN d{\cal H}\wedge \omega = (-1)^nd{\cal H}$. Then (\ref{2.2.4.hamilton-}) is
equivalent to (\ref{2.2.4.hamilton}).\bbox

\subsection{Some examples}
We pause to study on some simple examples how the Legendre correspondence and the
Hamilton work. In particular in the construction of ${\cal M}$ we let a large freedom
in the dimension of the fibers ${\cal M}_q$, having just the constraint
that $\dim {\cal M}_q\leq \dim {\cal P}_q={(n+k)!\over n!k!}$. This leads to a large choice
of approaches between two opposite ones: the first one consists in using as less variables as
possible, i.e.\,to choose ${\cal M}$ to be of minimal dimension 
(for example the de Donder--Weyl theory),
the other one consists in using the largest number of
variables, i.e.\,to choose ${\cal M}$ to be equal to the interior of ${\cal P}$ (the advantage
will be that in some circumstances we avoid degenerate situations).\\

\noindent We focus here on special cases of Example 2 of the previous Section:
we consider maps $u:{\cal X}\longrightarrow {\cal Y}$. We denote by $q^\mu=x^\mu$,
if $1\leq \mu\leq n$, coordinates on ${\cal X}$ and by $q^{n+i}=y^i$, if $1\leq i\leq k$,
coordinates on ${\cal Y}$. Recall that $\forall x\in {\cal X}$, $\forall y\in {\cal Y}$,
the set of linear maps $v$ from $T_x^*{\cal X}$ to $T_y{\cal Y}$ can be identified with
$T_y{\cal Y}\otimes T_x^*{\cal X}$. And coordinates representing some
$v\in T_y{\cal Y}\otimes T_x^*{\cal X}$ are denoted by $v^i_\mu$, in such a way that 
$v=\sum_\alpha\sum_i v^i_\mu{\partial \over \partial y^i}\otimes dx^\mu$.
Then through the diffeomorphism
$T_y{\cal Y}\otimes T_x^*{\cal X}\ni v\longmapsto T(v)\in  Gr^\omega_{(x,y)}{\cal N}$
(where ${\cal N}={\cal X}\times {\cal Y}$)
we obtain coordinates on $Gr^\omega_q{\cal N}\simeq D^\omega_q{\cal N}$.
We also denote by $e:=p_{1\cdots n}$, $p^{\mu}_i:=p_{1\cdots (\mu-1)i(\mu+1)\cdots n}$,
$p^{\mu_1\mu_2}_{i_1i_2}:=p_{1\cdots (\mu_1-1)i_1(\mu_1+1)\cdots
(\mu_2-1)i_2(\mu_2+1)\cdots n}$, etc, so that 
\[
\Omega=de\wedge \omega + \sum_{j=1}^n\sum_{\mu_1<\cdots <\mu_j}
\sum_{i_1<\cdots <i_j}dp^{\mu_1\cdots \mu_j}_{i_1\cdots i_j}\wedge
\omega_{\mu_1\cdots \mu_j}^{i_1\cdots i_j},
\]
where, for $1\leq p\leq n$,
\[
\begin{array}{ccl}
\omega & := & dx^1\wedge \cdots  \wedge dx^n\\
\omega^{i_1\cdots i_p}_{\mu_1\cdots \mu_p} & := & dy^{i_1}\wedge \cdots  \wedge 
dy^{i_p}\wedge \left( {\partial \over \partial x^{\mu_1}}\wedge \cdots  \wedge 
{\partial \over \partial x^{\mu_p}}\iN \omega\right) .
\end{array}
\]
{\bf Remark} --- It can be checked (see for instance \cite{HeleinKouneiher}) that,
by denoting by $p^*$ all coordinates
$p^{\mu_1\cdots \mu_j}_{i_1\cdots i_j}$ for $j\geq 1$, the Hamiltonian function has
always the form ${\cal H}(q,e,p^*) = e+H(q,p^*)$.

\subsubsection{The de Donder--Weyl formalism}
In the special case of the de Donder--Weyl theory, ${\cal M}^{dDW}_q$ is the submanifold of
$\Lambda^nT^*_q{\cal N}$ defined by the constraints
$p^{\mu_1\cdots \mu_j}_{i_1\cdots i_j} = 0$, for all $j\geq 2$
(Observe that these constraints are invariant by a change of coordinates, so that
they have an intrinsic meaning.) We thus have
\[
\Omega^{dDW}=de\wedge \omega + \sum_\mu \sum_idp^\mu_i\wedge \omega_\mu^i....
\]
Then the equation $\partial W/ \partial z(q,z,p) = 0$
is equivalent to $p^\mu_i=\partial l/ \partial v^i_\mu(q,v)$, so that
the Legendre Correspondence Hypothesis holds if and only if
$(q,v)\longmapsto (q,\partial l /\partial v(q,p))$ is an invertible map.
Note that then the enlarged pseudofibers $P_q(z)$ intersect ${\cal M}^{dDW}_q$ along lines
$\{e\omega + \partial l/ \partial v^i_\mu(q,v)\omega_\mu^i/e\in \Bbb{R}\}$.
So since dim$\Lambda^nT^*_q{\cal N}= {(n+k)!\over n!k!}$, dim${\cal M}^{dDW}_q=nk+1$ and 
dim$P_q(z)={(n+k)!\over n!k!}-nk$, the Legendre Correspondence Hypothesis can be rephrased by
saying that each $P_q(z)$ meets ${\cal M}^{dDW}_q$
transversaly along a line. Moreover $Z_q(e\omega + p^\mu_i\omega_\mu^i)$ is then
reduced to one point, namely $T(v)$, where
$v$ is the solution to $p^\mu_i={\partial l\over \partial v^i_\mu}(q,v)$.\\

\noindent For more details and a description using local coordinates, see
\cite{HeleinKouneiher}.

\subsubsection{Maps from $\Bbb{R}^2$ to $\Bbb{R}^2$ via the Lepage--Dedecker point of view}
Let us consider a simple situation where ${\cal X} = {\cal Y} = \Bbb{R}^2$ and
${\cal M}\subset \Lambda^2T^{\star}\Bbb{R}^4$. It corresponds to variational problems on maps
$u:\Bbb{R}^2\longrightarrow \Bbb{R}^2$. For any point $(x,y)\in \Bbb{R}^4$, we denote
by $(e,p^i_\mu,r)$ the coordinates on $\Lambda^2T_{(x,y)}\Bbb{R}^4$, such that
\[
\theta = e\,dx^1\wedge dx^2 + p^1_idy^i\wedge dx^2 + p^2_idx^1\wedge dy^i + r\,dy^1\wedge dy^2.
\]
An explicit parametrization of $\{z\in D^2_{(x,y)}\Bbb{R}^4/\omega(z)>0\}$ is given by the coordinates
$(t,v^i_\mu)$ through

\[z = t^2{\partial \over \partial x^1}\wedge {\partial \over \partial x^2} +
t\,\epsilon^{\mu\nu} v^i_\mu{\partial \over \partial y^i}\wedge {\partial \over \partial x^\nu}
+ (v^1_1v^2_2-v^1_2v^2_1){\partial \over \partial y^1}\wedge {\partial \over \partial y^2},
\]
where $\epsilon^{12} = - \epsilon^{21} =1$ and $\epsilon^{11} = \epsilon^{22} = 0$.
(Note that $z\in D^\omega_{(x,y)}\Bbb{R}^4$ if and only if $t=1$.)
So elements $\delta z\in T_zD^2_q\Bbb{R}^4$ are parametrized by coordinates
$\delta t$ and $\delta v^i_\mu$ through the relation
\[
\delta z = \delta t \left( 2t{\partial \over \partial x^1}\wedge {\partial \over \partial x^2} +
\epsilon^{\mu\nu}v^i_\mu{\partial \over \partial y^i}\wedge {\partial \over \partial x^\nu}\right)
+ \delta v^i_\mu\left( t\,\epsilon^{\mu\nu}{\partial \over \partial y^i}\wedge {\partial \over \partial x^\nu}
+ \epsilon^{\mu\nu}\epsilon_{ij}v^j_\nu {\partial \over \partial y^1}\wedge {\partial \over \partial y^2}
\right) .
\]
In the following we assume that $z\in D^\omega_{(x,y)}\Bbb{R}^4$, which means that we
specialize the above relations by letting $t=1$ (similarly if we were assuming that
$\delta z\in T_zD^\omega_q\Bbb{R}^4$, then we would have to set $\delta t=0$).\\

\noindent For any 2-form $p=edx^1\wedge dx^2 + \epsilon_{\mu\nu}p^\mu_idy^i\wedge dx^\nu +
r dy^1\wedge dy^2\in \Lambda^2T^{\star}_{(x,y)}\Bbb{R}^4$
and for all $\delta z\in T_zD^2_q\Bbb{R}^4$, we have
\[
\langle \delta z, p\rangle = \delta t
\left( 2e + v_\mu^ip_i^\mu\right)
+ \delta v^i_\mu\left( p_i^\mu + \epsilon^{\mu\nu}\epsilon_{ij}v_\nu^j r\right) .
\]
Hence $\left(T_zD^2_q\Bbb{R}^4\right)^\perp$ is the line spanned by
\[
 \left( v_1^1v_2^2 -v_1^2v_2^1 \right) dx^1\wedge dx^2 -
\epsilon_{ij}v_\nu^j dy^i\wedge dx^\nu + dy^1\wedge dy^2.
\]
And $\left(T_zD^\omega_q\Bbb{R}^4\right)^\perp$ is the plane
spanned by the above 2-form and $dx^1\wedge dx^2$.
Recall that the sets $P_q(z)$ and $P^h_q(z)$ are affine subspace respectively
parallel to $\left(T_zD^\omega_q\Bbb{R}^4\right)^\perp$ and
$\left(T_zD^2_q\Bbb{R}^4\right)^\perp$. We immediately see that they form a family
of non parallel affine subspaces so we expect that on the one hand these subspaces will
intersect, causing obstructions there for the invertibility of the Legendre mapping,
and on the other hand they will fill ``almost'' all of $\Lambda^2T^{\star}_{(x,y)}\Bbb{R}^4$,
giving rise to the phenomenon that the Legendre correspondence is ``generically
everywhere'' well defined.\\

\noindent {\bf Example 5} --- {\em The trivial variational problem ---
We just take $l =0$, so that any map map from $\Bbb{R}^2$ to $\Bbb{R}^2$
is a critical point of $\ell$ ! This example is motivated by gauge theories where the
gauge invariance gives rise to constraints. Here the situation is extreme in the
sense that the set of symmetries is maximal. Then
\[
W(q,z,p) = W(z,p) = e+ p^1_1v^1_1 + p^2_1v^1_2 + p^1_2v^2_1 + p^2_2v^2_2
+ r(v^1_1v^2_2 - v^1_2v^2_1).
\]
Then the equation $\partial W(z,p)/\partial z = 0$ leads to the relations
\[
\left\{ 
\begin{array}{ccccccc}
p^1_1+r v^2_2 & = & 0, & \quad & p^1_2-r v^1_2 & = & 0\\
p^2_1-r v^2_1 & = & 0, & \quad & p^2_2+r v^1_1 & = & 0.
\end{array}\right.
\]
[So that $\forall z$, $P_q(z)=\left(T_zD^\omega_q\Bbb{R}^4\right)^\perp$.] The geometrical
interpretation is that the set of all $P_q(z)$'s fills
${\cal P}_q:=\{(e,p^\mu_i,r)\in \Lambda^2T^*_q\Bbb{R}^4/
r\neq 0 \}\cup \{(e,0,0)/e\in \Bbb{R}\}$. In particular they meet along the line $\{(e,0,0)/e\in \Bbb{R}\}$.\\
$\bullet$ If we decide to work with the constraint $r=0$
(it corresponds to the de Donder--Weyl choice
of multisymplectic theory), then we are obliged to assume the extra constraints $p^\mu_i=0$.
Thus we are led to ${\cal M}_q=\{(e,0,0)\in \Lambda^2T^*_q\Bbb{R}^4\}$;
then the Legendre Correspondence
holds and in particular $Z_q(e,0,0)=D^\omega_q{\cal N}$ and ${\cal H} = e$.
But then ${\cal M}=\Bbb{R}^5$ with the variables $x^\mu,y^i$ and $e$ and with
$\Omega=de\wedge dx^1\wedge dx^2$ which is not a multisymplectic form because it is
degenerate (see Definition \ref{2.1.def1}).\\
$\bullet$ However, if we just assume that $r\neq 0$, then
we can choose ${\cal M}_q= \{(e,p^\mu_i,r)\in \Lambda^2T^*_q\Bbb{R}^4/r\neq 0\}$.
Then we compute ${\cal H}$ on ${\cal M}$:
\[
{\cal H}(q,p)=e - {p^1_1p^2_2 - p^1_2p^2_1\over r}.
\]
\begin{figure}[h]
\begin{center}
\includegraphics[scale=1]{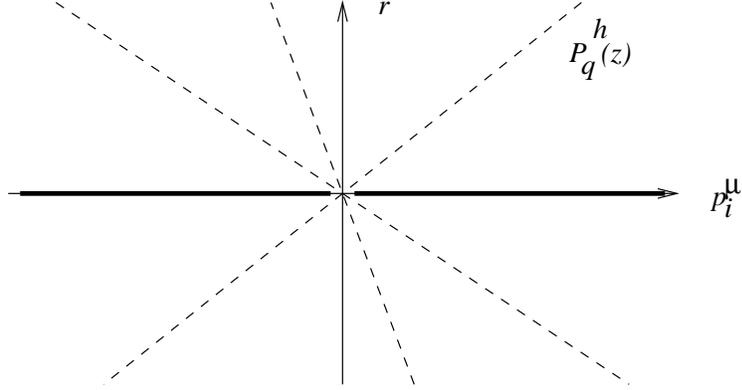}
\caption{${\cal P}^h_q$ (i.e.\,${\cal P}_q\cap {\cal H}^{-1}(h)$) is the subset
of $\Lambda^2T^*_q{\cal N}$ which is the reunion of the pseudo-fibers $P_q^h(z)$}
\end{center}
\end{figure}
One can then check that all Hamiltonian 2-curves are of the form
\[
\Gamma=\left\{\left(x,u(x),e(x)dx^1\wedge dx^2+ \epsilon_{\mu\nu}p^\mu_i(x)dy^i\wedge dx^\nu +
r(x) dy^1\wedge dy^2 \right)/x\in \Bbb{R}^2 \right\},
\]
where $u:\Bbb{R}^2\longrightarrow \Bbb{R}^2$ is an {\em arbitrary} smooth function,
$r:\Bbb{R}^2\longrightarrow \Bbb{R}^*$ is also an arbitrary smooth function and
\[
e(x) = r(x)\left( {\partial u^1\over \partial x^1}(x){\partial u^2\over \partial x^2}(x)
- {\partial u^1\over \partial x^2}(x){\partial u^1\over \partial x^2}(x)\right) + h,
\]
for some constant $h\in \Bbb{R}$, and}
\[
\left\{ \begin{array}{ccccccc}
p^1_1(x) & = & \displaystyle -r(x){\partial u^2\over \partial x^2}(x),  & \quad &
p^1_2(x) & = & \displaystyle r(x){\partial u^1\over \partial x^2}(x)\\
&&&&&&\\
p^2_1(x) & = & \displaystyle r(x){\partial u^2\over \partial x^1}(x),  & \quad &
p^2_2(x) & = & \displaystyle -r(x){\partial u^1\over \partial x^1}(x).
\end{array}\right.
\]
\noindent {\bf Example 6} --- {\em The elliptic Dirichlet integral ---
The Lagrangian is $l (x,y,v) = {1\over 2}|v|^2$ where
$|v|^2:=(v^1_1)^2 + (v^1_2)^2 +(v^2_1)^2 +(v^2_2)^2$. Then one can compute that
$\partial W(z,p)/\partial z = 0$ if and only if
$v^i_\mu = \left(p^\mu_i + \epsilon_{\mu\nu}\epsilon^{ij} p^\nu_j r\right) /
\left( 1-r^2\right)$.
Thus ${\cal P}_q:= \{(e,p^\mu_i,r)\in \Lambda^2T^*_q\Bbb{R}^4/
r\neq \pm 1 \}\cup \{(e,p_\mu^i,1)\in \Lambda^2T^*_q\Bbb{R}^4/p^1_1-p^2_2 = p^1_2+p^2_1 = 0\}
\cup \{(e,p_\mu^i,-1)\in \Lambda^2T^*_q\Bbb{R}^4/p^1_1+p^2_2 = p^1_2-p^2_1 = 0\}$.\\
$\bullet$ The de Donder--Weyl theory corresponds again to the choice $r=0$
and leads to an everywhere well defined Legendre transform.\\
$\bullet$ We could work as well on any hyperplane defined by $r=r_0$, for some
real constant $r_0$. If $r_0\neq \pm 1$, it is defined everywhere, if we choose
however $r_0=\pm 1$, then we again find extra constraints.\\
$\bullet$ Lastly we could work in an open subset
${\cal M}$ of ${\cal P}$. Then the Hamiltonian function is
\[
{\cal H}(q,p) = e + {1\over 1-r^2}
\left( {|p|^2\over 2} + r(p^1_1p^2_2 - p^1_2p^2_1)\right) .
\]
For more details on this kind of exemple (or more generally the action of a bosonic string)
see \cite{HeleinKouneiher}.}\\
\noindent {\bf Example 7} --- {\em Maxwell equations in two dimensions ---
We take $l(x,y,v) = -{1\over 2}\left( v^1_2-v^2_1\right)^2$. Note that if
we identify the components $(u^1,u^2)$ with the components $(A_1,A_2)$ of a
Maxwell gauge potential, we recover the usual Lagrangian
$l (dA)=-{1\over 4}\sum_{\mu,\nu}\left(
{\partial A_\nu\over \partial x^\mu} - {\partial A_\mu\over \partial x^\nu}\right)^2$
for Maxwell fields without charges.
Here relation $\partial W(z,p)/\partial z = 0$ is equivalent to
\[
\left\{ 
\begin{array}{ccccccc}
p^1_1+r v^2_2 & = & 0, & \quad & p^1_2-r v^1_2 & = & v^1_2 - v^2_1\\
p^2_1-r v^2_1 & = & v^2_1 - v^1_2, & \quad & p^2_2+r v^1_1 & = & 0.
\end{array}\right.
\]
Thus we have
${\cal P}_q=\{(e,p_\mu^i,r)\in \Lambda^2T^*_q\Bbb{R}^4/
r\neq 0, -2\}\cup \{(e,p_\mu^i,0)\in \Lambda^2T^*_q\Bbb{R}^4/
p^1_1 = p^2_2 = p^1_2 + p^2_1 = 0\}\cup \{(e,p_\mu^i,-2)\in \Lambda^2T^*_q\Bbb{R}^4/
 p^1_2 - p^2_1 = 0\}$.\\
$\bullet$ If we assume that $r=0$,
we find an intermediate situation, between the
trivial and the Dirichlet Lagrangians. We are again obliged to assume the extra constraints
$p^1_1 = p^2_2 = p^1_2 + p^2_1 = 0$. This reflects the gauge invariance of the problem.
Then the Legendre Correspondence Hypothesis is satisfied and the $Z_q(p)$'s are
three-dimensional submanifolds.\\
$\bullet$ Alternatively working in an open subset ${\cal M}$ of ${\cal P}$
the Hamiltonian is}
\[
{\cal H}(q,p) = e + {(p^1_2 + p^2_1)^2 - 4p^1_1p^2_2\over 4r} 
-{1\over 4} {(p^1_2 - p^2_1)^2\over 2+r}.
\]
\subsection{Gauge theories}
In this Section we show how to adapt the Legendre correspondence for gauge theories.
Our approach here is the most naive one, based on a local trivialisation and
we discuss only the example of the Yang--Mills--Higgs action on a ``space-time'' ${\cal X}$.
\\

\noindent
Let $\mathfrak{G}$ be a smooth compact Lie group of dimension $r$ with unit $\1$
and $\mathfrak{g}$ be its Lie algebra. Let ${\cal X}$ be a smooth $n$-dimensional
manifold. We denote by $(x^1,\cdots ,x^n)$ local coordinates on ${\cal X}$.
Similarly we let $(\mathfrak{u}_1,\cdots ,\mathfrak{u}_r)$ be a basis of $\mathfrak{g}$.
For simplicity we treat only $\mathfrak{g}$-connections on a trivial bundle over ${\cal X}$. Then
using a reference connection $\nabla^0$ on such a bundle,
a Yang--Mills field $\nabla$ can be identified with the $\mathfrak{g}$-valued 1-form $A$ on
${\cal X}$ such that $\nabla = \nabla^0 + A$. In local coordinates we write
$A=A_\mu(x)dx^\mu = \mathfrak{u}_IA^I_\mu(x)dx^\mu$.
We may couple $A$ to a Higgs field $\varphi:{\cal X}\longrightarrow \Phi$, where $\Phi$
is a vector space on which $\mathfrak{G}$ is acting. We denote by
$F:=dA+A\wedge A = \mathfrak{u}_IF^I$ the curvature 2-form of $A$ and
by $\nabla \varphi := d\varphi + A\varphi$ the covariant derivative on Higgs fields.
We are given a Riemannian metric $\eta$ on ${\cal X}$. Then the
Yang--Mills--Higgs action can be written
\begin{equation}\label{8.0.ym}
{\cal YM}[A,\varphi]=
\int_{\cal X} \left( -{1\over 4}|F|^2 +
{1\over 2}|\nabla^A\varphi|^2 + V(\varphi)\right) \omega,
\end{equation}
where $\omega$ is the Riemannian volume form on ${\cal X}$. The Lagrangian density
here is computed using $\mathfrak{G}$-invariant metrics $h$ and $g$ on $\mathfrak{g}$
and $\Phi$ respectively and reads : $|F|^2 = \eta^{\mu\lambda}\eta^{\nu\sigma}h_{IJ}F^I_{\mu\nu}
F^J_{\lambda\sigma}$, where $F^I_{\mu\nu} = {\partial A^I_\nu\over \partial x^\mu} -
{\partial A^I_\mu\over \partial x^\nu} + [A_\mu,A_\nu]^I$ and
$|\nabla\varphi|^2 = \eta^{\mu\nu}g_{ij}\nabla_\mu\varphi^i \nabla_\nu\varphi^j$,
where $\nabla_\mu\varphi^i = {\partial \varphi^i\over\partial x^\mu} +
\left(A_\mu\varphi\right)^i$.\\

%
%

\noindent
Translating in the setting expounded at the beginning of this section,
any choice of a field $(A,\varphi)$ is equivalent to the data of an
$n$-dimensional submanifold $\Gamma$ in ${\cal M}:= \left(\mathfrak{g}\otimes T^*{\cal X}\right)\times \Phi$
which is a section of this fiber bundle over ${\cal X}$. We will denote by $(x,a,\phi)$ a point
in ${\cal M}$, where $a = \mathfrak{u}_Ia^I_\mu dx^\mu\in \mathfrak{g}\otimes T_x^*{\cal X}$
and $\phi\in \Phi$.\\

\noindent
Now let us look at the Legendre correspondence for the Yang--Mills--Higgs action. For simplicity
we restrict ourself to the de Donder-Weyl approach (note that for main
purposes this theory is sufficient, unless we would be interested in a modified action of the kind
${\cal YM}_\tau[A,\varphi]:= {\cal YM}[A,\varphi] + \tau \int_{\cal X}\hbox{tr}F\wedge F$).
The Poincar\'e--Cartan form reads
\[
\theta^{dDW} := e\omega\wedge \tau + 
p^\mu_id\phi^i\wedge \omega_\mu + \pi^{\mu\nu}_Ida^I_\mu \wedge \omega_\nu,
\]
where $(e,p^\mu_i, \pi^{\mu\nu}_I)$ are the coordinates on the dual first jet bundle
of the fiber bundle ${\cal M}$.
The Legendre transform gives the relations
\[
\pi^{\mu\nu}_I = \eta^{\mu\lambda}\eta^{\nu\sigma}h_{IJ}F^J_{\lambda\sigma}
\quad \hbox{and}\quad
p^\mu_i = \eta^{\mu\nu}g_{ij}\nabla_\nu\varphi^j.
\]
The Hamiltonian function is
\[
{\cal H}(x,a,\phi,e,p,\pi) = 
e - {1\over 4}|\pi|^2 + {1\over 2}|p|^2 + {1\over 2}\pi_I^{\mu\nu}[a_\mu,a_\nu]^I
- p^\mu_i\left( a_\mu\phi\right)^i - V(\phi),
\]
where $|\pi|^2:= \eta_{\mu\lambda}\eta_{\nu\sigma}h^{IJ}\pi_I^{\mu\nu}\pi_J^{\lambda\sigma}$
and $|p|^2:= \eta_{\mu\nu}g^{ij}p^\mu_ip^\nu_j$.
The multisymplectic form is $\Omega^{dDW}= d\theta^{dDW}$ where we can also write
\[
\theta^{dDW}= e\omega+d\phi^i\wedge p_i + da^I\wedge \pi_I,\quad
\hbox{with}\quad
p_i:= p^\mu_i\omega_\mu,\, a^I:= a^I_\mu dx^\mu \hbox{ and }
\pi_I:= - {1\over 2}\pi^{\mu\nu}_I\omega_{\mu\nu}.
\]
Examples of algebraic observable $(n-1)$-forms are $p_i$, $dx^\mu\wedge \pi_I$,
$\phi^i\omega_\mu$ and $a^I\wedge \omega_{\mu\nu}$. As we will see in section 5 we
can also make sense of observable 0-forms like for instance $\phi^i$,
observable 1-forms like $a^I$ and observable $(n-2)$-form 
like $\pi_I$. Then it is not difficult to see that
the $(n-1)$-form $p_i$ is canonically conjugate to the 0-form $\phi^i$
and the $(n-2)$-form $\pi_I$ is canonically conjugate to the 1-form $A^I$, where
the meaning of ``canonically conjugate'' will be precised in Section 5.5.\\

\noindent
If we wish to study more general gauge theories and in particular
connections on a non trivial bundle we need a more general and more covariant framework.
Such a setting can consist in viewing a connection as a $\mathfrak{g}$-valued
1-form $a$ on a principal bundle ${\cal F}$ over the space-time satisfying some
equivariance conditions (under some action of the group $\mathfrak{G}$).
Similarly the Higgs field, a section of an associated bundle, can
be viewed as an equivariant map $\phi$ on ${\cal F}$ with values in a fixed space.
Thus the pair $(a,\phi)$ can be pictured geometrically as a section $\Gamma$, i.e.\,a submanifold
of some fiber bundle ${\cal N}$ over ${\cal F}$, satisfying two kinds of constraints:
\begin{itemize}
\item
$\Gamma$ is contained in a submanifold ${\cal N}_{\mathfrak{g}}$ (a geometrical
translation of the constraints ``the restriction of $a_f$ to the subspace tangent to the
fiber ${\cal F}_f$ is $-dg\cdot g^{-1}$'') and
\item $\Gamma$ is invariant
by an action of $\mathfrak{G}$ on ${\cal N}$ which preserves ${\cal N}_{\mathfrak{g}}$.
\end{itemize}
Within this more abstract framework we are reduced to a situation similar to the one
studied in the beginning of this section, but we need to understand what are the consequence
of the two equivariance conditions. (In particular this will imply that there is a canonical
distribution of subspaces which is tangent to all pseudofibers).
This will be done in details in \cite{HeleinKouneiher1.1}.
In particular we compare this abstract point of view with the more naive one expounded
above.

\section{Multisymplectic manifolds}
We now set up a general framework extending the situation encountered in the previous Section.
\subsection{Definitions}
\begin{defi}\label{2.1.def1}
Let ${\cal M}$ be a differential manifold. Let $n\in \Bbb{N}$ be some positive integer.
A smooth $(n+1)$-form $\Omega$ on ${\cal M}$, is a {\em multisymplectic} form
if and only if
\begin{enumerate}
\item [(i)] $\Omega$ is non degenerate, i.e.\,$\forall m\in {\cal M}$,
$\forall \xi \in T_m{\cal M}$, if $\xi \iN \Omega_m = 0$, then $\xi = 0$
\item [(ii)] $\Omega$ is closed, i.e.\,$d\Omega = 0$.
\end{enumerate}
Any manifold ${\cal M}$ equipped with a multisymplectic form $\Omega$ will be called a {\em multisymplectic} manifold.
\end{defi}
\noindent
In the following, $N$ denotes the dimension of ${\cal M}$.
For any $m\in {\cal M}$ we define the set
\[
D^n_m{\cal M}:= \{X_1\wedge \cdots \wedge X_n\in \Lambda^nT_m{\cal M}/
X_1,\ldots , X_n\in T_m{\cal M}\},
\]
of {\em decomposable} $n$-vectors and denote by $D^n{\cal M}$ the associated bundle.

\begin{defi}\label{2.1.def2}
Let ${\cal H}$ be a smooth real valued funtion defined over a multisymplectic manifold
$({\cal M}, \Omega)$.
A Hamiltonian $n$-curve $\Gamma$ is a $n$-dimensional submanifold of ${\cal M}$ such that for
any $m\in \Gamma$, there
exists a $n$-vector $X$ in $\Lambda^nT_m\Gamma$ which satisfies
$$X\iN \Omega = (-1)^nd{\cal H}.$$
We denote by ${\cal E}^{\cal H}$ the set of all such Hamiltonian $n$-curves.... We shall
also write for all $m\in {\cal M}$, $[X]^{\cal H}_m:=\{X\in D^n_m{\cal M}/X\iN \Omega = (-1)^nd{\cal H}_m\}$.
\end{defi}
Note that a Hamiltonian $n$-curve is automatically oriented by the $n$-vector $X$
involved in the Hamilton equation. Remark also that
it may happen that no Hamiltonian $n$-curve exist. An example is
${\cal M}:=\Lambda^2T^{\star}\Bbb{R}^4$ with
$\Omega = \sum_{1\leq\mu<\nu\leq 4}dp_{\mu\nu}\wedge dq^\mu\wedge dq^\nu$ for the
case ${\cal H}(q,p)=p_{12}+p_{34}$. Assume that a Hamiltonian 2-curve $\Gamma$ would exist and
let $X:(t^1,t^2)\longmapsto X(t^1,t^2)$ be a parametrisation of $\Gamma$ such that
${\partial X\over \partial t^1}\wedge {\partial X\over \partial t^2}\iN \Omega = (-1)^2d{\cal H}$.
Then, denoting by $X_\mu:={\partial X\over \partial t^\mu}$, we would have
$dx^\mu\wedge dx^\nu(X_1,X_2) = {\partial {\cal H}\over \partial p_{\mu\nu}}$, which is equal
to $\pm 1$ if $\{\mu,\nu\}=\{1,2\}$ or $\{3,4\}$ and to 0 otherwise. But this would contradict the
fact that $X_1\wedge X_2$ is decomposable. Hence there is no Hamiltonian
2-curve in this case.\\

\noindent
{\bf Example 8} --- {\em The basic example is the Lepage--Dedecker multisymplectic
manifold $(\Lambda^nT^*{\cal N}, \Omega)$ studied in the previous section (see also the
next section). Other examples are all smooth submanifolds of $\Lambda^nT^*{\cal N}$
on which the restriction of $\Omega$ is non degenerate,
like for instance the de Donder--Weyl manifold.}\\
{\bf Example 9} --- {\em Another example (see also \cite{Rovelli})
is provided by the Palatini or the Ashtekar formulation of pure gravity in 4-dimensional space-time.
Let us describe the Riemannian (non Minkowskian) version of it.
We consider $\Bbb{R}^4$ equipped with its standard metric $\eta_{IJ}$ and with
the standard volume 4-form $\epsilon_{IJKL}$. Let
$\mathfrak{p}\simeq \left\{(a,v)\simeq \left(\begin{array}{cc}a & v\\ 0 & 1\end{array}\right)/
a\in so(4),v\in \Bbb{R}^4\right\}$ $\simeq so(4)\ltimes \Bbb{R}^4$
be the Lie algebra of the Poincar\'e group acting on $\Bbb{R}^4$. Now let ${\cal X}$ be a
4-dimensional manifold, the ``space-time'', and consider ${\cal M}:=\mathfrak{p}\otimes T^*{\cal X}$,
the fiber bundle over ${\cal X}$ of 1-forms with coefficients in $\mathfrak{p}$.
We denote by $(x,e,A)$ a point in ${\cal M}$, where $x\in {\cal X}$, $e\in \Bbb{R}^4\otimes T^*_x$ and
$A\in so(4)\otimes T^*_x$. We shall work is the open subset of ${\cal M}$ where $e$ is
rank 4 (so that the 4 components of $e$ define a coframe on $T_x{\cal X}$).
First using the canonical projection $\Pi:{\cal M}\longrightarrow {\cal X}$
one can define a $\mathfrak{p}$-valued 1-form $\theta^{\mathfrak{p}}$ on ${\cal M}$
(similar to the Poincar\'e--Cartan 1-form) by
\[
\forall (x,e,A)\in {\cal M}, \forall X\in T_{(x,e,A)}{\cal X}, \quad
\theta^{\mathfrak{p}}_{(x,e,A)}(X) := (e(\Pi^*X),A(\Pi^*X)).
\]
Denoting (for $1\leq I,J\leq 4$) by $T^I:\mathfrak{p}\longrightarrow \Bbb{R}$, 
$(a,v)\longmapsto v^I$ and by
$R^I_J:\mathfrak{p}\longrightarrow \Bbb{R}$, $(a,v)\longmapsto a^I_J$,
the coordinate mappings we can define a 4-form on ${\cal M}$ by
\[
\theta_{Palatini}:= {1\over 4!}\epsilon_{IJKL}\eta^{LN}(T^I\circ \theta^{\mathfrak{p}}) \wedge 
(T^J\circ \theta^{\mathfrak{p}})\wedge \left(R^K_N\circ d\theta^{\mathfrak{p}}+
(R^K_M\circ \theta^{\mathfrak{p}}) \wedge (R^M_N\circ \theta^{\mathfrak{p}})\right).
\]
Now consider any section of ${\cal M}$
over ${\cal X}$. Write it as $\Gamma:=\{(x,e_x,A_x)/x\in {\cal X}\}$ where now
$e$ and $A$ are {\em 1-forms} on $x$ (and not coordinates anymore). Then
\[
\int_\Gamma \theta_{Palatini} = \int_{\cal X}{1\over 4!}\epsilon_{IJKL}\eta^{LN}
e^I\wedge e^J\wedge F^K_L,
\]
where $F^I_J:= dA^I_J + A^I_K\wedge A^K_J$ is the curvature of the connection 1-form $A$. We recognize the
Palatini action for pure gravity in 4 dimensions: this functional has the property
that a critical point of it provides us with
a solution of Einstein gravity equation $R_{\mu\nu} -{1\over 2}g_{\mu\nu} = 0$ by setting
$g_{\mu\nu}:= \eta_{IJ}e^I_\mu e^J_\nu$. By following the same steps as in the proof of Theorem
\ref{2.2.4.theo} one proves that a 4-dimensional submanifold $\Gamma$ which is a critical point of this action,
satisfies the Hamilton equation $X\iN \Omega_{Palatini} = 0$, where $\Omega_{Palatini}:= d\theta_{Palatini}$. Thus
$({\cal M},\Omega_{Palatini})$ is a multisymplectic manifold naturally associated to gravitation.
In the above construction, by replacing $A$ and $F$ by their self-dual parts $A_+$ and
$F_+$ (and so reducing the gauge group to $SO(3)$) one obtains the Ashtekar action.\\
Remark also that a similar construction can be done for the Chern--Simon action in dimension 3.}
\begin{defi}\label{2.1.def3}
A {\em symplectomorphism} $\phi$ of a multisymplectic manifold $({\cal M},\Omega)$
is a smooth diffeomorphism $\phi:{\cal M}\longrightarrow {\cal M}$ such that
$\phi^*\Omega = \Omega$. An {\em infinitesimal symplectomorphism} is a vector
field $\xi\in \Gamma({\cal M},T{\cal M})$ such that $L_\xi\Omega =0$.
We denote by $\mathfrak{sp}_0{\cal M}$ the set of infinitesimal symplectomorphisms
of $({\cal M}, \Omega)$.
\end{defi}
Note that, since $\Omega$ is closed, $L_\xi\Omega=d(\xi\iN \Omega)$, so that a vector field
$\xi$ belongs to $\mathfrak{sp}_0{\cal M}$ if and only if $d(\xi\iN \Omega)=0$. Hence
if the homology group $H^n({\cal M})$ is trivial there exists an $(n-1)$-form $F$ on
${\cal M}$ such that $dF + \xi\iN \Omega = 0$: such an $F$ is then an algebraic observable $(n-1)$-form
(see Section 3.3).\\

\subsection{Observable $(n-1)$-forms}
We now define the concept of observable
$(n-1)$-forms $F$.
The idea is that given a point $m\in {\cal M}$
and a Hamiltonian function ${\cal H}$, if $X(m)\in [X]^{\cal H}_m$, then
$\langle X(m),dF_m\rangle$
should not depend on the choice of $X(m)$ but only on $d{\cal H}_m$.

\subsubsection{Definitions}
\begin{defi}\label{3.2.1.def}
Let $m\in {\cal M}$ and $a\in \Lambda^nT^{\star}_m{\cal M}$; $a$ is called a {\em copolar}
$n$-form if and only if there exists an open dense subset
${\cal O}^a_m{\cal M}\subset D^n_m{\cal M}$ such that
\[
\forall X,\tilde{X}\in {\cal O}^a_m{\cal M},\quad
X\iN \Omega = \tilde{X}\iN \Omega\ \Longrightarrow \ a(X) = a(\tilde{X}).
\]
We denote by $P^n_mT^{\star}{\cal M}$ the set of copolar $n$-forms at $m$.
A $(n-1)$-form $F$ on ${\cal M}$ is called {\em observable} if and only if for every
$m\in {\cal M}$, $dF_m$ is copolar i.e.\,$dF_m\in P^n_mT^{\star}{\cal M}$.
We denote by $\mathfrak{P}^{n-1}{\cal M}$ the set of observable $(n-1)$-forms on
${\cal M}$.
\end{defi}
{\bf Remark} --- (i) The reason for the terminology ``copolar'' will become clear
in Section 5.\\
(ii) For any $m\in {\cal M}$, $P^nT^{\star}_m{\cal M}$ is a vector space (in particular
if $a,b\in P^nT^{\star}_m{\cal M}$ and $\lambda,\mu\in \Bbb{R}$ then $\lambda a + \nu b\in 
P^nT^{\star}_m{\cal M}$ and we can choose ${\cal O}^{\lambda a + \nu b}_m{\cal M}
= {\cal O}^a_m{\cal M}\cap {\cal O}^b_m{\cal M}$) and so
it is possible to construct a basis $(a_1,\cdots ,a_r)$ of it. Note also that for any
$a\in P^nT^{\star}_m{\cal M}$ we can write $a = t^1a_1+\cdots +t^ra_r$ which
implies that we can choose ${\cal O}^a_m{\cal M} = \cap_{s=1}^r{\cal O}^{a_s}_m{\cal M}$.
So having choosing such a basis $(a_1,\cdots ,a_r)$ we will denote by
${\cal O}_m{\cal M}:=\cap_{s=1}^r{\cal O}^{a_s}_m{\cal M}$ (it is still open and dense in $D^n_m{\cal M}$)
and in the following we will replace ${\cal O}^a_m{\cal M}$ by ${\cal O}_m{\cal M}$ in the above definition.
We will also denote by ${\cal O}{\cal M}$ the associated bundle.

\begin{lemm}\label{3.2.1.lemme}
Let $\phi:{\cal M}\longrightarrow {\cal M}$ be a symplectomorphism and
$F\in \mathfrak{P}^{n-1}{\cal M}$. Then $\phi^*F\in \mathfrak{P}^{n-1}{\cal M}$.
As a corollary, if $\xi\in \mathfrak{sp}_0{\cal M}$ (i.e.\,is an infinitesimal symplectomorphism)
and $F\in \mathfrak{P}^{n-1}{\cal M}$, then $L_\xi F\in \mathfrak{P}^{n-1}{\cal M}$.
\end{lemm}
{\em Proof} --- For any $n$-vector fields $X$ and $\widetilde{X}$, which are sections
of ${\cal OM}$, and for any $F\in \mathfrak{P}^{n-1}{\cal M}$,
$$X\iN \Omega = \widetilde{X}\iN \Omega \Longleftrightarrow
X\iN \phi^*\Omega = \widetilde{X}\iN \phi^*\Omega \Longleftrightarrow
(\phi_*X)\iN \Omega = (\phi_*\widetilde{X})\iN \Omega$$
implies
$$dF(\phi_*X) = dF(\phi_*\widetilde{X}) \Longleftrightarrow 
\phi_*dF(X) = \phi_*dF(\widetilde{X}) \Longleftrightarrow
d(\phi_*F)(X) = d(\phi_*F)(\widetilde{X}).$$
Hence $\phi_*F\in \mathfrak{P}^{n-1}{\cal M}$. \bbox

\noindent Assume that a given Hamiltonian function ${\cal H}$ on ${\cal M}$ is such that
$[X]^{\cal H}_m\subset {\cal O}_m{\cal M}$.
Then we shall say that ${\cal H}$ is {\em admissible}. If ${\cal H}$ is so, we
define the {\em pseudobracket} for all observable $(n-1)$-form $F\in \mathfrak{P}^{n-1}{\cal M}$
$$\{{\cal H},F\} := X\iN dF = dF(X),$$
where $X$ is any $n$-vector in $[X]^{\cal H}_m$.
Remark that, using the same notations as in section 2.3.1, if ${\cal H}(x,u,e,p^*) = e + H(x,u,p^*)$,
then $\{{\cal H},x^1dx^2\wedge \cdots  \wedge dx^n\} = 1$.

\subsubsection{Dynamics equation using dynamical brackets}
Our purpose here is to generalize the classical well-known relation $dF/dt = \{H,F\}$ of the classical
mechanics.
\begin{prop}\label{3.2.2.propdyn}
Let ${\cal H}$ be a smooth admissible Hamiltonian on ${\cal M}$ and $F$, $G$ two observable
$(n-1)$-forms with ${\cal H}$. Then $\forall \Gamma\in {\cal E}^{\cal H}$,
$$\{{\cal H},F\}dG_{|\Gamma} = \{{\cal H},G\}dF_{|\Gamma}.$$
\end{prop}
{\em Proof} --- This result is equivalent to proving that, if $X\in D_m^n{\cal M}$
is different of 0 and is tangent to $\Gamma$ at $m$, then
\begin{equation}\label{3.2.1.FdG}
\{{\cal H},F\}dG(X) = \{{\cal H},G\}dF(X).
\end{equation}
Note that by rescaling, we can assume w.l.g.\,that $X\iN \Omega = (-1)^nd{\cal H}$,
i.e.\,$X\in [X]^{\cal H}_m$. But then (\ref{3.2.1.FdG}) is equivalent to the obvious relation
$\{{\cal H},F\}\{{\cal H},G\} = \{{\cal H},G\}\{{\cal H},F\}$. \bbox

\noindent This result immediately implies the following result.
\begin{coro}\label{3.2.2.corodyn}
Let ${\cal H}$ be a smooth admissible Hamiltonian function on ${\cal M}$.
Assume that $F$ and $G$ are observable $(n-1)$-forms with ${\cal H}$ and
that $\{{\cal H},G\} = 1$ (see the remark at the end of Paragraph 3.2.1).
Then denoting $\omega:=dG$ we have:
$$\forall \Gamma\in {\cal E}^{\cal H},\quad \{{\cal H},F\}\omega_{|\Gamma} = dF_{|\Gamma}.$$
\end{coro}

\subsection{Algebraic observable $(n-1)$-forms}
\subsubsection{Definitions}
\begin{defi}\label{3.3.1.def}
Let $m\in {\cal M}$ and $a\in \Lambda^nT^{\star}_m{\cal M}$; $a$ is called {\em algebraic
copolar} if and only if
$$\forall X,\widetilde{X}\in \Lambda^nT_m{\cal M},\quad
X\iN \Omega = \widetilde{X}\iN \Omega\quad \Longrightarrow \quad a(X)=a(\widetilde{X}).$$
We denote by $P^n_0T^{\star}_m{\cal M}$ the set of algebraic copolar $n$-forms.\\
\noindent
A $(n-1)$-form $F$ on $({\cal M}, \Omega)$ is called {\em algebraic observable}
$(n-1)$-form if and only if for all
$m\in {\cal M}$, $dF_m\in P^n_0T^{\star}_m{\cal M}$.
We denote by $\mathfrak{P}^{n-1}_0{\cal M}$ the set of all algebraic observable $(n-1)$-forms.
\end{defi}
We invite the reader to compare this definition with definition \ref{3.2.1.def}:
the requirements of definition \ref{3.3.1.def} are stronger
than those of definition \ref{3.2.1.def}. Hence
$\mathfrak{P}^{n-1}_0{\cal M}\subset \mathfrak{P}^{n-1}{\cal M}$. In general the converse inclusion
is false.
\begin{defi}
A multisymplectic manifold $({\cal M}, \Omega)$ is {\em pataplectic} if and only if the set of
observable $(n-1)$-forms coincides with the set of algebraic observable $(n-1)$-forms,
i.e.\,$\mathfrak{P}^{n-1}_0{\cal M}= \mathfrak{P}^{n-1}{\cal M}$.
\end{defi}

\noindent
We shall see in Paragraph 3.3.3 that the multisymplectic manifold corresponding
to the de Donder--Weyl theory is not pataplectic (if $k\geq 2$).
But any open subset of $\Lambda^nT^*{\cal N}$ is pataplectic, as proved in Section 4.3.
(Moreover we shall also characterize completely
the set of algebraic observable $(n-1)$-forms in Section 4.4.) \\

\noindent
We remark also that any $(n-1)$-form $F$ such that there exists a vector field $\xi$ satisfying
$dF+\xi\iN \Omega = 0$ is algebraic observable, for then $X\iN \Omega = \widetilde{X}\iN \Omega$ implies
$dF(X) = -\xi\iN \Omega(X) = - (-1)^nX\iN \Omega(\xi) = - (-1)^n\widetilde{X}\iN \Omega(\xi) = dF(\widetilde{X})$.
The converse is true:
\begin{lemm}\label{3.3.1.lemme}
Let $m\in {\cal M}$ and $\phi\in P^n_0T^{\star}_m{\cal M}$.
Then there exists a unique $\xi\in T_m{\cal M}$ such that
$\phi + \xi\iN \Omega =0$.\\
\noindent 
As a corollary, if $F$ is an algebraic observable $(n-1)$-form,
then there exists an unique tangent vector field
$\xi_F$ on ${\cal M}$ such that $dF+\xi_F\iN \Omega = 0$ everywhere.
\end{lemm}
{\em Proof} --- Let us fix some point $m\in {\cal M}$ and define the equivalence
relation $\sim$ in $\Lambda^nT_m{\cal M}$ by
\[
X\;\sim\;\widetilde{X}\quad \Longleftrightarrow \quad X\iN \Omega = \widetilde{X}\iN \Omega.
\]
Set $V_m:=\{(-1)^nX\iN \Omega/X\in \Lambda^nT_m{\cal M}\}\subset T^{\star}_m{\cal M}$
and consider the linear mapping
\[
\begin{array}{cccc}
L: & \Lambda^nT_m{\cal M}/\sim & \longrightarrow & V_m\\
 & [X] & \longmapsto & (-1)^nX\iN \Omega ,\end{array}
\]
where $[X]$ is the class of $X\in \Lambda^nT_m{\cal M}$ modulo $\sim$.
It is clear that this map is well defined, one-to-one and onto, hence a vector
space isomorphism. Also $\hbox{dim}V_n\leq \hbox{dim}T^{\star}_m{\cal M} = N$.\\

\noindent 
Now observe that for any vector $\xi\in T_m{\cal M}$ a linear form
$\alpha_\xi \in \left( \Lambda^nT_m{\cal M}\right) ^{\star}$ can be constructed by
$\Lambda^nT_m{\cal M}\ni X\longmapsto \alpha_\xi(X):= -\xi\iN \Omega(X)\in \Bbb{R}$.
This form is constant on each class modulo $\sim$, since
$X$ $\sim$ $\widetilde{X}$ implies
\[
\alpha_\xi(\widetilde{X}) = - \xi\iN \Omega(\widetilde{X})
= -(-1)^n\widetilde{X}\iN \Omega(\xi) = -(-1)^nX\iN \Omega(\xi)
= - \xi\iN \Omega(X)
= \alpha_\xi(X).
\]
This hence defines the following linear map
\[
\begin{array}{cccccc}
T_\xi: & T_m{\cal M} & \longrightarrow & \Lambda^nT^{\star}{\cal M} & \longrightarrow &
\left( \Lambda^nT{\cal M}/\sim\right) ^{\star}\\
& \xi & \longmapsto & - \xi\iN \Omega & \longmapsto & \left[ [X]\longmapsto  \alpha_\xi(X)\right] .
\end{array}
\]
The linear map $T_\xi$ is also one-to-one because $\Omega$ is non degenerate.
But (recall that $\hbox{dim}T_m{\cal M}=N$)
on the other hand the dimension of its target space is
\[
\hbox{dim}\left( \Lambda^nT{\cal M}/\sim\right) ^{\star}
= \hbox{dim}\left( \Lambda^nT{\cal M}/\sim\right)
= \hbox{dim}V_m \leq \hbox{dim}T^{\star}_m{\cal M} =N.
\]
So we deduce that
$\hbox{dim}\left( \Lambda^nT{\cal M}/\sim\right) ^{\star}=N$ and
$V_m=T^{\star}_m{\cal M}$. Hence $T_\xi$ is in fact a vector space isomorphism.
Now we can restate the hypothesis of the Lemma by saying that
$X\longmapsto \phi(X)$ belongs to
$\left( \Lambda^nT{\cal M}/\sim\right) ^{\star}$, so that we can represent
this linear form by an unique $\xi\in T_m{\cal M}$ as claimed. \bbox

\subsubsection{Poisson brackets between observable $(n-1)$-forms}
There is a natural way to construct a Poisson bracket
$\{.,.\}:\mathfrak{P}^{n-1}_0{\cal M}$ $\times$
$\mathfrak{P}^{n-1}_0{\cal M}$
$\longmapsto$ $\mathfrak{P}^{n-1}_0{\cal M}$. To each algebraic observable forms
$F,G \in \mathfrak{P}^{n-1}_0{\cal M}$ we associate first the vector fields
$\xi_F$ and $\xi_G$ such that $\xi_F\iN \Omega + dF = \xi_G\iN \Omega + dG = 0$ and then
the $(n-1)$-form
$$\{F,G\} := \xi_F\wedge \xi_G\iN \Omega.$$
It can be shown (see \cite{HeleinKouneiher}) that
$\{F,G\}\in \mathfrak{P}^{n-1}_0{\cal M}$ and that
$$d\{F,G\} + [\xi_F,\xi_G]\iN \Omega = 0,$$
where $[.,.]$ is the Lie bracket on vector fields. Moreover this bracket
satisfies the Jacobi condition modulo an exact term\footnote{Note that in case where
the multisymplectic manifold $({\cal M},\Omega)$ is
{\em exact} in the sense of M. Forger, C. Paufler and H. R\"omer \cite{ForgerPauflerRomer},
i.e.\,if there exists an $n$-form $\theta$ such that $\Omega = d\theta$ (beware that our
sign conventions differ with \cite{ForgerPauflerRomer}), an alternative Poisson bracket can be defined:
\[
\{F,G\}_\theta:= \{F,G\} + d(\xi_G\iN F - \xi_F\iN G + \xi_F\wedge \xi_G\iN \theta).
\]
Then this bracket satisfies the Jacobi identity (in particular with a right hand side
equal to 0), see \cite{ForgerRomer}, \cite{ForgerPauflerRomer}.} (see \cite{HeleinKouneiher})
\[
\{\{F,G\},H\} + \{\{G,H\},F\} + \{\{H,F\},G\} =
d(\xi_F\wedge \xi_G\wedge \xi_H\iN \Omega).
\]
This bracket can be extended to forms in $\mathfrak{P}^{n-1}{\cal M}$ through
different strategies:
\begin{itemize}
\item By exploiting the relation
\[ 
\{F,G\} = \xi_F\iN dG = - \xi_G\iN dF,
\]
which holds for all $F,G\in \mathfrak{P}^{n-1}_0{\cal M}$. A natural definition
is to set: 
\[
\forall F\in \mathfrak{P}^{n-1}_0{\cal M},\,
\forall G\in \mathfrak{P}^{n-1}{\cal M},\quad
\{F,G\} = - \{G,F\} := \xi_F\iN dG.
\]
In \cite{HeleinKouneiher} we call this operation an {\em external} Poisson bracket.
\item If we know that there is an embedding $\iota :{\cal M}\longrightarrow \widehat{\cal M}$,
into a higher dimensional pataplectic manifold $(\widehat{\cal M},\widehat{\Omega})$, and that
(i) $\mathfrak{P}^{n-1}_0\widehat{\cal M} =\mathfrak{P}^{n-1}\widehat{\cal M}$,
(ii) the pull-back mapping $\mathfrak{P}^{n-1}_0\widehat{\cal M}\longrightarrow
\mathfrak{P}^{n-1}{\cal M}: \widehat{F}\longmapsto \iota^*\widehat{F}$ is --- modulo
the set of closed $(n-1)$-forms on $\widehat{\cal M}$ which vanish on ${\cal M}$ ---
an isomorphism. Then there exists a unique Poisson bracket on
$\mathfrak{P}^{n-1}{\cal M}$ which is the image of the Poisson bracket on
$\mathfrak{P}^{n-1}_0\widehat{\cal M}$.\\
This situation is achieved for instance if ${\cal M}$ is a submanifold
of $\Lambda^nT^*{\cal N}$, a situation which arises after a Legendre transform.
This will lead basically to the same structure as the external Poisson bracket. In more
general cases the question of extending ${\cal M}$ into $\widehat{\cal M}$ is relatively
subtle and is discussed in the paper \cite{HeleinKouneiher1.2}.

\end{itemize}

\subsubsection{Example of observable $(n-1)$-forms which are not algebraic observable $(n-1)$-forms}
In order to picture the difference between algebraic and non algebraic observable
$(n-1)$-forms, let us consider the example of the de Donder--Weyl theory
here corresponding to a submanifold of ${\cal M}=\Lambda^nT^{\star}(\Bbb{R}^n\times \Bbb{R}^k)$
(for $n,k\geq 2$) defined in Paragraph 2.3.1. We use the same notations as in Paragraph 2.3.1.
As a straightforward consequence of Lemma \ref{3.3.1.lemme}, the set $\mathfrak{P}^{n-1}_0{\cal M}^{dDW}$
of algebraic observable $(n-1)$-forms coincides with the set of $(n-1)$-forms $F$ on
${\cal M}^{dDW}$ such that, at each point $m\in {\cal M}^{dDW}$, $dF_m$ has the form
\[
dF_m = \left( a^\mu{\partial \over \partial x^\mu} + b^i{\partial \over \partial y^i}\right)
\iN \Omega + f\omega + f_i^\mu\omega^i_\mu
\]
(where we assume summation over all repeated indices).
Now we observe that, by the Pl\"ucker relations,
\[
\forall 1\leq p\leq n,\ \forall X\in D^n{\cal M}^{dDW},\quad
\left(\omega (X)\right)^{p-1}\omega^{i_1\cdots i_p}_{\mu_1\cdots \mu_p}(X) = 
\det \left( \omega^{i_\beta}_{\mu_\alpha}(X)\right) _{1\leq \alpha,\beta\leq p},
\]
so it turns out that, if $X\in D^n{\cal M}^{dDW}$ is such that $\omega(X)\neq 0$, then
all the values $\omega^{i_1\cdots i_p}_{\mu_1\cdots \mu_p}(X)$ can be computed from
$\omega(X)$ and $\left( \omega^i_\mu(X)\right) _{1\leq \mu\leq n;1\leq i\leq k}$.\\

\noindent 
Hence we deduce that the set of (non algebraic) observable $(n-1)$-forms on
${\cal M}^{dDW}$ contains the set of $(n-1)$-forms $F$ on ${\cal M}^{dDW}$ such that, at each point
$m\in {\cal M}^{dDW}$, $dF_m$ has the form
\[
dF_m = \left( a^\mu{\partial \over \partial x^\mu} + b^i{\partial \over \partial y^i}\right)
\iN \Omega + \sum_{j=1}^n\sum_{i_1<\cdots <i_j}\sum_{\mu_1<\cdots <\mu_j}
f_{i_1\cdots i_p}^{\mu_1\cdots \mu_p}\omega^{i_1\cdots i_p}_{\mu_1\cdots \mu_p}.
\]
Let us denote by $\mathfrak{P}^{n-1}_0\Lambda^nT^*({\cal X}\times {\cal Y})_{|{\cal M}^{dDW}}$
this set. An equivalent definition could be that
$\mathfrak{P}^{n-1}_0\Lambda^nT^*({\cal X}\times {\cal Y})_{|{\cal M}^{dDW}}$
is the set of the restrictions of algebraic observable forms
$\widetilde{F}\in \mathfrak{P}^{n-1}_0\Lambda^nT^*({\cal X}\times {\cal Y})$ on
${\cal M}^{dDW}$ (and this is the reason for this notation). Hence
$\mathfrak{P}^{n-1}{\cal M}^{dDW}\supset 
\mathfrak{P}^{n-1}_0\Lambda^nT^*({\cal X}\times {\cal Y})_{|{\cal M}^{dDW}}$. We
will see in Section 4.3 that the reverse inclusion holds, so that actually
$\mathfrak{P}^{n-1}{\cal M}^{dDW} =
\mathfrak{P}^{n-1}_0\Lambda^nT^*({\cal X}\times {\cal Y})_{|{\cal M}^{dDW}}$, with
${\cal O}_m{\cal M}^{dDW} = \{ X\in D^n_m{\cal M}^{dDW}/\omega(X)\neq 0\}$.

\subsubsection{Observable functionals}
The physical observed quantities should correspond to functionals on the set
${\cal E}^{\cal H}$. The simplest way to
obtain these is to integrate an observable $(n-1)$-form over a submanifold of codimension $1$
of a Hamiltonian $n$-curve. Since we shall need later to integrate forms of degree
$p-1$, where $1\leq p\leq n$, we give a more general definition.
\begin{defi}\label{2.1.def40}
Let ${\cal H}$ be a smooth real valued funtion defined over a multisymplectic manifold
$({\cal M}, \Omega)$. A {\em slice of codimension} $n-p+1$ is a cooriented
submanifold $\Sigma$ of ${\cal M}$ of codimension $n-p+1$ such that for any
$\Gamma\in {\cal E}^{\cal H}$, $\Sigma$ is transverse to $\Gamma$.
By {\em cooriented} we mean that for each $m\in \Sigma$, the quotient space $T_m{\cal M}/T_m\Sigma$ is
oriented continuously in function of $m$.
\end{defi}
{\bf Example 10} --- {\em Assume that ${\cal M} = \Lambda^nT^{\star}({\cal X}\times {\cal Y})$ and
that ${\cal H}(x,y,p) = e + H(x,y,p^*)$ as in Section 2.1.
Then the inverse image of any submanifold of ${\cal X}$ by the projection
mapping ${\cal M}\longrightarrow {\cal X}$, $(x,y,p)\longmapsto x$
is a slice of codimension 1. For instance if $t:{\cal M}\longrightarrow \Bbb{R}$ is a smooth
function which depends only on $x$ and such that $dt\neq 0$ everywhere
(a time coordinate), then any level set of $t$ is a codimension 1
slice (constant time hypersurface) and a (class of) vector $\tau\in T_m{\cal M}/T_m\Sigma$
is positively oriented if and only if $dt(\tau)>0$.}\\

\noindent 
Using a slice $\Sigma$ of codimension $1$ and an observable $(n-1)$-form $F$
we can define the {\em observable functional} denoted symbolically by
$\int_{\Sigma}F:{\cal E}^{\cal H}\longmapsto \Bbb{R}$ by:
$$\Gamma\quad \longmapsto \int_{\Sigma\cap \Gamma} F.$$
Here the intersection $\Sigma\cap \Gamma$ is oriented as follows: assume that
$\alpha\in T_m^{\star}{\cal M}$ is such that $\alpha_{|T_m\Sigma}=0$ and $\alpha>0$
on $T_m{\cal M}/T_m\Sigma$ and let $X\in \Lambda^nT_m\Gamma$ be positively oriented.
Then we require that $X\Ni \alpha\in \Lambda^{n-1}T_m(\Sigma\cap \Gamma)$ is positively
oriented.\\

\noindent
Lastly, for any slice $\Sigma$ of codimension 1 we can define a Poisson bracket
between the observable functionals $\int_{\Sigma}F$ and $\int_{\Sigma}G$ by
$\forall \Gamma \in {\cal E}^{\cal H}$,
$$\left\{ \int_{\Sigma}F, \int_{\Sigma}G\right\} (\Gamma) :=
\int_{\Sigma\cap \Gamma}\{F,G\}.$$
If $\partial \Gamma = \emptyset$, it is clear that this Poisson bracket satisfies the
Jacobi identity. Computations in \cite{Kijowski1}, \cite{Kanatchikov1}, \cite{HeleinKouneiher}
show that this Poisson bracket coincides with the Poisson bracket of
the standard canonical formalism used for quantum field theory.

\subsubsection{Pataplectic invariant Hamiltonian functions}
We have seen in section 2 the role of the pseudofibers inside $\Lambda^nT^*{\cal N}$
in the Legendre correspondence. We shall prove in Section 4.5 that the tangent spaces
to these pseudofibers can be characterised intrinsically. This motivates the following definition.\\

\noindent
For all Hamiltonian function ${\cal H}:{\cal M}\longrightarrow \Bbb{R}$ and for all
$m\in {\cal M}$ we define the {\em generalized pseudofiber direction} to be
\begin{equation}\label{3.3.4.lh}
\begin{array}{ccl}
L^{\cal H}_m & := & \{\xi \in T_m{\cal M}/\forall X\in [X]^{\cal H}_m,
\forall \delta X\in T_XD^n_m{\cal M}, \xi\iN \Omega(\delta X) = 0\}\\
 & = & \displaystyle \left(T_{[X]_m^{\cal H}}D^n_m{\cal M}\iN \Omega\right)^{\perp}.
\end{array}
\end{equation}
And we write $L^{\cal H}:=\cup_{m\in {\cal M}}L^{\cal H}_m\subset T{\cal M}$ for
the associated bundle.
The next lemma illustrates this definition. In the following if $\xi$ is a
smooth vector field, we denote by $e^{s\xi}$ (for $s\in I$, where $I$ is an
interval of $\Bbb{R}$) its
flow mapping. And if $E$ is any subset of ${\cal M}$, we denote by
$E_s:=e^{s\xi}(E)$ its image by $e^{s\xi}$.
\begin{lemm}\label{3.3.4.lem}
Let $\Gamma\in {\cal E}^{\cal H}$ be a Hamiltonian $n$-curve and $\xi$ be a vector field
which is a smooth section of $L^{\cal H}$. Suppose that, for all $s\in I$,
$\Gamma_s$ is a Hamiltonian $n$-curve. Let $\Sigma$ be a smooth
$(n-1)$-dimensional submanifold of $\Gamma$ and $F\in \mathfrak{P}^{n-1}_0{\cal M}$.
If one of the two following hypotheses is satisfied: either\\
(a) $\partial \Sigma =\emptyset$, or\\
(b) $\xi\iN F= 0$ everywhere, then
\begin{equation}\label{3.3.4.inv}
\forall s\in I,\quad
\int_{\Sigma}F = \int_{\Sigma_s}F.
\end{equation}
i.e.\,the integral of $F$ on the image of $\Sigma$ by $e^{s\xi}$
does not depend on $s$.
\end{lemm}
\begin{figure}[h]
\begin{center}
\includegraphics[scale=1]{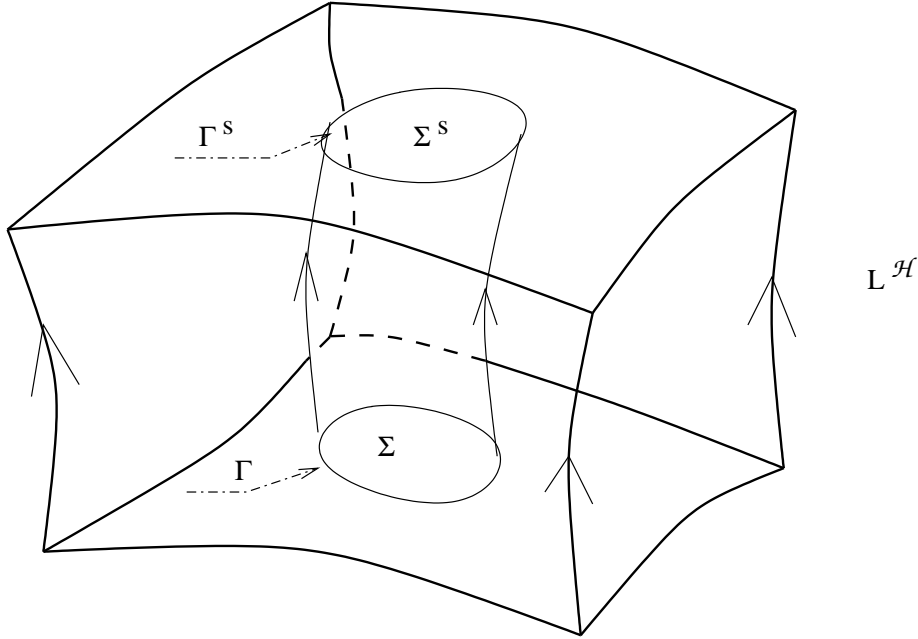}
\caption{Invariance of an observable functional along the generalized pseudofiber directions as in
Lemma \ref{3.3.4.lem}}
\end{center}
\end{figure}
{\em Proof} --- Let us introduce some extra notations:
$\sigma:I\times \Gamma\longrightarrow {\cal M}$ is
the map $(s,m)\longmapsto \sigma(s,m):=e^{s\xi}(m)$.
Moreover for all $m\in \Sigma_s\cup \partial \Sigma_s$ we consider
a basis $(X_1,\cdots ,X_n)$ of $T_m\Gamma_s$ such that
$X:=X_1\wedge \cdots \wedge X_n\in [X]^{\cal H}_m$, $(X_2,\cdots ,X_n)$
is a basis of $T_m\Sigma_s$ and, if $m\in \partial \Sigma_s$, $(X_3,\cdots ,X_n)$
is a basis of $T_m\partial \Sigma_s$.
Lastly we let $(\theta^1,\cdots ,\theta^n)$ be a basis of $T^*_m\Gamma_s$, dual
of $(X_1,\cdots ,X_n)$. We first note that
\[
\int_{(0,s)\times \partial \Sigma}\sigma^*F = 0,
\]
either because $\partial \Sigma=\emptyset$ (a) or because, if $\partial \Sigma\neq \emptyset$,
this integral is equal to
\[
\int_{ (0,s)\times \partial \Sigma} F_{\sigma(s,m)}
(\xi,X_3,\cdots ,X_n)ds\wedge \theta^3\wedge \cdots \wedge \theta^n
\]
which vanishes by (b). Thus
\[
\begin{array}{ccl}
\displaystyle 
\int_{\Sigma_s}F - \int_{\Sigma}F  = \int_{\Sigma}\left(e^{s\xi}\right)^*F-F
& = & \displaystyle \int_{\Sigma}\left(e^{s\xi}\right)^*F-F
- \int_{(0,s)\times \partial \Sigma}\sigma^*F\\
\displaystyle =\ 
\int_{\partial \left( (0,s)\times \Sigma\right)} \sigma^*F & = &
\displaystyle \int_{ (0,s)\times \Sigma} d\left(\sigma^*F \right)\\
\displaystyle =\ \int_{ (0,s)\times \Sigma} \sigma^*dF & = &
\displaystyle \int_{ \sigma((0,s)\times \Sigma)} dF_{\sigma(s,m)}
(\xi,X_2,\cdots ,X_n)ds\wedge \theta^2\wedge \cdots \wedge \theta^n.
\end{array}
\]
But since $F\in \mathfrak{P}^{n-1}_0{\cal M}$, we have that
\[
\begin{array}{ccc}
dF_{\sigma(s,m)}(\xi,X_2,\cdots ,X_n) & = & - \Omega(\xi_F,\xi,X_2,\cdots ,X_n)\\
& = & \Omega(\xi,\xi_F,X_2,\cdots ,X_n)\\
& =& \langle \xi_F\wedge X_2\wedge \cdots \wedge X_n,\xi\iN \Omega \rangle.
\end{array}
\]
Now the key observation is that $\xi_F\wedge X_2\wedge \cdots \wedge X_n\in T_XD^n_m{\cal M}$
and so the hypothesis of the Lemma implies that
$\langle \xi_F\wedge X_2\wedge \cdots \wedge X_n,\xi\iN \Omega \rangle=0$.... Hence
(\ref{3.3.4.inv}) is satisfied. \bbox
\noindent
The above result leads to the question whether the image of a Hamiltonian $n$-curve
by the flow of a vector fields with values in $L^{\cal H}$ is also a Hamiltonian
$n$-curve. This motivates the following definition.
\begin{defi}\label{3.3.4.def}
We say that ${\cal H}$ is {\em pataplectic invariant} if
\begin{itemize}
\item $\forall \xi\in L^{\cal H}_m$, $d{\cal H}_m(\xi)=0$
\item for all Hamiltonian $n$-curve $\Gamma\in {\cal E}^{\cal H}$, for all
vector field $\xi$ which is a smooth section of $L^{\cal H}$, then, 
for $s\in \Bbb{R}$ sufficiently small, $\Gamma_s:=e^{s\xi}(\Gamma)$ is also
a Hamiltonian $n$-curve.
\end{itemize}
\end{defi}
We shall prove in Section 4.5 (Theorem \ref{4.4.theo}) that, if ${\cal M}$ is
an open subset of $\Lambda^nT^*{\cal N}$, any Hamiltonian function on ${\cal M}$
obtained by means of a Legendre correspondence from a Lagrangian problem is
pataplectic invariant. Another consequence of Lemma \ref{3.3.4.lem} and of
Theorem \ref{4.4.theo} will be derived in Proposition \ref{4.4.prop2}, in Section 4.5.

\section{The study of $\Lambda^nT^{\star}{\cal N}$}
In this section we analyze in details the special case where
${\cal M}$ is an open subset of $\Lambda^nT^{\star}{\cal N}$. Since we are
interested here in local properties of ${\cal M}$, we will use local
coordinates $m=(q,p)=(q^\alpha,p_{\alpha_1\cdots \alpha_n})$ on ${\cal M}$, and the
multisymplectic form reads $\Omega=\sum_{\alpha_1<\cdots <\alpha_n}dp_{\alpha_1\cdots \alpha_n}\wedge
dq^{\alpha_1}\wedge \cdots  \wedge dq^{\alpha_n}$.
For $m=(q,p)$, we write
\[
d_q{\cal H}:= \sum_{1\leq \alpha\leq n+k}{\partial {\cal H}\over \partial q^\alpha}dq^\alpha,\quad
d_p{\cal H}:= \sum_{1\leq \alpha_1<\cdots <\alpha_n\leq n+k}
{\partial {\cal H}\over \partial p_{\alpha_1\cdots \alpha_n}}dp_{\alpha_1\cdots \alpha_n},
\]
so that $d{\cal H} = d_q{\cal H} + d_p{\cal H}$.
\subsection{The structure of $[X]^{\cal H}_m$}
Here we are given some Hamiltonian function ${\cal H}:{\cal M}\longrightarrow \Bbb{R}$ and
a point $ m\in {\cal M}$ such that $[X]^{\cal H}_m\neq \emptyset$ and\footnote{observe that,
although the splitting $d{\cal H} = d_q{\cal H} + d_p{\cal H}$ depends on a trivialisation
of $\Lambda^nT^*{\cal N}$, the condition $d_p{\cal H}_m \neq 0$ is intrinsic: indeed it is
equivalent to $d{\cal H}_{m|\hbox{Ker}d\Pi_m} \neq 0$, where $\Pi:\Lambda^nT^*{\cal N}
\longrightarrow {\cal N}$.} $d_p{\cal H}_m\neq 0$.
Given any $X=X_1\wedge \cdots \wedge X_n\in D^n_m{\cal M}$ and any form $a\in T^*_m{\cal M}$
we will write that $a_{|X}\neq 0$ (resp. $a_{|X}= 0$) if and only if
$(a(X_1),\cdots ,a(X_n))\neq 0$ (resp. $(a(X_1),\cdots ,a(X_n))= 0$).
We will say that a form $a\in T^*_m{\cal M}$ is {\em proper on} $[X]^{\cal H}_m$
if and only if we have either 
\begin{equation}\label{4.1/2.differe}
\forall X\in [X]^{\cal H}_m,\quad a_{|X} \neq 0,
\end{equation}
or
\begin{equation}\label{4.1/2.nul}
\forall X\in [X]^{\cal H}_m,\quad a_{|X} = 0.
\end{equation}
In the first case (\ref{4.1/2.differe}) we call $a$ a {\em point-slice}.
We are interested in characterizing all proper 1-forms on $[X]^{\cal H}_m$.
We show in this section the following.
\begin{lemm}\label{4.1/2.lemme}
Let ${\cal M}$ be an open subset of $\Lambda^nT^{\star}{\cal N}$ endowed with its
standard multisymplectic form $\Omega$, let ${\cal H}:{\cal M}\longrightarrow \Bbb{R}$
be a smooth Hamiltonian function. Let $m\in {\cal M}$ such that $d_p{\cal H}_m\neq 0$ and
$[X]^{\cal H}_m\neq \emptyset$. Then\\
(i) the $n+k$ forms $dq^1,\cdots ,dq^{n+k}$
are proper on $[X]^{\cal H}_m$ and satisfy the following property:
$\forall X\in [X]^{\cal H}_m$ and
for all $Y,Z\in T_m{\cal M}$ which are in the vector space spanned by $X$,
if $dq^\alpha(Y) = dq^\alpha(Z)$, $\forall \alpha=1,\cdots ,n+k$, then $Y=Z$.\\
(ii) Moreover for all $a\in T^*_m{\cal M}$ which is proper on $[X]^{\cal H}_m$ we have
\begin{equation}\label{4.1/2.base}
\exists ! \lambda\in \Bbb{R},
\exists !(a_1,\cdots ,a_{n+k})\in \Bbb{R}^{n+k},\quad 
a=\lambda d{\cal H}_m + \sum_{\alpha=1}^{n+k}a_\alpha dq^\alpha.
\end{equation}
(iii) Up to a change of coordinates on ${\cal N}$ we can assume that $dq^1,\cdots ,dq^n$
are point-slices and that $dq^{n+1},\cdots ,dq^{n+k}$ satisfy (\ref{4.1/2.nul}). Then
$a\in T^*{\cal M}$ is a point-slice if and only if (\ref{4.1/2.base}) occurs
with $(a_1,\cdots ,a_n)\neq 0$.

\end{lemm}
{\em Proof} --- {\em First step --- analysis of $[X]^{\cal H}_m$}.
We start by introducing some extra notations:
each vector $Y\in T_m{\cal M}$ can be decomposed
into a ``vertical'' part $Y^V$ and a ``horizontal'' part $Y^H$ as follows: for any
$Y=\sum_{1\leq \alpha\leq n+k}Y^\alpha {\partial \over \partial q^\alpha} +
\sum_{1\leq \alpha_1<\cdots <\alpha_n\leq n+k}
Y_{\alpha_1\cdots \alpha_n}{\partial \over \partial p_{\alpha_1\cdots \alpha_n}}$,
set $Y^H:= \sum_{1\leq \alpha\leq n+k}Y^\alpha {\partial \over \partial q^\alpha}$ and
$Y^V:= \sum_{1\leq \alpha_1<\cdots <\alpha_n\leq n+k} Y_{\alpha_1\cdots \alpha_n}
{\partial \over \partial p_{\alpha_1\cdots \alpha_n}}$. Let $X=X_1\wedge \cdots \wedge X_n
\in D^n_m\left( \Lambda^nT^{\star}{\cal N}\right)$ and let us use this decomposition
to each $X_\mu$: then $X$ can be split as $X=\sum_{j=0}^nX_{(j)}$,
where each $X_{(j)}$ is homogeneous of degree
$j$ in the variables $X_\mu^V$ and homogeneous of degree $n-j$ in the
variables $X_\mu^H$.\\

\noindent 
Recall that a decomposable $n$-vector $X$
is in $[X]^{\cal H}_m$ if and only if $X\iN \Omega = (-1)^nd{\cal H}$. This equation actually splits as 
\begin{equation}\label{3.1.dph}
X_{(0)}\iN \Omega = (-1)^nd_p{\cal H}
\end{equation}
and 
\begin{equation}\label{3.1.dqh}
X_{(1)}\iN \Omega = (-1)^nd_q{\cal H}.
\end{equation}
Equation (\ref{3.1.dph}) determines in an unique way $X_{(0)}\in D^n_q{\cal N}$. The condition 
$d_p{\cal H}\neq 0$ implies that necessarily\footnote{Note also that (\ref{3.1.dph}) implies
that $d_p{\cal H}$ must satisfy some compatibility conditions since $X_{(0)}$ is decomposable.}
$X_{(0)}\neq 0$. At this stage we can choose a family of $n$ linearly
independant vectors
$X^0_1,\cdots ,X^0_n$ in $T_q{\cal N}$ such that $X^0_1\wedge \cdots \wedge X_n^0=X_{(0)}$.
Thus the forms $dq^\alpha$ are proper on $[X]^{\cal H}_m$, since their
restriction on $X$ are fully determined by their restriction on the vector subspace
spanned by $X^0_1,\cdots ,X^0_n$.
Furthermore the subspace of $T_m{\cal M}$ spanned by $X$ is a graph over the subspace of
$T_q{\cal N}$ spanned by $X_{(0)}$. This proves the part (i) of the Lemma.\\

\noindent Proving (ii) and (iii) requires more work. First
we deduce that there exists a unique family
$(X_1,\cdots ,X_n)$ of vectors in $T_m{\cal M}$ such that $\forall \mu$,
$X_\mu^H=X_\mu^0$ and $X_1\wedge \cdots \wedge X_n=X$.
And Equation (\ref{3.1.dqh}) consists in further underdetermined
conditions on the vertical components $X_{\mu,\alpha_1\cdots \alpha_n}$
of the $X_\mu$'s, namely
\[
\sum_{\mu}\sum_{\alpha_1<\cdots <\alpha_n}C_\beta^{\mu,\alpha_1\cdots \alpha_n}X_{\mu,\alpha_1\cdots \alpha_n}
= - {\partial {\cal H}\over \partial q^\beta},
\]
where
\[
C_\beta^{\mu,\alpha_1\cdots \alpha_n}:= \sum_\nu \delta^{\alpha_\nu}_\beta(-1)^{\mu+\nu}
\Delta^{\alpha_1\cdots \widehat{\alpha_\nu}..\alpha_n}_{1\cdots \widehat{\mu}\cdots n}
\]
and
\[
\Delta^{\alpha_1\cdots \alpha_{n-1}}_{\mu_1\cdots \mu_{n-1}}:= \left|
\begin{array}{ccc}
X^{\alpha_1}_{\mu_1} & \dots & X^{\alpha_1}_{\mu_{n-1}}\\
\vdots & & \vdots \\
X^{\alpha_n}_{\mu_1} & \dots & X^{\alpha_{n-1}}_{\mu_{n-1}}
\end{array}
\right| .
\]
{\em Step2 --- Local coordinates}.
To further understand these relations we choose suitable coordinates $q^\alpha$
in such a way that $d_p{\cal H}_m=dp_{1\cdots n}$ and
\begin{equation}\label{3.1.dph1}
X_\mu^H={\partial \over \partial q^\mu}\quad \hbox{for}\quad \mu =1,...,n,
\end{equation}
so that (\ref{3.1.dph}) is automatically satisfied. In this setting we also have
$$(-1)^nX_{(1)}\iN \Omega = -\sum_\mu X_{\mu,1\cdots n}dq^\mu
- (-1)^n\sum_\mu \sum_{n<\beta}(-1)^\mu X_{\mu,1\cdots \widehat{\mu}\cdots n\beta}dq^\beta,$$
and so (\ref{3.1.dqh}) is equivalent to 
\begin{equation}\label{3.1.dqh1}
\left\{ \begin{array}{ccl}
X_{\mu,1\cdots n} & =  & \displaystyle -{\partial {\cal H}\over \partial q^\mu},\;
\hbox{for } 1\leq \mu \leq n\\
 & & \\
\displaystyle 
(-1)^n\sum_\mu  (-1)^\mu X_{\mu,1\cdots \widehat{\mu}\cdots n\beta} & = &
\displaystyle - {\partial {\cal H}\over \partial q^\beta},\;
\hbox{for } n+1\leq \beta \leq n+k.
\end{array}\right.
\end{equation}
Let us introduce some notations:
$I:=\{(\alpha_1,\cdots ,\alpha_n)/1\leq \alpha_1<\cdots \leq \alpha_n\leq n+k\}$,
$I^0:=\{(1,\cdots ,n)\}$,
$I^*:=\{(\alpha_1,\cdots ,\alpha_{n-1},\beta)/1\leq \alpha_1<\cdots <\alpha_{n-1}\leq n,
n+1\leq \beta\leq n+k\}$,
$I^{**}:=I\setminus \left( I^0 \cup I^*\right).$
We note also $M_\mu:= \sum_{(\alpha_1,\cdots ,\alpha_n)\in I^*}X_{\mu,\alpha_1\cdots \alpha_n}
\partial ^{\alpha_1\cdots \alpha_n}$,
$R_\mu:= \sum_{(\alpha_1,\cdots ,\alpha_n)\in I^{**}}X_{\mu,\alpha_1\cdots \alpha_n}
\partial ^{\alpha_1\cdots \alpha_n}$ and
$M^\nu_{\mu,\beta}:=(-1)^{n+\nu}X_{\mu,1\cdots \widehat{\nu}\cdots n\beta}$.
Then the set of solutions
of (\ref{3.1.dph}) and (\ref{3.1.dqh}) satisfying (\ref{3.1.dph1}) is
\begin{equation}\label{3.1.xmu}
X_\mu = {\partial \over \partial q^\mu} - {\partial {\cal H}\over \partial q^\mu}
{\partial \over \partial p_{1\cdots n}} + M_\mu +R_\mu,
\end{equation}
where the components of $R_\mu$ are arbitrary, and the coefficients of
$M_\mu$ are only subject to the constraint
\begin{equation}\label{3.1.constraint}
\sum_\mu M^\mu_{\mu,\beta} =
- {\partial {\cal H}\over \partial q^\beta},\quad \hbox{ for } n+1\leq \beta \leq n+k.
\end{equation}

\noindent {\em Step 3 --- The search of all proper 1-forms on $[X]^{\cal H}_m$}.
Now let $a\in T^*_m{\cal M}$ and let us look at
necessary and sufficient conditions for $a$ to be a proper 1-form on $[X]^{\cal H}_m$.
We write
\[
a = \sum_\alpha a_\alpha dq^\alpha + \sum_{\alpha_1<\cdots <\alpha_n}
a^{\alpha_1\cdots \alpha_n}dp_{\alpha_1\cdots \alpha_n}.
\]
Let us write $a^*:=
\left( a^{\alpha_1\cdots \alpha_n}\right)_{(\alpha_1,\cdots ,\alpha_n)\in I^*}$,
$a^{**}:=\left( a^{\alpha_1\cdots \alpha_n}\right)_{(\alpha_1,\cdots ,\alpha_n)\in I^{**}}$
and
\[
\left\langle M_\mu,a^*\right\rangle :=\sum_\nu\sum_{n<\beta}(-1)^{n+\nu}
M^\nu_{\mu,\beta}a^{1\cdots \widehat{\nu}\cdots n\beta},
\]
and
\[
\left\langle R_\mu,a^{**}\right\rangle :=
\sum_{(\alpha_1,\cdots ,\alpha_n)\in I^{**}}X_{\mu,\alpha_1\cdots \alpha_n}
a^{\alpha_1\cdots \alpha_n}.
\]
Using (\ref{3.1.xmu}) we obtain that
\[
a(X_\mu) = a_\mu -
{\partial {\cal H}\over \partial q^\mu}
a^{1\cdots n} +
\left\langle M_\mu,a_*\right\rangle +
\left\langle R_\mu,a_{**}\right\rangle .
\]
\begin{lemm}\label{4.1/2.lemme2}
Condition (\ref{4.1/2.differe}) (resp. (\ref{4.1/2.nul}))
is equivalent to the two following conditions:
\begin{equation}\label{3.1.nul}
a^* =a^{**} = 0
\end{equation}
and
\begin{equation}\label{3.1.nonnul}
\left( a_1 - {\partial {\cal H}\over \partial q^1} a^{1\cdots n},
\; \cdots \; ,
a_n - {\partial {\cal H}\over \partial q^n}
a^{1\cdots n} \right)
\neq 0 \quad \hbox{(resp. }= 0\hbox{)}.
\end{equation}
\end{lemm}
{\em Proof} --- We first look at necessary and sufficient conditions on for $a$
to be a point-slice, i.e.\,to satisfy (\ref{4.1/2.differe}).
Let us denote by $\vec{A}:=\left( a_\mu - {\partial {\cal H}\over \partial q^\mu}
a^{1\cdots n}\right)_\mu$ and
$\vec{M}:=\left( M_\mu\right)_\mu$, $\vec{R}:=\left( R_\mu\right)_\mu$.
We want conditions on $a^{\alpha_1\cdots \alpha_n}$
in order that the image of the affine map $(\vec{M},\vec{R})\longmapsto
\vec{\cal A}(\vec{M},\vec{R}):= \vec{A} +
\langle \vec{M},a_*\rangle + \langle \vec{R},a_{**}\rangle$ does not
contain 0 (assuming that $\vec{M}$ satisfies the constraint (\ref{3.1.constraint})).
We see immediately that if $a^{**}$
would be different from 0, then by choosing $\vec{M}=0$ and $\vec{R}$
suitably, we could have $\vec{\cal A}(\vec{M},\vec{R})=0$. Thus $a^{**}=0$.
Similarly, assume by contradiction that $a^*$ is different from 0. Up
to a change of coordinates, we can assume that
$\left( a^{1\cdots \widehat{\nu}\cdots n(n+1)}\right)_{1\leq \nu\leq n}\neq 0$.
And by another change of coordinates, we can further assume that
$a^{2\cdots n(n+1)} = \lambda \neq 0$ and
$a^{1\cdots \widehat{\nu}\cdots n(n+1)} = 0$, if $\nu\geq 1$.
Then choose $M^\nu_{\mu,\beta}= 0$ if $\beta\geq n+2$, and
\[
\left(\begin{array}{ccccc}
M^1_{1,n+1} & M^1_{2,n+1} & M^1_{3,n+1} & \cdots & M^1_{n,n+1} \\
M^2_{1,n+1} & M^2_{2,n+1} & M^2_{3,n+1} & \cdots & M^2_{n,n+1} \\
\vdots & \vdots & \vdots &  & \vdots \\
M^n_{1,n+1} & M^n_{2,n+1} & M^n_{3,n+1} & \cdots & M^n_{n,n+1}
\end{array}\right) =
\left(\begin{array}{ccccc}
t_1 & t_2 & t_3 & \cdots & t_n\\
0 & s & 0 & \cdots & 0\\
\vdots & \vdots & \vdots &  & \vdots \\
0 & 0 & 0 & \cdots & 0
\end{array}\right),
\]
where $s = - t_1 -\partial {\cal H}/\partial q^{n+1}$. Then we find that
${\cal A}_\mu(\vec{M},\vec{R}) = A_\mu +(-1)^{n+1}\lambda t_\mu$, so that this expression 
vanishes for a suitable choice of the $t_\mu$'s. Hence we get a contradiction. Thus
we conclude that $a^*=0$ and $\vec{A}\neq 0$. The analysis of 1-forms which
satisfies (\ref{4.1/2.nul}) is similar: this condition is equivalent to $a^*=0$ and
$\vec{A}=0$.
\bbox
\noindent
{\em Conclusion}. We translate the conclusion of Lemma \ref{4.1/2.lemme2} without
using local coordinates: it gives relation (\ref{4.1/2.base}). \bbox 

\subsection{Slices of codimension 1}

We consider a smooth function $f:{\cal M}\longrightarrow \Bbb{R}$,
we fix some $s\in \Bbb{R}$ and we are looking for necessary and sufficent conditions
for the level set $f^{-1}(s):=\{m\in {\cal M}/f(m)=s\}$ to be a slice of codimension $1$.
It just means that $\forall m\in f^{-1}(s)$, $df_m$ is a point-slice. Using Lemma
\ref{4.1/2.lemme2} we obtain two conditions on $df_m$:
the condition (\ref{3.1.nul}) can be restated as follows: for all $m\in {\cal M}$
there exists a real number $\lambda(m)$ such that $d_pf_m=\lambda(m)d_p{\cal H}_m$.
Condition (\ref{3.1.nonnul}) is equivalent to: $\exists (\alpha_1,\cdots ,\alpha_n)\in I$,
$\exists 1\leq \mu\leq n$,
\begin{equation}\label{4.2.supercrochet}
\{{\cal H},f\}^{\alpha_1\cdots \alpha_n}_{\alpha_\mu}(m) : = 
{\partial {\cal H}\over \partial p_{\alpha_1\cdots \alpha_n}}(m){\partial f\over \partial q^{\alpha_\mu}}(m)
-
{\partial f\over \partial p_{\alpha_1\cdots \alpha_n}}(m){\partial {\cal H}\over \partial q^{\alpha_\mu}}(m)
\neq 0.
\end{equation}
[Alternatively using Lemma \ref{4.1/2.lemme}, $df_m$ is a point-slice
if and only if $\exists \lambda(m)\in \Bbb{R}$, $\exists (a_1,\cdots ,a_{n+k})\in \Bbb{R}^{n+k}$
such that $df_m=\lambda(m)d{\cal H}_m + \sum_{\alpha=1}^{n+k}a_\alpha dq^\alpha$
and $(a_1,\cdots ,a_n)\neq 0$.]
Now we remark
that $d_pf_m=\lambda(m)d_p{\cal H}_m$ everywhere if and only if there exists a
function $\widehat{f}$ of the variables
$(q,h)\in {\cal N}\times \Bbb{R}$ such that $f(q,p)=\widehat{f}(q,{\cal H}(q,p))$.
So we deduce the following.
\begin{theo}\label{4.1/2.theo}
Let ${\cal M}$ be an open subset of $\Lambda^nT^{\star}{\cal N}$ endowed with its
standard multisymplectic form $\Omega$, let ${\cal H}:{\cal M}\longrightarrow \Bbb{R}$
be a smooth Hamiltonian function and let $f:{\cal M}\longrightarrow \Bbb{R}$ be
a smooth function. Assume that $d_p{\cal H}\neq 0$ and $[X]^{\cal H}\neq \emptyset$
everywhere. Then all level sets of $f$ are slices if and only if 
$\exists (q,h)\in {\cal N}\times \Bbb{R}$ such that $f(q,p)=\widehat{f}(q,{\cal H}(q,p))$
and $\forall m\in {\cal M}$, $\exists (\alpha_1,\cdots ,\alpha_n)\in I$,
$\exists 1\leq \mu\leq n$,
$\{{\cal H},f\}^{\alpha_1\cdots \alpha_n}_{\alpha_\mu}(m)\neq 0$.
\end{theo}
{\bf Example 11} --- {\em We come back here to the situation and the notations
expounded in Section 2.3, about the Legendre correspondence for maps
$u:{\cal X}\longrightarrow {\cal Y}$ which are critical points of a Lagrangian
functional $l$. Denoting by $p^*$ the set of coordinates
$p^{\mu_1\cdots \mu_j}_{i_1\cdots i_j}$ for $j\geq 1$, the Hamiltonian function has
always the form ${\cal H}(q,e,p^*) = e+H(q,p^*)$. Assume now that, 
$\forall q\in {\cal N}={\cal X}\times {\cal Y}$, there exists some value $p_0^*$ of $p^*$ such that
$\partial H/\partial p^*(q,p^*_0)=0$. Note that this situation
arises in almost all standard situation (if in particular the Lagrangian $l(x,u,v)$
has a quadratic dependence in $v$). Now let us assume the hypotheses of Theorem \ref{4.1/2.theo}
and consider a function $f$ whose level sets are slices.
Then since $f(q,p) = \widehat{f}(q,{\cal H}(q,p))$ we deduce that
$\{{\cal H},f\}^{\alpha_1\cdots \alpha_n}_{\alpha_\mu}(q,p) =
{\partial {\cal H}\over \partial p_{\alpha_1\cdots \alpha_n}}(q,p)
{\partial \widehat{f}\over \partial q^{\alpha_\mu}}(q,{\cal H}(q,p))$.
Now for all $(q,h)\in {\cal N}\times \Bbb{R}$, let $p^*_0$ be such that
$\partial H/\partial p^*(q,p^*_0)=0$ and let $e_0:= h -H(q,p_0^*)$. Since
${\partial {\cal H}\over \partial p^*}(q,e_0,p_0^*) = 0$ and ${\partial {\cal H}\over \partial e} = 1$,
condition (\ref{4.2.supercrochet}) at $m = (q,e_0,p_0^*)$ means that $\exists \mu$ with
$1\leq \mu\leq n$ such that ${\partial \widehat{f}\over \partial x^\mu}(q,h) = 
{\partial \widehat{f}\over \partial x^\mu}(q,{\cal H}(q,e_0,p_0^*))\neq 0$.
This singles out {\bf space-time coordinates}: they are the functions on ${\cal M}$
needed to build slices.}

\subsection{Algebraic and non algebraic observable $(n-1)$-forms coincide}
We show here that $\left(\Lambda^nT^{\star}{\cal N},\Omega\right)$
is a pataplectic manifold.
\begin{theo}\label{4.3.theopata}
If ${\cal M}$ is an open subset of $\Lambda^nT^{\star}{\cal N}$, then
$\mathfrak{P}^{n-1}_0{\cal M}=\mathfrak{P}^{n-1}{\cal M}$.
\end{theo}
{\em Proof} --- We already know that $\mathfrak{P}^{n-1}_0{\cal M}\subset \mathfrak{P}^{n-1}{\cal M}$.
Hence we need to prove the reverse inclusion. So in the following we consider some $m\in {\cal M}$ and
a form $a\in P_m^n{\cal M}$ and we will prove that there exists a vector
field $\xi$ on ${\cal M}$ such that $a=\xi\iN \Omega$. We write ${\cal O}_m{\cal M}:=
{\cal O}^a_m{\cal M}$.\\

\noindent
{\em Step 1} --- We show that given $m=(q,p)\in {\cal M}$ it is
possible to find $n+k$ vectors $(\widetilde{Q}_1,\cdots ,\widetilde{Q}_{n+k})$ of $T_q{\cal M}$ such that,
if $\Pi_*(\widetilde{Q}_\alpha):= Q_\alpha$ (the image of $\widetilde{Q}_\alpha$ by the
map $\Pi:{\cal M}\longrightarrow {\cal N}$), then $(Q_1,\cdots ,Q_{n+k})$ is
a basis of $T_q{\cal N}$ and $\forall (\alpha_1,\cdots , \alpha_n)$ such that
$1\leq \alpha_1<\cdots < \alpha_n\leq n+k$,
$\widetilde{Q}_{\alpha_1}\wedge \cdots \wedge \widetilde{Q}_{\alpha_n}\in {\cal O}_m{\cal M}$.\\

\noindent
This can be done by induction by using the fact that ${\cal O}_m{\cal M}$ is dense
in $D^n_m{\cal M}$. We start from any family of vectors $(\widetilde{Q}^0_1,\cdots ,\widetilde{Q}^0_{n+k})$
of $T_q{\cal N}$ such that $(Q^0_1,\cdots ,Q^0_{n+k})$ is a basis of $T_q{\cal N}$
(where $Q^0_\alpha:= \Pi_*(\widetilde{Q}^0_\alpha)$). We then order the ${(n+k)!\over n!k!}$
multi-indices $(\alpha_1,\cdots ,\alpha_n)$ such that $1\leq \alpha_1<\cdots < \alpha_n\leq n+k$
(using for instance the dictionary rule).
Using the density of ${\cal O}_m{\cal M}$ we can perturb slightly
$(\widetilde{Q}^0_1,\cdots ,\widetilde{Q}^0_{n+k})$ into
$(\widetilde{Q}^1_1,\cdots ,\widetilde{Q}^1_{n+k})$ in such a way that for instance
$\widetilde{Q}^1_1\wedge \cdots \wedge \widetilde{Q}^1_n\in {\cal O}_m{\cal M}$
(assuming that $(1,\cdots ,n)$ is the smallest index). Then we perturb further
$(\widetilde{Q}^1_1, \cdots ,\widetilde{Q}^1_{n+k})$ into
$(\widetilde{Q}^2_1, \cdots , \widetilde{Q}^2_{n+k})$ in such a way
that $\widetilde{Q}^2_1\wedge \cdots \wedge \widetilde{Q}^2_{n-1}\wedge
\widetilde{Q}^2_{n+1}\in {\cal O}_m{\cal M}$ (assuming that $(1,\cdots ,n-1,n+1)$ is the next
one). Using the fact that ${\cal O}_m{\cal M}$ is open we can do it in such a way
that we still have $\widetilde{Q}^2_1\wedge \cdots \wedge \widetilde{Q}^2_n\in {\cal O}_m{\cal M}$.
We proceed further until the conclusion is reached.\\

\noindent
In the following we choose local coordinates around $m$ in such a way that
$\widetilde{Q}_\alpha = \partial _\alpha + \sum_{1\leq \alpha_1<\cdots < \alpha_n\leq n+k}
P_{\alpha,\alpha_1\cdots \alpha_n}\partial ^{\alpha_1\cdots \alpha_n}$.\\

\noindent
{\em Step 2} --- We choose a multi-index $(\alpha_1,\cdots ,\alpha_n)$
with $1\leq \alpha_1<\cdots < \alpha_n\leq n+k$ and define the
set ${\cal O}_m^{\alpha_1\cdots \alpha_n}{\cal M}:={\cal O}_m{\cal M}
\cap D_m^{\alpha_1\cdots \alpha_n}{\cal M}$, where
\[
D_m^{\alpha_1\cdots \alpha_n}{\cal M}:=
\left\{ X_1\wedge \cdots \wedge X_n\in D^n_m{\cal M}/
\forall \mu, X_\mu = {\partial \over \partial q^\mu}
+ \sum_{1\leq \beta_1<\cdots < \beta_n\leq n+k}
X_{\mu,\beta_1\cdots \beta_n}{\partial \over \partial p^{\beta_1\cdots \beta_n}}\right\}.
\]
We want to understand the consequences of the relation
\begin{equation}\label{4.3.property}
\forall X,\widetilde{X}\in {\cal O}_m^{\alpha_1\cdots \alpha_n}{\cal M},\quad
X\iN \Omega = \widetilde{X}\iN \Omega\quad \Longrightarrow \quad a(X) = a(\widetilde{X}).
\end{equation}
Note that ${\cal O}_m^{\alpha_1\cdots \alpha_n}{\cal M}$ is open and non empty (since by the previous
step, $\widetilde{Q}_{\alpha_1}\wedge \cdots \wedge \widetilde{Q}_{\alpha_n}\in
{\cal O}_m^{\alpha_1\cdots \alpha_n}{\cal M}$).
We also observe that, on $D_m^{\alpha_1\cdots \alpha_n}{\cal M}$, $X\longmapsto X\iN \Omega$
and $X\longmapsto a(X)$ are respectively an affine function and a polynomial
function of the coordinates variables $X_{\mu,\beta_1\cdots \beta_n}$. Thus the
following result implies that actually ${\cal O}_m^{\alpha_1\cdots \alpha_n}{\cal M} =
D_m^{\alpha_1\cdots \alpha_n}{\cal M}$.
\begin{lemm}\label{4.3.technique1}
Let $N\in \Bbb{N}$ and let $P$ be a polynomial on $\Bbb{R}^N$ and $f_1,\cdots, f_p$ be
affine functions on $\Bbb{R}^N$. Assume that there exists some $x_0\in \Bbb{R}^N$ and
a neighbourhood $V_0$ of $x_0$ in $\Bbb{R}^N$ such that
\[
\forall x,\widetilde{x}\in V_0,\quad \hbox{if }\forall j = 1,\cdots ,p,
\ f_j(x) = f_j(\widetilde{x}), \quad \hbox{then }P(x) = P(\widetilde{x}).
\]
Then this property is true on $\Bbb{R}^N$.
\end{lemm}
{\em Proof} --- We can assume without loss of generality that the functions
$f_j$ are linear and also choose coordinates on $\Bbb{R}^N$ such that
$f_j(x) = x^j$, $\forall j = 1,\cdots ,p$. Then the assumption means that,
on $V_0$, $P$ does not depend on $x^{p+1},\cdots x^N$. Since $P$ is a polynomial
we deduce that $P$ is a polynomial on the variables $x^1,\cdots ,x^p$ and so the
property is true everywhere. \bbox
\noindent
{\em Step 3} --- Without loss of generality we will also assume in the following
that $(\alpha_1,\cdots ,\alpha_n) = (1,\cdots, n)$ for simplicity. We
shall denote by $m^I$ all coordinates $q^\alpha$ and $p_{\alpha_1\cdots \alpha_n}$,
so that we can write
\[
a = \sum_{I_1<\cdots <I_n}A_{I_1\cdots I_n}dm^{I_1}\wedge \cdots \wedge dm^{I_n}.
\]
We will prove that if $(I_1,\cdots ,I_n)$ is a multi-index such that
\begin{itemize}
\item $\{m^{I_1},\cdots ,m^{I_n}\}$ contains at least two distinct coordinates
of the type $p_{\alpha_1\cdots \alpha_n}$ and
\item $\{m^{I_1},\cdots ,m^{I_n}\}$ does not contain any $q^\alpha$, for
$n+1\leq\alpha \leq n+k$
\end{itemize}
then $A_{I_1\cdots I_n} = 0$. Without loss of generality we can suppose that $\exists p\in \Bbb{N}$
such that $1\leq p\leq n-2$ and
\[
m^{I_1} = q^1,\cdots ,m^{I_p} = q^p\quad \hbox{and}\quad
m^{I_{p+1}},\cdots ,m^{I_n}\in \{p_{\alpha_1\cdots \alpha_n}/1\leq\alpha_1<\cdots <\alpha_n\leq n+k\}.
\]
We test property (\ref{4.3.property}) specialized to the case where $X=X_1\wedge \cdots \wedge X_n$ with
\[
X_\mu = {\partial \over \partial q^\mu} + \sum_{j=p+1}^nX_\mu^{I_j}
{\partial \over \partial m^{I_j}},\quad \forall \mu = 1,\cdots ,n.
\]
Then
\begin{equation}\label{4.3.determinant}
a(X) = A_{I_1\cdots I_n}\left|\begin{array}{cccccc}
1 & \cdots & 0 & 0 & \cdots & 0\\
\vdots & \ddots & \vdots & \vdots & & \vdots \\
0 & \cdots & 1 & 0 & \cdots & 0\\
X_1^{I_{p+1}} & \cdots & X_p^{I_{p+1}} & X_{p+1}^{I_{p+1}} & \cdots & X_n^{I_{p+1}}\\
\vdots & & \vdots & \vdots &  & \vdots \\
X_1^{I_n} & \cdots & X_p^{I_n} & X_{p+1}^{I_n} & \cdots & X_n^{I_n}
\end{array}\right|
 = A_{I_1\cdots I_n}\left|\begin{array}{ccc}
X_{p+1}^{I_{p+1}} & \cdots & X_n^{I_{p+1}}\\
\vdots &  & \vdots \\
X_{p+1}^{I_n} & \cdots & X_n^{I_n}
\end{array}\right| .
\end{equation}
Remind that using the notations of Paragraph
4.1 we can write any $X\in D_m^{1\cdots n}{\cal M}$ as $X=X_1\wedge \cdots \wedge X_n$
with
\[
X_\mu = {\partial \over \partial q^\mu} + E_\mu{\partial \over \partial p_{1\cdots n}}
+ \sum_{\beta = n+1}^{n+k}M_{\mu,\beta} + R_\mu,
\]
where $M_{\mu,\beta}:= \sum_{\nu = 1}^n(-1)^{n+\nu}M^\nu_{\mu,\beta}
\partial ^{1\cdots \widehat{\nu}\cdots n\beta}$ and
$R_\mu:= \sum_{(\alpha_1,\cdots ,\alpha_n)\in I^{**}}X_{\mu,\alpha_1\cdots \alpha_n}
\partial ^{\alpha_1\cdots \alpha_n}$. And then
\[
(-1)^nX\iN \Omega = dp_{1\cdots n} - \sum_{\mu=1}^nE_\mu dq^\mu -
\sum_{\beta=n+1}^{n+k}\left(\sum_{\mu=1}^nM_{\mu,\beta}^\mu\right) dq^\beta.
\]
Within our specialization this leads to the following {\em key observation}\footnote{Remark
that each of the $n-p$ last lines in the $n\times n$ determinant in (\ref{4.3.determinant})
is either $(E_1,\cdots ,E_n)$ or of the type $(M_{1,\beta}^\nu,\cdots ,M_{m,\beta}^\nu)$
or $(X_{1,\alpha_1\cdots \alpha_n},\cdots ,X_{n,\alpha_1\cdots \alpha_n})$,
for $(\alpha_1,\cdots ,\alpha_n)\in I^{**}$.}: {\bf at most one} line
$(X_1^{I_j},\cdots ,X_n^{I_j})$ ( for $p+1\leq j\leq n$) in the $n\times n$
determinant in (\ref{4.3.determinant}) is a function of $X\iN \Omega$ (for
$m^{I_j}=p_{1\cdots n}$). In all other lines number $\nu$, where $p+1\leq \nu\leq n$ and $\nu\neq j$, there is
{\bf at most} one component $X^{I_\nu}_\mu$ which is a function of $X\iN \Omega$. All the
other components are independant of $X\iN \Omega$. Thus we have the following alternative.
\begin{enumerate}
\item $\{m^{I_{p+1}},\cdots ,m^{I_n}\}$ does not contain $p_{1\cdots n}$
(i.e.\,the line $(E_1,\cdots ,E_n)$ does not appear in the $n\times n$ determinant in (\ref{4.3.determinant})),
or
\item $\{m^{I_{p+1}},\cdots ,m^{I_n}\}$ contains $p_{1\cdots n}$
(i.e.\,one of the lines is $(E_1,\cdots ,E_n)$)
\end{enumerate} 
{\em Case} (i) --- Then the right hand side
determinant in (\ref{4.3.property}) is a polynomial of degree $n-p\geq 2$. Thus we can find
a monomial in this determinant of the form $X^{I_{p+1}}_{\sigma(p+2)}\cdots X^{I_n}_{\sigma(n)}$
(where $\sigma$ is a substitution of $\{p+1,\cdots ,n\}$)
where each variable is independant of $X\iN \Omega$.
Hence in order to achieve (\ref{4.3.property}) we must have $A_{I_1\cdots I_n} = 0$.\\
{\em Case} (ii) --- We assume w.l.g.\,that $m^{I_{p+1}} = p_{1\cdots n}$. We shall
freeze the variables $X^{I_{p+1}}_\mu$ (i.e.\,$E_\mu$) suitably and specialize again property
(\ref{4.3.property}) by letting free only the variables $X^{I_j}_\mu$ for $p+2\leq j\leq n$
and $1\leq \mu \leq n$.
Two subcases occur: if $p < n-2$ then we choose $X^{I_{p+1}}_\mu = \delta^{p+1}_\mu$. Then we are reduced
to a situation quite similar to the first case and we can conclude using the same
argument (this time with a determinant which is a monomial of degree $n-1-p\geq 2$).\\
If $p= n-2$ then $a(X) = A_{I_1\cdots I_n}\left(
X_n^{I_n} X_{n-1}^{I_{n-1}} - X_{n-1}^{I_n} X_n^{I_{n-1}}\right)$. If the knowledge of
$X\iN \Omega$ prescribes $X_n^{I_n}$ then by the key observation $X_{n-1}^{I_n}$ is free and by
choosing $X^{I_{n-1}}_\mu = \delta^n_\mu$ we obtain $a(X) = - A_{I_1\cdots I_n}X_{n-1}^{I_n}$.
If $X\iN \Omega$ prescribes $X_{n-1}^{I_n}$ then $X_n^{I_n}$ is free and by
choosing $X^{I_{n-1}}_\mu = \delta^{n-1}_\mu$ we obtain $a(X) = A_{I_1\cdots I_n}X_n^{I_n}$.
In both cases we must have $A_{I_1\cdots I_n} = 0$ in order to have (\ref{4.3.property}).\\

\noindent
{\em Conclusion} --- Steps 2 and 3 show that, on ${\cal O}_m^{1\cdots n}{\cal M}$,
$X\longmapsto a(X)$ is an affine function on the variables $X_{\mu,\beta_1\cdots \beta_n}$. Then
by standard results in linear algebra (\ref{4.3.property}) implies that,
$\forall X\in {\cal O}_m^{1\cdots n}{\cal M}$,
$a(X)$ is an affine combination of the components of $X\iN \Omega$. By repeating this
step on each ${\cal O}_m^{\alpha_1\cdots \alpha_n}{\cal M}$ we deduce the conclusion.
\bbox
\begin{theo}\label{4.3.theodDW}
Assume that ${\cal N} = {\cal X}\times {\cal Y}$, ${\cal M} = \Lambda^nT^*{\cal N}$ and consider
${\cal M}^{dDW}$ to be the submanifold of $\Lambda^nT^*{\cal N}$ as defined
in Paragraph 2.3.1 equipped with the multisymplectic form $\Omega^{dDW}$
which is the restriction of $\Omega$ to ${\cal M}^{dDW}$. Then $\mathfrak{P}^{n-1}{\cal M}^{dDW}$
coincides with $\mathfrak{P}^{n-1}_0\Lambda^nT^*({\cal X}\times {\cal Y})_{|{\cal M}^{dDW}}$,
the set of the restrictions of algebraic observable $(n-1)$-forms of
$({\cal M},\Omega)$ to ${\cal M}^{dDW}$.
\end{theo}
{\em Proof} ---
The fact that $\mathfrak{P}^n{\cal M}^{dDW}$ contains all the restrictions
of algebraic observable $(n-1)$-forms of $({\cal M},\Omega)$ to ${\cal M}^{dDW}$
was observed in Paragraph 3.3.3.
The proof of the reverse inclusion follows the same strategy as the proof of Theorem
\ref{4.3.theopata} and is left to the reader. \bbox

\subsection{All algebraic observable $(n-1)$-forms}
\begin{prop}\label{4.3.prop}
If ${\cal M}$ is an open subset of $\Lambda^nT^{\star}{\cal N}$, then
the set of all infinitesimal symplectomorphisms
$\Xi$ on ${\cal M}$ are of the form $\Xi = \chi + \overline{\xi}$, where
\begin{equation}\label{4.3.chixi}
\chi := \sum_{\beta_1<\cdots <\beta_n}\chi_{\beta_1\cdots \beta_n}(q)
{\partial \over \partial p_{\beta_1\cdots \beta_n}} \quad \hbox{ and }\quad
\overline{\xi}:= \sum_\alpha \xi^\alpha(q){\partial \over \partial q^\alpha}
- \sum_{\alpha,\beta}{\partial \xi^\alpha\over \partial q^\beta}(q)\Pi^\beta_\alpha,
\end{equation}
where
\begin{itemize}
\item the coefficients $\chi_{\beta_1\cdots \beta_n}$ are so that
$d(\chi\iN \Omega)= 0$,
\item $\xi:= \sum_\alpha \xi^\alpha(q){\partial \over \partial q^\alpha}$
is an arbitrary vector field on ${\cal N}$,
\item
$\displaystyle
\Pi^\beta_\alpha:= \sum_{\beta_1<\cdots <\beta_n}\sum_\mu \delta^\beta_{\beta_\mu}
p_{\beta_1\cdots \beta_{\mu-1}\alpha\beta_{\mu+1}\cdots \beta_n}
{\partial \over \partial p_{\beta_1\cdots \beta_n}}.$
\end{itemize}
As a consequence any algebraic observable $(n-1)$-form $F$ can be written as
$F=Q^\zeta + P_\xi$, where
\[
Q^\zeta = \sum_{\beta_1<\cdots <\beta_{n-1}}
\zeta_{\beta_1\cdots \beta_{n-1}}(q)dq^{\beta_1}\wedge \cdots  \wedge dq^{\beta_{n-1}}
\quad \hbox{and}\quad P_\xi = \xi\iN \theta.
\]
Then $\chi\iN \Omega = -dQ^\zeta$ and
$\overline{\xi}\iN \Omega = - dP_\xi$.
\end{prop}
{\em Proof} --- The fact that $P_\xi$ and $Q^\zeta$ belong to $\mathfrak{P}^{n-1}{\cal M}$
was already proved in \cite{HeleinKouneiher}. Here we need to prove that any infinitesimal
symplectomorphism $\Xi$ can be written as above. Let us write
$$\Xi=\sum_\alpha\Xi^\alpha(q,p){\partial \over \partial q^\alpha}
+ \sum_{\alpha_1<\cdots <\alpha_n}\Xi_{\alpha_1\cdots \alpha_n}(q,p)
{\partial \over \partial p_{\alpha_1\cdots \alpha_n}},$$
and analyse the equation $d(\Xi\iN \Omega) = 0$. We can write 
$d(\Xi\iN \Omega) = A+B+C+D$, where
$$A:= \sum_{\alpha_1<\cdots <\alpha_n}\sum_\beta
{\partial \Xi_{\alpha_1\cdots \alpha_n}\over \partial q^\beta}
dq^\beta\wedge dq^{\alpha_1}\wedge \cdots  \wedge dq^{\alpha_n},$$
$$B:= \sum_{\alpha_1<\cdots <\alpha_n}\sum_{\beta_1<\cdots <\beta_n}
{\partial \Xi_{\alpha_1\cdots \alpha_n}\over \partial p_{\beta_1\cdots \beta_n}}
dp_{\beta_1\cdots \beta_n}\wedge dq^{\alpha_1}\wedge \cdots  \wedge dq^{\alpha_n},$$
$$C:= \sum_\mu\sum_\beta\sum_{\alpha_1<\cdots <\alpha_n}
{\partial \Xi^{\alpha_\mu}\over \partial q^\beta}
dp_{\alpha_1\cdots \alpha_n}\wedge dq^{\alpha_1}\wedge \cdots  
\wedge dq^{\alpha_{\mu-1}}\wedge dq^\beta\wedge dq^{\alpha_{\mu+1}}\wedge \cdots 
\wedge dq^{\alpha_n}$$
and
$$D:= \sum_\alpha\sum_{\beta_1<\cdots <\beta_n}
{\partial \Xi^\alpha\over \partial p_{\beta_1\cdots \beta_n}}
dp_{\alpha_1\cdots \alpha_n}\wedge \left( {\partial \over \partial q^\alpha}\iN \Omega
- {\partial \over \partial q^\alpha}\iN \left( 
dp_{\beta_1\cdots \beta_n}\wedge dq^{\beta_1}\wedge \cdots  \wedge dq^{\beta_n}\right)
\right) .$$
The equation $d(\Xi\iN \Omega) = 0$ can be split into three equations
$A=0$, $B+C=0$ and $D=0$, according to the homogeneity in the $dp_*$'s.\\
{\em Relation} $D=0$ --- fix $\beta_1<\cdots <\beta_n$  and $\alpha$, then choose
any $\alpha_1<\cdots <\alpha_n$ and $\mu$ such that $\alpha_\mu=\alpha$ and
$\partial_{\alpha_1}\wedge \cdots  \wedge \widehat{\partial_{\alpha_\mu}} 
\wedge \cdots  \wedge \partial_{\alpha_n}\iN
dq^{\beta_1}\wedge \cdots  \wedge dq^{\beta_n}=0$. Then
\[
{\partial \Xi^\alpha\over \partial p_{\beta_1\cdots \beta_n}}=
(-1)^\mu{\partial \over \partial p_{\beta_1\cdots \beta_n}}\wedge
{\partial \over \partial p_{\alpha_1\cdots \alpha_n}}\wedge
{\partial \over \partial q^{\alpha_1}}\wedge \cdots  \wedge
\widehat{{\partial \over \partial q^{\alpha_\mu}}} 
\wedge \cdots \wedge
{\partial \over \partial q^{\alpha_n}}\iN D= 0.
\]
Hence $\Xi^\alpha(q,p) = \xi^\alpha(q)$.\\
{\em Relation} $B+C=0$ --- it implies that, if the cardinal of
$\{\alpha_1,\cdots  ,\alpha_n\}\cap \{\beta_1,\cdots  ,\beta_n\}$ is less or
equal than $n-2$, then
${\partial \Xi_{\alpha_1\cdots \alpha_n}\over \partial p_{\beta_1\cdots \beta_n}}(q,p)=0$. In
other cases, $\exists \mu, \beta$ such that
$\{\beta_1,\cdots  ,\beta_n\}=
\{\alpha_1,\cdots  \alpha_{\mu-1}, \beta,\alpha_{\mu+1},\cdots  ,\alpha_n\}$ and it
gives
$${\partial \Xi_{\alpha_1\cdots  \alpha_{\mu-1}\beta\alpha_{\mu+1}\cdots  \alpha_n}
\over \partial p_{\alpha_1\cdots \alpha_n}}(q,p) +
{\partial \xi^{\alpha_\mu}\over \partial q^\beta}(q) = 0.$$
Hence there exists coefficients $\chi_{\beta_1\cdots  \beta_n}$ which depends only on
$q$ such that
$$\Xi_{\beta_1\cdots  \beta_n}(q,p) = 
\chi_{\beta_1\cdots  \beta_n}(q) - \sum_\mu\sum_\alpha
{\partial \xi^\alpha\over\partial q^{\beta_\mu}}(q)\ 
p_{\beta_1\cdots  \beta_{\mu-1}\alpha\beta_{\mu+1}\cdots  \beta_n},$$
and we recover $\Xi=\overline{\xi}+ \chi$, where $\overline{\xi}$ and $\chi$ are given
by (\ref{4.3.chixi}).\\
{\em Relation} $A=0$ --- it gives then $d(\chi\iN \Omega) = 0$.\bbox

\noindent We let\footnote{recall that $\mathfrak{sp}_0{\cal M}$ is the set of all
symplectomorphisms of $({\cal M},\Omega)$ (see Definition \ref{2.1.def3})}
$\mathfrak{sp}_Q{\cal M}$ to be the set of infinitesimal
pataplectomorphisms of the form $\chi$ (with $\chi\iN \Omega$ closed)
and $\mathfrak{sp}_P{\cal M}$ those of the form $\overline{\xi}$
(for all vector fields $\xi\in \Gamma({\cal M},T{\cal M})$)
as defined in (\ref{4.3.chixi}). Then $\mathfrak{sp}_0{\cal M} = 
\mathfrak{sp}_Q{\cal M}\oplus \mathfrak{sp}_P{\cal M}$.
We also denote by $\mathfrak{P}^{n-1}_Q{\cal M}$ (as in \cite{HeleinKouneiher})
the set of algebraic observable $(n-1)$-forms $Q^\zeta$ such that $\xi_{Q^\zeta}\in \mathfrak{sp}_Q{\cal M}$
and we denote by $\mathfrak{P}^{n-1}_P{\cal M}$ the set of algebraic observable $(n-1)$-forms $P_\xi$
such that $\xi_{P_\xi}\in \mathfrak{sp}_P{\cal M}$.
The Lie bracket relations are:
$$\left\{
\begin{array}{ccl}
[\chi_1,\chi_2] & = & 0 \\
{[}\overline{\xi_1},\overline{\xi_2}{]} & = & \overline{ {[}\xi_1,\xi_2 {]}}\\
{[}\overline{\xi}, \chi{]} & = & \displaystyle \sum_{\beta_1<\cdots <\beta_n}\psi_{\beta_1\cdots \beta_n}(q)
{\partial \over \partial p_{\beta_1\cdots \beta_n}},
\end{array}\right.$$
where
$$\psi_{\beta_1\cdots \beta_n}:= \sum_\alpha \left( \xi^\alpha 
{\partial \chi_{\beta_1\cdots \beta_n}\over \partial q^\alpha} + \sum_\mu
\chi_{\beta_1\cdots \beta_{\mu-1}\alpha\beta_{\mu+1}\cdots \beta_n}
{\partial \xi^\alpha\over \partial q^{\beta_\mu}}\right) .$$
As a consequence we have that $[\overline{\xi}, \chi]\in \mathfrak{sp}_Q{\cal M}$,
$\forall \overline{\xi}\in \mathfrak{sp}_P{\cal M}$ and
$\forall \chi\in \mathfrak{sp}_Q{\cal M}$. Thus we conclude that $\mathfrak{sp}_0{\cal M}$
is the semidirect product $\mathfrak{sp}_P{\cal M}\ltimes \mathfrak{sp}_Q{\cal M}$. In
particular, $\mathfrak{sp}_Q{\cal M}$ is an Abelian ideal of $\mathfrak{sp}_0{\cal M}$. 
Note that the Poisson brackets of the
corresponding algebraic observable $(n-1)$-forms has been computed in \cite{HeleinKouneiher}.

\subsection{Invariance of the Hamiltonian in the pataplectic point of view}
We come back here to some of the properties of the Legendre correspondence described
in sections 2.2 and 2.3: for all $q\in {\cal N}$
the fiber ${\cal P}_q\subset \Lambda^nT^*_q{\cal N}$ is foliated by a family
of affine subspaces, the pseudofibers $P_q^h(z)$, for $h\in \Bbb{R}$ and $z\in D^\omega_q{\cal N}$, 
with the property that
\[
P_q^h(z) = p + \left(T_zD^n_q{\cal N}\right)^\perp, \quad \forall p\in P^h_q(z).
\]
We work in this Section on an open subset ${\cal M}_q$ of ${\cal P}_q$: then the Legendre Correspondence
Hypothesis implies that $Z_q(p)$ is reduced to one point that we shall denote by
$Z(q,p)$. Moreover
\begin{itemize}
\item ${\cal H}$ is constant and equal to $h$ on each $P_q^h(z)$ (Lemma 
\ref{2.2.2.dernierlemme})
\item for all Hamiltonian $n$-curve $\Gamma\in \widehat{\cal G}^\omega$ and for all
section $\Gamma\ni m\longmapsto \pi(m)$ of the pull-back of the bundle $\Lambda^nT^*{\cal N}$
by the canonical projection $\Gamma\longrightarrow {\cal N}$, such that
$\pi(m)\in \left(T_zD^n_q{\cal N}\right)^\perp$, the submanifold
$\{m+\pi(m)/ m\in \Gamma\}$ is also a Hamiltonian $n$-curve.
(Corollary \ref{2.2.3.coro1}).
\end{itemize}

\noindent We now wish to rephrase the content of Lemma \ref{2.2.2.dernierlemme}
and Corollary \ref{2.2.3.coro1} in
terms which would make sense on an arbitrary pataplectic manifold. Given any smooth
function ${\cal H}:{\cal M}\longrightarrow \Bbb{R}$ we recall the definition given
by (\ref{3.3.4.lh}):
\[
\begin{array}{ccl}
L^{\cal H}_{(q,p)} & := & \{ \xi\in T_{(q,p)}{\cal M}/\ \forall X\in [X]^{\cal H}_{(q,p)},
\forall \delta X\in T_XD^n_{(q,p)}{\cal M},\ \xi\iN \Omega (\delta X) = 0\}.
\end{array}
\]
We will prove that each subspace $\left(T_zD^n_q{\cal N}\right)^\perp$ can be
identified with $L^{\cal H}_{(q,p)}$, where $p\in P_q(z)$.
We first need a preliminary result.

\begin{lemm}\label{4.4.lemm2}
Let ${\cal M}$ be an open subset of $\Lambda^nT^{\star}{\cal N}$ and let ${\cal H}$ be
an {\em arbitrary} smooth function from ${\cal M}$ to $\Bbb{R}$, such that $d_p{\cal H}$
never vanishes.
Let $\xi\in L^{\cal H}_{(q,p)}$, then $dq^\alpha(\xi) = 0$, $\forall \alpha$, i.e.
$$\xi = \sum_{\alpha_1<\cdots <\alpha_n}\xi_{\alpha_1\cdots \alpha_n}
{\partial \over \partial p_{\alpha_1\cdots \alpha_n}}.$$
\end{lemm}
{\em Proof} --- We use the results proved in Section 4.1: we know that we can assume
w.l.g.\,that $d_p{\cal H}=dp_{1\cdots n}$. Then any $n$-vector $X\in D^n_{(q,p)}{\cal M}$
such that $(-1)^nX\iN \Omega = d{\cal H}$ can be written $X=X_1\wedge \cdots \wedge X_n$, where
each vector $X_\mu$ is given by (\ref{3.1.xmu}) with the conditions on $M^\nu_{\mu,\beta}$
and $R_\mu$ described in Section 4.1. We construct a solution  $X$ of
$(-1)^nX\iN \Omega = d{\cal H} = \sum_\alpha{\partial {\cal H}\over \partial q^\alpha}
dq^\alpha + dp_{1\cdots n}$ by choosing
\begin{itemize}
\item $R_\mu = 0$, $\forall 1\leq \mu\leq n$
\item $M^\nu_{\mu,\beta} = 0$ if $(\mu,\nu) \neq (1,1)$
\item $M^1_{1,\beta} = - {\partial {\cal H}\over \partial q^\beta}$,
$\forall n+1\leq \beta\leq n+k$
\end{itemize}
in relations (\ref{3.1.xmu}). It corresponds to
$$\left\{ \begin{array}{ccl}
X_1 & = & \displaystyle {\partial \over \partial q^1} - {\partial {\cal H}\over \partial q^1}
{\partial \over \partial p_{1\cdots n}} + (-1)^n \sum_{\beta=n+1}^{n+k}
{\partial {\cal H}\over \partial q^\beta}{\partial \over \partial p_{2\cdots n\beta}}\\
X_\mu & = & \displaystyle {\partial \over \partial q^\mu} -
{\partial {\cal H}\over \partial q^\mu}
{\partial \over \partial p_{1\cdots n}},\quad \hbox{if } 2\leq \mu \leq n.
\end{array}\right.$$
We first choose $\delta X^{(1)}\in T_XD^n_{(q,p)}{\cal M}$ to be
$\delta X^{(1)} := \delta X^{(1)}_1\wedge X_2\wedge \cdots  \wedge X_n$, where
$\delta X^{(1)}_1:={\partial \over \partial p_{1\cdots n}}$. It gives
$$\delta X^{(1)} = {\partial \over \partial p_{1\cdots n}}\wedge {\partial \over \partial q^2}
\wedge \cdots  \wedge {\partial \over \partial q^n}.$$
Now let $\xi\in L^{\cal H}_{(q,p)}$, we must have $\xi\iN \Omega(\delta X^{(1)}) = 0$. But a
computation gives
$$\xi\iN \Omega(\delta X^{(1)}) = (-1)^n\delta X^{(1)}\iN \Omega(\xi)
= -dq^1(\xi),$$
so that $dq^1(\xi) = 0$.\\

\noindent For $n+1\leq \beta\leq n+k$, consider $\delta X^{(\beta)} :=
\delta X^{(\beta)}_1\wedge X_2\wedge \cdots  \wedge X_n\in T_XD^n_{(q,p)}{\cal M}$,
where $\delta X^{(\beta)}_1:={\partial \over \partial p_{2\cdots n\beta}}$. Then
we compute that $\delta X^{(\beta)}\iN \Omega = dq^\beta$. Hence,
by a similar reasoning, the relation
$\xi\iN \Omega(\delta X^{(\beta)}) = 0$ is equivalent to $dq^\beta(\xi) = 0$.\\

\noindent Lastly by considering another solution $X\in D^n_{(q,p)}{\cal M}$ to the Hamilton equation,
where the role of $X_1$ has been exchanged with the role of $X_\mu$, for some
$2\leq \mu\leq n$, we can prove that $dq^\mu(\xi) = 0$, as well. \bbox

\noindent Recall that the tangent space $T_{(q,p)}\left( \Lambda^nT^*{\cal N}\right)$
possesses a canonical ``vertical'' subspace 
$\{0\}\times T_p\left(\Lambda^nT^*_q{\cal N}\right)\simeq
\Lambda^nT^*_q{\cal N}$:
Lemma \ref{4.4.lemm2} can be rephrased by saying that,
if $d_p{\cal H}\neq 0$ everywhere, then $L^{\cal H}_{(q,p)}$ can be identified
with a vector subspace of this vertical subspace.

\begin{prop}\label{4.4.prop}
Let ${\cal M}$ be an open subset of $\Lambda^nT^{\star}{\cal N}$ and let ${\cal H}$ be a Hamiltonian
function on ${\cal M}$ built from a Lagrangian density $L$ by means of the
Legendre correspondence. Then, through the identification
$\{0\}\times T_p{\cal M}_q\simeq
\Lambda^nT^*_q{\cal N}$, $L^{\cal H}_{(q,p)}$ coincides with
$\left(T_{Z(q,p)}D^n_q{\cal N}\right)^\perp$.
\end{prop}
{\em Proof} --- First we remark that the hypotheses imply that
$d_p{\cal H}$ never vanishes (because $d{\cal H}(0,\omega)=1$).
Let $\xi\in L^{\cal H}_{(q,p)}$, using the preceeding remark
we can associate a $n$-form $\pi \in \Lambda^nT^{\star}_q{\cal N}$
to $\xi$ with coordinates $\pi_{\alpha_1\cdots \alpha_n} = \xi_{\alpha_1\cdots \alpha_n}$.
We also observe that $\pi = \xi\iN \Omega$. Now let us look at the condition:
\begin{equation}\label{4.4.implication}
\forall X\in [X]^{\cal H}_{(q,p)},\quad \forall \delta X\in T_XD^n_{(q,p)}{\cal M},
\quad  \xi\iN \Omega(\delta X)=0.
\end{equation}
By the analysis of section 4.1 we know that the ``horizontal'' part $X_{(0)}$ of $X$
is fully determined by ${\cal H}$: it is actually $X_{(0)} = Z(q,p)$. Now take
any $\delta X\in T_XD^n_{(q,p)}{\cal M}$ and split it into its horizontal part
$\delta z\in T_{Z(q,p)}D^n_q{\cal N}$ and a vertical part $\delta X^V$. We remark that
\begin{itemize}
\item $\delta z\in T_{Z(q,p)}D^n_q{\cal N}$
\item $\xi\iN \Omega(\delta X) = \pi(\delta X) = \pi(\delta z)$.
\end{itemize}
Hence (\ref{4.4.implication}) means that $\pi\in \left( T_{Z(q,p)}D^n_q{\cal N}\right)^\perp$.
So the result follows. \bbox

\begin{theo}\label{4.4.theo}
Let  ${\cal M}$ be an open subset of $\Lambda^nT^{\star}{\cal N}$ and let ${\cal H}$ be a Hamiltonian
function on ${\cal M}$ built from a Lagrangian density $L$ by means of the
Legendre correspondence. Then
\begin{equation}\label{4.4.dh=0}
\forall (q,p)\in {\cal M},
\forall \xi\in L^{\cal H}_{(q,p)},\quad  d{\cal H}_{(q,p)}(\xi) = 0.
\end{equation}
And if $\Gamma\in \widehat{\cal G}^\omega$ is a Hamiltonian $n$-curve and if
$\xi$ a vector field which is a smooth section of $L^{\cal H}$, then denoting by
$e^{s\xi}$ the flow mapping of $\xi$
\begin{equation}\label{4.4.stability}
\forall s\in \Bbb{R},\hbox{ small enough },
e^{s\xi}(\Gamma)\hbox{ is a Hamiltonian }n{-curve}.
\end{equation}
\end{theo}
{\em Proof} --- Through Proposition \ref{4.4.prop} (\ref{4.4.dh=0}) and
(\ref{4.4.stability}) are the translations of the infinitesimal
versions of Lemma \ref{2.2.2.dernierlemme} and Corollary \ref{2.2.3.coro1}
respectively.\bbox
\noindent
Note that Theorem \ref{4.4.theo} is the motivation for Definition \ref{3.3.4.def}.
Below is a consequence of Lemma \ref{3.3.4.lem} and of Theorem \ref{4.4.theo}.
\begin{prop}\label{4.4.prop2}
$\bullet$ Let ${\cal C}\subset \Lambda^nT^*{\cal N}$
be a smooth subbundle, ${\cal P}\subset \Lambda^nT^*{\cal N}$ 
and ${\cal H}:{\cal P}\longrightarrow \Bbb{R}$ be a Hamiltonian
function obtained by means of a Legendre correspondence. (Remind
that ${\cal P}_q=\cup_{z,h}P^h_q(z)$.)\\
$\bullet$ Let $h\in \Bbb{R}$ and
assume that all the pseudofibers $P^h_q(z)$ contained in ${\cal H}^{-1}(h)$
intersect ${\cal C}$ (so necessarily along ${\cal C}\cap {\cal P}$).\\
$\bullet$ Let $F\in
\mathfrak{P}_0^{n-1}\left( \Lambda^nT^*{\cal N}\right)$ be an algebraic observable form
of the form $\xi\iN \theta$, where $\xi$ a tangent vector field to ${\cal N}$ and
$\theta$ is the Poincar\'e--Cartan form (i.e.\,$F$ is similar to an energy or momentum observable
form).\\
If the restriction $F_{|{\cal C}\cap {\cal P}}$
of $F$ to ${\cal C}\cap {\cal P}$ vanishes, then for all Hamiltonian $n$-curve $\Gamma$
such that the value of ${\cal H}$ on $\Gamma$ is $h$ and
for any smooth $(n-1)$-dimensional submanifold  $\Sigma$ of $\Gamma$,
\begin{equation}\label{4.4.cycle}
\int_{\Sigma} F = 0.
\end{equation}
\end{prop}
\begin{figure}[h]
\begin{center}
\includegraphics[scale=1]{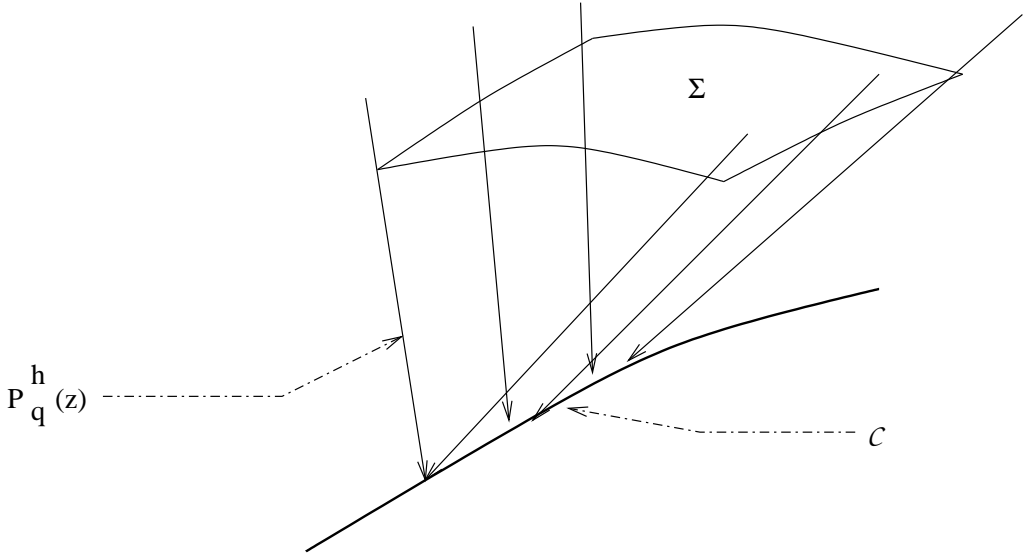}
\caption{The deformation of an $(n-1)$-dimensional submanifold $\Sigma$ along
pseudofibers towards ${\cal C}$.}
\end{center}
\end{figure}
{\em Proof} --- We consider a family of deformations of $\Gamma$ and
$\Sigma$ by the flow of a vector field $\zeta$ which is a section of $L^{\cal H}$ and which
pushes $\Gamma$ towards ${\cal C}$, hence towards ${\cal C}\cap {\cal P}$.
By Theorem \ref{4.4.theo} all $\Gamma_s$'s are Hamiltonian $n$-curves. Moreover Lemma
\ref{4.4.lemm2} implies that $\zeta$ is ``vertical'', i.e.\,$dq^\alpha(\zeta)=0$, and
so $\zeta\iN \theta=0$ which implies $\zeta\iN F=0$.
Thus we can apply Lemma \ref{2.2.2.dernierlemme}: 
$\int_{\Gamma_s}F$ does not depend on $s$. But on the other hand,
since $F_{|{\cal C}\cap {\cal P}}=0$ and $F$ is continuous
$\int_{\Gamma_s}F$ tends to 0 when $\Gamma_s$ converges to ${\cal C}\cap {\cal P}$.
So the result follows. \bbox
\noindent
{\bf Example 12} --- {\em Consider the trivial variational problem on maps from $\Bbb{R}^2$
to $\Bbb{R}^2$ discussed in Example 5, Section 2.3.2. Take (in the notations of
Example 5) ${\cal C}:=\{(x,y,e,p,r)\in \Lambda^2T^*\Bbb{R}^4/ e=r=0\}$. Then all
pseudofibers $P^0_q(z)$ cross ${\cal C}$ along
${\cal C}\cap {\cal P}:=\{(x,y,e,p,r)\in \Lambda^2T^*\Bbb{R}^4/ e=p=r=0\}\simeq \Bbb{R}^4$.
We then observe that all algebraic observable $(n-1)$-forms $F$ of the form $\xi\iN \theta$,
where $\xi$ is a tangent vector field in $\Bbb{R}^4$,
vanishes on ${\cal C}\cap {\cal P}$. These forms corresponds
to the momentum and the energy-momentum of the field.
So Proposition \ref{4.4.prop2} tells us that the corresponding observable functionals
--- although there are formally observable in our theory --- vanishes everywhere
in this example.
Thanks to that there is no contradiction since the trivial variational problem does
not carry any dynamical information. A similar inspection of Example 7 of Section 2.3.2
(the Maxwell field in 2 dimensions) teaches us that combinations $p^1_2+p^2_1$,
$p^1_1$ and $p^2_2$ does not carry information too in this case.}\\

\section{Observable $(p-1)$-forms}
We now introduce 
observable $(p-1)$-forms, for $1\leq p< n$. The simplest situation where
such forms play some role occurs when studying variational problems on
maps $u:{\cal X}\longrightarrow {\cal Y}$: any coordinate function $y^i$ on ${\cal Y}$
is an observable functional, which at least in a classical context can be measured.
This observable 0-form can be considered as canonically conjugate with the
momentum observable form $\partial /\partial y^i\iN \theta$.
A more complex situation is given by Maxwell equations: as proposed for the first
time by I.\,Kanatchikov in \cite{Kanatchikov1} (see also \cite{HeleinKouneiher}),
the electromagnetic gauge potential and the Faraday fields can be modelled in
an elegant way by observable 1-forms and $(n-2)$-forms respectively.\\

\noindent
{\bf Example 13} --- {\em Maxwell equations on Minkowski space-time ---
Assume here for simplicity that ${\cal X}$ is
the four-dimensional Minkowski space. Then the gauge field is a 1-form
$A(x) = A_{\mu}(x)dx^{\mu}$
defined over ${\cal X}$, i.e.\,a section of the bundle $T^{\star}{\cal X}$.
The action functional in the presence of
a (quadrivector) current field $j(x)= j^{\mu}(x)\partial /\partial x^{\mu}$ is
$\int_{\cal X}l(x,A,dA)\omega$, where
$\omega = dx^0\wedge dx^1\wedge dx^2\wedge dx^3$ and
$$l(x,A,dA) = - {1\over 4}F_{\mu\nu}F^{\mu\nu} - j^{\mu}(x)A_{\mu},$$
where $F_{\mu\nu}:= \partial _{\mu}A_{\nu} - \partial _{\nu}A_{\mu}$ and
$F^{\mu\nu}:= \eta^{\mu\lambda}\eta^{\nu\sigma}F_{\lambda\sigma}$ (see \cite{HeleinKouneiher}).
The associated multisymplectic
manifold is then ${\cal M}:= \Lambda^4T^{\star}(T^{\star}{\cal X})$ with the multisymplectic form
$$\Omega = de\wedge \omega + \sum_{\mu,\nu}dp^{A_{\mu}\nu}\wedge
da_{\mu}\wedge \omega_\nu + \cdots $$
For simplicity we restrict ourself to the de Donder--Weyl submanifold (where all momentum
coordinates excepted $e$ and $p^{A_{\mu}\nu}$ are set to 0). This implies automatically
the further constraints $p^{A_{\mu}\nu} + p^{A_{\nu}\mu} = 0$, because the Legendre
correspondence degenerates when restricted to the de Donder--Weyl submanifold.
We shall hence denote
\[
p^{\mu\nu}:= p^{A_{\mu}\nu} = - p^{A_{\nu}\mu}.
\]
Let us call ${\cal M}^{Max}$ the resulting multisymplectic manifold.
Then the multisymplectic form can be written as
\[
\Omega = de\wedge \omega + d\pi\wedge da\quad \hbox{where }
a:=a_\mu dx^\mu\hbox{ and }
\pi:= -{1\over 2}\sum_{\mu,\nu}p^{\mu\nu}\omega_{\mu\nu}.
\]
(We also have $d\pi\wedge da = \sum_{\mu,\nu}dp^{\mu\nu}\wedge da_\mu\wedge \omega_\nu$.)
Note that here $a_{\mu}$ is not anymore a function of $x$ but a fiber coordinate.
The Hamiltonian is then
\[
{\cal H}(x,a,p) = e - {1\over 4}
\eta_{\mu\lambda}\eta_{\nu\sigma}p^{\mu\nu}p^{\lambda\sigma}
+ j^{\mu}(x)a_{\mu}.
\] }

\noindent
However the dynamics and the Poisson bracket structure for $(p-1)$-forms
is then more subtle for $1\leq p<n$ than for $p=n$,
since if $F$ is such a $(p-1)$-form then there is no way a priori to ``evaluate''
$dF$ along a Hamiltonian $n$-vector $X$ and a fortiori no way to
make sense that ``$dF_{|X}$ should not depend on $X$ but on $d{\cal H}_m$''..
This situation is in some sense reminiscent from the problem of measuring
a distance in relativity: we actually never measure the distance between
two points (finitely or infinitely close) but we do {\em compare} observable
quantities (distance, time) between themselves. This analogy suggests us the conclusion
that we should define observable $(p-1)$-forms {\em collectively}. The idea
is naively that if for instance $F_1,\cdots ,F_n$ are 0-forms, then
they are observable forms if $dF_1\wedge \cdots \wedge dF_n$ can be ``evaluated'' in the
sense that $dF_1\wedge \cdots \wedge dF_n(X)$ does not depend on the choice
of the Hamiltonian $n$-vector $X$ but on $d{\cal H}$. So
it just means that $dF_1\wedge \cdots \wedge dF_n$ is copolar.
Keeping this in mind we shall define first what are the exterior differentials of
observable $(p-1)$-forms (copolar $p$-forms, see below), before
defining observable forms themselves.

\subsection{Copolarisation}
\begin{defi}\label{6.1.def4}
Let ${\cal M}$ be a multisymplectic manifold.
A {\em copolarisation} on ${\cal M}$ is a smooth
vector subbundle denoted by
$P^*T^{\star}{\cal M}$ of $\Lambda^*T^{\star}{\cal M}$
satisfying the following properties
\begin{itemize}
\item $P^*T^{\star}{\cal M}:=\oplus_{j=1}^NP^jT^{\star}{\cal M}$, where
$P^jT^{\star}{\cal M}$ is a subbundle of $\Lambda^jT^{\star}{\cal M}$
\item for each $m\in {\cal M}$, $(P^*T^{\star}_m{\cal M},+,\wedge )$ is a subalgebra of
$(\Lambda^*T^{\star}_m{\cal M},+,\wedge )$
\item $\forall m\in {\cal M}$ and $\forall a\in \Lambda^nT^{\star}_m{\cal M}$,
$a\in P^nT^{\star}_m{\cal M}$ if and only if $\forall X,\widetilde{X}\in {\cal O}_m$,
$X\iN \Omega = \widetilde{X}\iN \Omega \Longrightarrow a(X) = a(\widetilde{X})$.
\end{itemize}
\end{defi}
\begin{defi}\label{6.1.def5}
Let ${\cal M}$ be a multisymplectic manifold with a copolarisation $P^*T^{\star}{\cal M}$.
Then for $1\leq p\leq n$, the set of
{\em observable $(p-1)$-forms} associated to $P^*T^{\star}{\cal M}$ is the set of smooth
$(p-1)$-forms $F$ (sections of $\Lambda^{p-1}T^{\star}{\cal M}$) such that for any
$m\in {\cal M}$, $dF_m\in P^pT^{\star}_m{\cal M}$. This set is denoted by
$\mathfrak{P}^{p-1}{\cal M}$. We shall write
$\mathfrak{P}^*{\cal M}:= \oplus_{p=1}^n\mathfrak{P}^{p-1}{\cal M}$.
\end{defi}

\noindent
Recall the equivalence relation $\sim$ between $n$-vectors introduced in the proof of
Lemma \ref{3.3.1.lemme} and consider its restriction to decomposable
$n$-vectors: $\forall X,\widetilde{X}\in D^n_m{\cal M}$, $X\sim \widetilde{X}
\Longleftrightarrow X\iN \Omega = \widetilde{X}\iN \Omega$.
Note that an equivalent characterisation of $\sim$ is the following: for
any $n$-vectors $X$ and $\widetilde{X}\in {\cal O}_m\subset D^n_m{\cal M}$,
$X\sim \widetilde{X}$ if and only if $\langle X, \phi\rangle = \langle \widetilde{X}, \phi\rangle $,
$\forall \phi\in P^nT^{\star}_m{\cal M}$. Indeed on the one hand we have:
$X\sim \widetilde{X} \Longleftrightarrow X\iN \Omega = \widetilde{X}\iN \Omega
\Longrightarrow \phi(X) = \phi(\widetilde{X})$. On the other hand if
$\forall \phi\in P^nT^{\star}_m{\cal M},
\langle X, \phi\rangle = \langle \widetilde{X}, \phi\rangle$ then it is true in particular for
$\phi = \xi\iN \Omega$, so
\[
\begin{array}{cc}
& \forall \xi\in T_m{\cal M},\;
(-1)^n\xi\iN \Omega(X) = (-1)^n\xi\iN \Omega(\widetilde{X})\\
\Longleftrightarrow  & \forall \xi\in T_m{\cal M},\;
X\iN \Omega(\xi) = \widetilde{X}\iN \Omega(\xi)\\
\Longleftrightarrow  & X\iN \Omega = \widetilde{X}\iN \Omega\\
\Longleftrightarrow  & X\sim \widetilde{X}.
\end{array}
\]
Hence, using a given copolarisation, we can enlarge the equivalence relation $\sim$
to $p$-vectors, for $1\leq p\leq n$, as follows.

\begin{defi}\label{6.1.def6}
Let ${\cal M}$ be a multisymplectic manifold with a copolarisation $P^*T^{\star}{\cal M}$.
For each $m\in {\cal M}$
and $1\leq p\leq n$, consider the equivalence relation in $\Lambda^pT_m{\cal M}$ defined
by $X \sim \widetilde{X}$ if and only if
$\langle X,a\rangle = \langle \widetilde{X},a\rangle$, $\forall a\in P^pT^{\star}_m{\cal M}$.
Then the quotient set $P^pT_m{\cal M}:=\Lambda^pT_m{\cal M}/\sim$ is called
a {\em polarisation} of ${\cal M}$.
If $X\in \Lambda^*T_m{\cal M}$, we denote by $[X]\in P^*T_m{\cal M}$ its equivalence class.\\
Equivalentely a {\em polarisation} can be defined as being the dual bundle of 
the copolarisation $P^*T^{\star}{\cal M}$.
\end{defi}
{\em Remark} ---
In the case of pataplectic manifolds where the set of algebraic observable $(n-1)$-forms
$\mathfrak{P}^{n-1}_0{\cal M}$
coincides with $\mathfrak{P}^{n-1}{\cal M}$, we observe that $\forall a\in P^nT^{\star}_m{\cal M}$,
there exists a unique vector $\xi\in T_m{\cal M}$ such that
$a + \xi\iN \Omega = 0$. Hence for any $1\leq p\leq n$, for any observable
$(p-1)$-forms $F\in \mathfrak{P}^{p-1}{\cal M}$ and for any
$\phi \in P^{n-p}T^{\star}_m{\cal M}$, there exists a unique vector
$\xi_F(\phi)\in T_m{\cal M}$ such that $\phi\wedge dF + \xi_F(\phi)\iN \Omega = 0$.
We thus obtain a linear mapping
\begin{equation}\label{6.1.xi}
\begin{array}{cccc}
\xi_F: & P^{n-p}T^{\star}_m{\cal M} & \longrightarrow & T_m{\cal M}\\
 & \phi & \longmapsto & \xi_F(\phi).
\end{array}
\end{equation}
Hence we can associate to $F$ a tensor field $\xi_F$ whose characterisation at each
point $m$ is described above. By duality between $P^{n-p}T^{\star}_m{\cal M}$ and
$P^{n-p}T_m{\cal M}$, $\xi_F$ can also be identified with a section of the bundle
$P^{n-p}T{\cal M}\otimes _{\cal M}T{\cal M}$.

\subsection{Examples of copolarisation}
On an open subset ${\cal M}$ of
$\Lambda^nT^*{\cal N}$ we can construct the following copolarisation, that
we will call {\em standard}: for each $(q,p)\in \Lambda^nT^*{\cal N}$ and
for $1\leq p\leq n-1$ we take
$P^p_{(q,p)}T^*{\cal M}$ to be the vector space spanned by
$\left(dq^{\alpha_1}\wedge \cdots \wedge dq^{\alpha_p}
\right)_{1\leq \alpha_1<\cdots <\alpha_p\leq n+k}$; and
$P^n_{(q,p)}T^*{\cal M}=\{\xi\iN \Omega/\xi\in T_m{\cal M}\}$. It means
that $P^n_{(q,p)}T^*{\cal M}$ contains all $dq^{\alpha_1}\wedge \cdots \wedge dq^{\alpha_n}$'s
plus forms of the type $\xi\iN \Omega$, for $\xi\in T_q{\cal N}$ (which
corresponds to differentials of momentum and energy-momentum observable $(n-1)$-forms).\\

\noindent We remark that, for this standard choice of copolarisation, $P^1_{(q,p)}T^*{\cal M}$
coincides with the set of 1-forms which are proper (see Section 4.5) on $[X]^{\cal H}_{(q,p)}$,
for all possible choices of ${\cal H}$ around $(q,p)$
provided that $d_p{\cal H}_m\neq 0$: this is a straightforward consequence of
Lemma \ref{4.1/2.lemme}. We believe that this property is more general, i.e....\,that
the standard copolarisation is characterised by the property of being proper.
This motivates the following definitions and conjectures.
\begin{defi}\label{5.3/2.defi1}
Assume that $({\cal M},\Omega)$ is a multisymplectic manifold and that we are given an open
dense subset ${\cal O}_m\subset D^n_m{\cal M}$. Let $1\leq p\leq n$.
We say that a $p$-form $a\in \Lambda^pT^*_m{\cal M}$
is {\em proper on} ${\cal O}_m$ if and only if $\forall X\in {\cal O}_m$,
\begin{itemize}
\item either $a_{|X}\neq 0$ and then $\forall \widetilde{X}\in {\cal O}_m$,
such that $\widetilde{X}\iN \Omega =X\iN \Omega$, $a_{|\widetilde{X}}\neq 0$
\item or $a_{|X} = 0$ and then $\forall \widetilde{X}\in {\cal O}_m$,
such that $\widetilde{X}\iN \Omega =X\iN \Omega$, $a_{|\widetilde{X}} = 0$.
\end{itemize}
\end{defi}
\noindent
{\bf Conjecture 1} --- {\em Assume that ${\cal M}$ is an open subset of $\Lambda^nT^*{\cal N}$
and ${\cal O}_m:=\{X\in D^n_m{\cal M}/\left( X\iN \Omega\right)_m\neq 0\}$. Then the set of $p$-forms
(for $1\leq p\leq n$) which are proper on ${\cal O}_m$ coincides with
the standard copolarisation.}\\
{\bf Conjecture 2} --- {\em Let $({\cal M},\Omega)$ be an arbitrary multisymplectic manifold.
If, for all $m\in {\cal M}$, ${\cal O}_m$ is an open dense
subset of $D^n_m{\cal M}$, then the set of $p$-forms
(for $1\leq p\leq n$) which are proper on ${\cal O}_m$ is a copolarisation.}\\

\noindent
Another situation is the following.\\

\noindent {\bf Example 13'} --- {\em Maxwell equations --- We continue Example 13 given at the
beginning of this section. In ${\cal M}^{Max}$ with the multisymplectic form
$\Omega = de\wedge \omega + d\pi\wedge da$ the more natural choice of copolarisation is:
\begin{itemize}
\item $\displaystyle P^1_{(q,p)}T^*{\cal M}^{Max} = \bigoplus_{0\leq \mu\leq 3}\Bbb{R}dx^\mu$.
\item $\displaystyle P^2_{(q,p)}T^*{\cal M}^{Max} = \bigoplus_{0\leq \mu_1<\mu_2\leq 3}\Bbb{R}
dx^{\mu_1}\wedge dx^{\mu_2}\oplus \Bbb{R}da$, where $da:=\sum_{\mu=0}^3da_\mu\wedge dx^\mu$.
\item $\displaystyle P^3_{(q,p)}T^*{\cal M}^{Max} = \bigoplus_{0\leq \mu_1<\mu_2<\mu_3\leq 3}\Bbb{R}
dx^{\mu_1}\wedge dx^{\mu_2}\wedge dx^{\mu_3}\oplus \bigoplus_{0\leq \mu\leq 3}\Bbb{R} dx^\mu\wedge da
\oplus \Bbb{R}d\pi$.
\item $\displaystyle P^4_{(q,p)}T^*{\cal M}^{Max} = \Bbb{R}\omega\oplus \bigoplus_{0\leq \mu_1<\mu_2\leq 3}\Bbb{R}
dx^{\mu_1}\wedge dx^{\mu_2}\wedge da\oplus \bigoplus_{0\leq \mu\leq 3}\Bbb{R}dx^\mu\wedge d\pi\oplus
\bigoplus_{0\leq \mu\leq 3} \Bbb{R}{\partial \over \partial x^\mu}\iN \theta$.
\end{itemize} 
It is worth stressing out the fact that we did not include the differential of
the coordinates $a_\mu$ of $a$ in $P^1_{(q,p)}T^*{\cal M}^{Max}$. There are
strong physical reasons for that since the gauge potential is not observable.
But another reason is that if we had included the $da_\mu$'s in $P^1_{(q,p)}T^*{\cal M}^{Max}$,
we would not have a copolarisation since $da_\mu\wedge d\pi$ does not satisfy the
condition $\forall X,\widetilde{X}\in {\cal O}_m$, $[X]=[\widetilde{X}]\Rightarrow
b(X)=b(\widetilde{X})$ required. This confirms the agreement of the definition of
copolarisation with physical purposes.}

\subsection{Results for the dynamics}
We wish here to generalize Proposition \ref{3.2.2.propdyn} to observable
$(p-1)$-forms for $1\leq p<n$. This result actually justifies the relevance of
Definitions \ref{6.1.def4}, \ref{6.1.def5} and \ref{6.1.def6}. Throughout this
section  we assume that $({\cal M},\Omega)$ is equipped with a copolarisation.
We start with some
technical results. If ${\cal H}$ is a Hamiltonian function, we recall that we denote by
$[X]^{\cal H}$ the class modulo $\sim$ of decomposable $n$-vector fields $X$ such that
$X\iN \Omega =(-1)^nd{\cal H}$.

\begin{lemm}\label{6.2.lem1}
Let $X$ and $\widetilde{X}$ be two decomposable $n$-vectors in $D^n_m{\cal M}$.
If $X \sim \widetilde{X}$ then $\forall 1\leq p\leq n$,
$\forall a\in P^pT^{\star}_m{\cal M}$, 
\begin{equation}\label{6.4.1}
X\Ni a \sim \widetilde{X}\Ni a.
\end{equation}
Hence we can define $[X]\Ni a := [X\Ni a]\in P^{n-p}T{\cal M}$.
\end{lemm}
{\em Proof} ---
This result amounts to the property that for all $0\leq p\leq n$,
$\forall a\in P^pT^{\star}_m{\cal M}$, $\forall b\in P^{n-p}T^{\star}_m{\cal M}$,
\[
\langle X\Ni a, b\rangle = \langle \widetilde{X}\Ni a, b\rangle
\quad \Longleftrightarrow \quad a\wedge b(X) = a\wedge b(\widetilde{X}),
\]
which is true because of $[X] = [\widetilde{X}]$ and $a\wedge b\in P^nT^{\star}_m{\cal M}$. \bbox
\noindent
As a consequence of Lemma \ref{6.2.lem1}, we have the following definition. 
\begin{defi}\label{6.2.def1}
Let $F\in \mathfrak{P}^{p-1}{\cal M}$ and ${\cal H}$ a Hamiltonian function....
The pseudobracket $\{{\cal H}, F\}$ is the section of $P^{n-p}T{\cal M}$ defined by
$$\{{\cal H}, F\} := (-1)^{(n-p)p}[X]^{\cal H}\Ni dF.$$
In case $p=n$, $\{{\cal H}, F\}$ is just the scalar function
$[X]^{\cal H}\iN dF = \langle [X]^{\cal H},dF\rangle$.
\end{defi}
In the case of pataplectic manifolds where $\mathfrak{P}^{n-1}_0{\cal M} = \mathfrak{P}^{n-1}{\cal M}$,
an alternative definition can be given using the tensor field $\xi_F$ defined by
(\ref{6.1.xi}).
\begin{lemm}\label{6.2.lem2}
Assume that $\mathfrak{P}^{n-1}_0{\cal M} = \mathfrak{P}^{n-1}{\cal M}$.
For any Hamiltonian function ${\cal H}$ and any $F\in \mathfrak{P}^{p-1}{\cal M}$,
we have 
\begin{equation}\label{6.2.2}
\{{\cal H}, F\} = - \xi_F\iN d{\cal H},
\end{equation}
where the right hand side is the section of $P^{n-p}T{\cal M}$ defined by
\begin{equation}\label{6.2.3}
\langle \xi_F\iN d{\cal H}, \phi\rangle :=
\xi_F(\phi)\iN d{\cal H},\quad \forall \phi \in P^{n-p}_0T^{\star}{\cal M}.
\end{equation}
\end{lemm}
{\em Proof} --- Starting from Definition \ref{6.2.def1},
we have $\forall \phi \in P^{n-p}_0T^{\star}{\cal M}$,
$$\begin{array}{ccl}
\langle \{{\cal H},F\}, \phi\rangle & = &
(-1)^{(n-p)p}\langle [X]^{\cal H}\Ni dF, \phi\rangle \\
 & = & (-1)^{(n-p)p}\langle [X]^{\cal H}, dF\wedge \phi\rangle \\
 & = & \langle [X]^{\cal H}, \phi\wedge dF\rangle  \\
 & = & - \langle [X]^{\cal H}, \xi_F(\phi)\iN \Omega\rangle \\
 & = & - (-1)^n\xi_F(\phi)\iN [X]^{\cal H}\iN \Omega\\
 & = & - \xi_F(\phi)\iN d{\cal H}.
\end{array}$$
\bbox
\noindent {\bf Example 14} --- {\em Assume that ${\cal M}=\Lambda^nT^{\star}({\cal X}\times \Bbb{R})$
with $\Omega^{dDW} = de\wedge \omega + dp^i_\mu\wedge dy\wedge \omega_\mu$.
Consider the Hamiltonian ${\cal H}(x,y,e,p)=e + \eta_{\mu\nu}p^\mu p^\nu/2
+ V(y)$. Then the solutions of $X\iN \Omega = (-1)^nd{\cal H}$ can be described by
$X=X_1\wedge \cdots \wedge X_n$ with
$$X_\mu = {\partial \over \partial x^\mu} + {\partial {\cal H}\over \partial p^\mu}
{\partial \over\partial y} + P^\nu_\mu{\partial \over \partial p^\nu},$$
where $\sum_\mu P^\mu_\mu = -{\partial {\cal H}\over \partial y}$.
Then if $F = x^1$,
$$\{{\cal H},x^1\} = (-1)^{n-1} \left[ {\partial \over \partial x^2}\wedge \cdots \wedge
{\partial \over \partial x^n}
+ \sum_{\mu=2}^n{\partial \over \partial x^2}\wedge \cdots \wedge
{\partial \over \partial x^{\mu-1}}\wedge
{\partial \over \partial y}
\wedge {\partial \over \partial x^{\mu+1}}\wedge \cdots \wedge
{\partial \over \partial x^n} + \cdots \right] $$
represents (modulo the relation $\sim$) the hyperplane in $T_m\Gamma$
on which $dx^1$ vanishes.}\\

\noindent We now prove the basic result relating these notions to the dynamics.

\begin{theo}\label{6.2.thm1}
Let $({\cal M},\Omega)$ be a multisymplectic manifold.
Assume that $1\leq p\leq n$, $1\leq q\leq n$ and $n\leq p+q$.
Let $F\in \mathfrak{P}^{p-1}{\cal M}$ and $G\in \mathfrak{P}^{q-1}{\cal M}$....
Let $\Sigma$ be a slice of codimension $2n-p-q$ and $\Gamma$ a Hamiltonian $n$-curve.
Then for any $(p+q-n)$-vector $Y$ tangent to $\Sigma\cap \Gamma$, we have
\begin{equation}\label{6.2.4}
\{{\cal H}, F\}\iN dG(Y) = (-1)^{(n-p)(n-q)}\{{\cal H}, G\}\iN dF(Y),
\end{equation}
which implies
$$\{{\cal H}, F\}\iN dG_{|\Gamma} = (-1)^{(n-p)(n-q)}
\{{\cal H}, G\}\iN dF_{|\Gamma}.$$
\end{theo}
{\em Proof} --- Proving (\ref{6.2.4}) is equivalent to proving 
\begin{equation}\label{6.2.5}
\langle \{{\cal H},F\}\wedge Y,dG\rangle = (-1)^{(n-p)(n-q)}
\langle \{{\cal H},G\}\wedge Y,dF\rangle.
\end{equation}
We thus need to compute first $\{{\cal H},F\}\wedge Y$.
For that purpose, we use Definition \ref{6.2.def1}: 
$\{{\cal H}, F\} = (-1)^{(n-p)(n-q)}[X]^{\cal H}\Ni dF$. Of course it will be more suitable
to use the representant of $[X]^{\cal H}$ which is tangent to $\Gamma$: we let
$(X_1,\cdots ,X_n)$ to be a basis of $T_m\Gamma$ such that
$$X_1\wedge \cdots \wedge X_n =: X \in [X]^{\cal H}$$
Then we can write
$$Y = \sum_{\nu_1<\cdots <\nu_{p+q-n}}T^{\nu_1\cdots \nu_{p+q-n}}
X_{\nu_1}\wedge \cdots \wedge X_{\nu_{p+q-n}}$$

\noindent Now
$$\begin{array}{ccl}
\{{\cal H}, F\} & = & (-1)^{(n-p)p}[X]^{\cal H}\Ni dF\\
 & = & \displaystyle (-1)^{(n-p)p}\sum_{\tiny \begin{array}{c}\mu_1<\cdots <\mu_p\\
\mu_{p+1}<\cdots <\mu_n\end{array}}
\delta^{\mu_1\cdots \mu_n}_{1\cdots n}dF(X_{\mu_1},\cdots ,X_{\mu_p})
X_{\mu_{p+1}}\wedge \cdots \wedge X_{\mu_n},
\end{array}$$
so that\\

\noindent $\displaystyle \{{\cal H}, F\}\wedge Y = (-1)^{(n-p)(p+q-n)}Y\wedge \{{\cal H}, F\}$\\

\noindent $\displaystyle =(-1)^{(n-p)(n-q)}
\sum_{\nu_1<\cdots <\nu_{p+q-n}}
\sum_{\tiny \begin{array}{c}
\mu_1<\cdots <\mu_p\\
\mu_{p+1}<\cdots <\mu_n\end{array}}
T^{\nu_1\cdots \nu_{p+q-n}}
\delta^{\mu_1\cdots \mu_n}_{1\cdots n}$

$$dF(X_{\mu_1},\cdots ,X_{\mu_p})
X_{\nu_1}\wedge \cdots \wedge X_{\nu_{p+q-n}} \wedge
X_{\mu_{p+1}}\wedge \cdots \wedge X_{\mu_n}.
$$
Now $X_{\nu_1}\wedge \cdots \wedge X_{\nu_{p+q-n}} \wedge
X_{\mu_{p+1}}\wedge \cdots \wedge X_{\mu_n}\neq 0$ if and only if it is possible
to complete the family $\{X_{\nu_1},\cdots ,X_{\nu_{p+q-n}}\}$ by
$\{X_{\lambda_1},\cdots ,X_{\lambda_{n-q}}\}$ in such a way that
$\{X_{\nu_1},\cdots ,X_{\nu_{p+q-n}},X_{\lambda_1},\cdots ,X_{\lambda_{n-q}}\}
= \{X_{\mu_1},\cdots ,X_{\mu_p}\}$ and
$\delta^{\nu_1\cdots \nu_{p+q-n}\lambda_1\cdots \lambda_{n-q}}_{\mu_1\cdots \mu_p}\neq 0$.
Hence\\

\noindent $\displaystyle \{{\cal H}, F\}\wedge Y$\\

\noindent $\displaystyle =(-1)^{(n-p)(n-q)}
\sum_{\tiny \begin{array}{c}
\mu_1<\cdots <\mu_p\\
\mu_{p+1}<\cdots <\mu_n\end{array}}
\sum_{\tiny \begin{array}{c}\nu_1<\cdots <\nu_{p+q-n}\\
\lambda_1<\cdots <\lambda_{n-q}\\
\end{array}}
\delta^{\nu_1\cdots \nu_{p+q-n}\lambda_1\cdots \lambda_{n-q}}_{\mu_1\cdots \mu_p}
T^{\nu_1\cdots \nu_{p+q-n}}\delta^{\mu_1..\mu_n}_{1\cdots n}$
$$dF(X_{\nu_1},\cdots ,X_{\nu_{p+q-n}},X_{\lambda_1},\cdots ,X_{\lambda_{n-q}})
X_{\nu_1}\wedge \cdots \wedge X_{\nu_{p+q-n}} \wedge
X_{\mu_{p+1}}\wedge \cdots \wedge X_{\mu_n}$$
\noindent $\displaystyle = (-1)^{(n-p)(n-q)}
\sum_{\tiny \begin{array}{c}
\mu_1<\cdots <\mu_p\\
\mu_{p+1}<\cdots <\mu_n\end{array}}
\sum_{\tiny \begin{array}{c}\nu_1<\cdots <\nu_{p+q-n}\\
\lambda_1<\cdots <\lambda_{n-q}\\
\end{array}}
\delta^{\nu_1\cdots \nu_{p+q-n}\lambda_1\cdots \lambda_{n-q}
\mu_{p+1}\cdots \mu_n}_{1\cdots n}T^{\nu_1\cdots \nu_{p+q-n}}$
$$(X_{\nu_1}\wedge \cdots \wedge X_{\nu_{p+q-n}}\iN dF)(X_{\lambda_1},\cdots ,X_{\lambda_{n-q}})
X_{\nu_1}\wedge \cdots \wedge X_{\nu_{p+q-n}} \wedge
X_{\mu_{p+1}}\wedge \cdots \wedge X_{\mu_n}$$
\noindent $\displaystyle = (-1)^{(n-p)(n-q)}
(-1)^{(n-q)(p+q-n)}X\Ni (Y\iN dF)$\\

\noindent $\displaystyle = (-1)^{(n-q)q}X\Ni (Y\iN dF).$\\

\noindent 
We conclude that
$$\begin{array}{ccl}
\langle \{{\cal H}, F\}\wedge Y, dG\rangle & = &
(-1)^{(n-q)q}\langle X,(Y\iN dF)\wedge dG\rangle \\
 & = & \langle X,dG \wedge (Y\iN dF)\rangle \\
 & = & \langle X\Ni dG,Y\iN dF\rangle \\
 & = & (-1)^{(n-q)q}\langle \{{\cal H},G\},Y\iN dF\rangle \\
 & = & (-1)^{(n-q)(n-p)}\langle \{{\cal H},G\}\wedge Y,dF\rangle .
\end{array}$$
So the result follows\footnote{Equation (\ref{6.2.4}) can be generalized for
$1\leq p\leq n$, $1\leq q\leq n$ and $p+q\leq n$ by
\begin{equation}\label{6.2.der}
\{{\cal H},F\}\Ni dG = (-1)^{p+q-n}(-1)^{(n-p)(n-q)}
\{{\cal H},G\}\Ni dF \quad \in P^{n-p-q}T_m{\cal M}.
\end{equation}
However in constrast with (\ref{6.2.4}) for $p+q>n$, this relation does not contain
any information on the dynamics, since it does not involve any Hamiltonian $n$-curve. Indeed
we have: for all $a\in P^{n-p-q}T^{\star}_m{\cal M}$,
$$\begin{array}{ccl}
\langle \{{\cal H},F\}\Ni dG,a\rangle & = & \langle \{{\cal H},F\}, dG\wedge a\rangle \\
 & = & (-1)^{(n-p)p}\langle [X]^{\cal H}\Ni dF, dG\wedge a\rangle \\
 & = & (-1)^{(n-p)p}\langle [X]^{\cal H}, dF\wedge dG\wedge a\rangle \\
 & = & (-1)^{(n-p)p}(-1)^{pq}\langle [X]^{\cal H}, dG\wedge dF\wedge a\rangle \\
 & = & (-1)^{(n-p)p+pq}\langle [X]^{\cal H}\Ni dG, dF\wedge a\rangle \\
 & = & (-1)^{(n-p)p+pq}(-1)^{(n-q)q}\langle \{{\cal H},G\}, dF\wedge a\rangle \\
 & = & (-1)^{(n-p)(n-q)+p^2+q^2-n^2}\langle \{{\cal H},G\}\Ni dF,a\rangle .\\
\end{array}$$
from which (\ref{6.2.der}) follows.
}. \bbox

\begin{coro}\label{6.2.cor1}
Assume the same hypothesis as in Theorem \ref{6.2.thm1}, then we have the following
relations (by decreasing the generality)
\begin{enumerate}
\item If $F\in \mathfrak{P}^{p-1}{\cal M}$ and
$G\in \mathfrak{P}^{n-1}{\cal M}$, then
$$\{{\cal H}, F\}\iN dG_{|\Gamma} = \{{\cal H}, G\}dF_{|\Gamma}$$
\item If $F\in \mathfrak{P}^{p-1}{\cal M}$ and if
$G\in \mathfrak{P}^{n-1}{\cal M}$ is such that $\{{\cal H},G\} = 1$, then denoting
$\omega := dG$ (a ``volume form'')
$$\{{\cal H}, F\}\iN \omega_{|\Gamma} = dF_{|\Gamma}.$$
\item If $F, G\in \mathfrak{P}^{n-1}{\cal M}$, we recover proposition 1.
\end{enumerate}
\end{coro}
{\em Proof} --- It is a straightforward application of Theorem \ref{6.2.thm1}. \bbox
\noindent
{\bf Example 15} --- {\em Consider a variational problem on maps $u:{\cal X}\longrightarrow {\cal Y}$
as in Example 2, Section 2.2.1. Take $F=y^i$ (a 0-form) and $G=x^1dx^2\wedge \cdots 
\wedge dx^n$, in such a way that $dG=\omega$, the volume form. Then we are in case (ii)
of the corollary: we can compute that $\{{\cal H}, y^i\}\iN \omega = \sum_\mu 
\partial {\cal H}/\partial p^\mu_idx^\mu$ and $\{{\cal H},G\}dy^i=dy^i$.
Hence this implies the relation $dy^i_{|\Gamma}=\sum_\mu 
\partial {\cal H}/\partial p^\mu_idx^\mu_{|\Gamma}$.}

\subsection{Observable functionals}
Using a slice $\Sigma$ of codimension $n-p+1$ as introduced in definition \ref{2.1.def40} 
and an observable $(p-1)$-form $F$ we can define 
an {\em observable functional} denoted symbolically by
$\int_{\Sigma}F:{\cal E}^{\cal H}\longrightarrow \Bbb{R}$ by:
\[
\Gamma\quad \longmapsto \int_{\Sigma\cap \Gamma} F.
\]
Here the intersection $\Sigma\cap \Gamma$ is oriented as follows: assume that
$\alpha^1,\cdots ,\alpha^{n-p+1}\in T_m^{\star}{\cal M}$ are such that
$\alpha^1\wedge \cdots \wedge \alpha^{n-p+1}$ vanishes on $T_m\Sigma$ and is
positively oriented on $T_m{\cal M}/T_m\Sigma$ and let $X\in \Lambda^nT_m\Gamma$
be positively oriented.
Then we require that $X\Ni \alpha^1\wedge \cdots \wedge \alpha^{n-p+1}\in
\Lambda^{p-1}T_m(\Sigma\cap \Gamma)$ is positively
oriented.\\

\noindent 
A natural question is to try to understand the slices in the basic situation
where ${\cal M}$ is an open subset of $\Lambda^nT^*{\cal N}$.
Let us consider a map $f=(f^1,\cdots ,f^{n-p+1})$ from ${\cal M}$
to $\Bbb{R}^{n-p+1}$ and look for necessary and sufficient conditions on $f$
in order that its level sets be slices. We use the same hypotheses
$d_p{\cal H}\neq 0$ and $[X]^{\cal H}\neq \emptyset$ as in section 4.1.
We first analyze the situation locally.
Given a point $m\in {\cal M}$, the property ``$X\in [X]^{\cal H}$ $\Longrightarrow$
$df_{m|X}$ is of rank $n-p+1$'' is equivalent to:
\[
\forall (t_1,\cdots ,t_{n-p+1})\in \Bbb{R}^{n-p+1}\setminus \{0\},\quad
X\in [X]^{\cal H} \Longrightarrow \sum_{i=1}^{n-p+1}t_idf^i_{m|X}\neq 0.
\]
Hence by using Lemma \ref{4.1/2.lemme} we deduce that the property 
$X\in [X]^{\cal H} \Longrightarrow$ rank $df_{m|X}=n-p+1$ is equivalent to
\begin{itemize}
\item $\forall (t_1,\cdots ,t_{n-p+1})\in \Bbb{R}^{n-p+1}\setminus \{0\}$,
$\exists \lambda(m)\in \Bbb{R}$,
$\sum_{i=1}^{n-p+1}t_id_pf^i_m=\lambda(m)d_p{\cal H}_m$.
And then one easily deduce that $\exists \lambda^1(m),\cdots ,\lambda^{n-p+1}(m)\in \Bbb{R}$,
such that $\lambda(m)=\sum_{i=1}^{n-p+1}t_i\lambda^i(m)$.
\item $\forall (t_1,\cdots ,t_{n-p+1})\in \Bbb{R}^{n-p+1}\setminus \{0\}$,
$\exists (\alpha_1,\cdots ,\alpha_n)\in I, \exists 1\leq \mu\leq n$,
$\{{\cal H},\sum_{i=1}^{n-p+1}t_if^i\}^{\alpha_1\cdots \alpha_n}_{\alpha_\mu}(m)\neq 0$.
\end{itemize}
Now the second condition translate as $\forall (t_1,\cdots ,t_{n-p+1})\in \Bbb{R}^{n-p+1}\setminus \{0\}$,
$\exists (\alpha_1,\cdots ,\alpha_n)\in I, \exists 1\leq \mu\leq n$,
\[
\sum_{i=1}^{n-p+1}t_i{\partial {\cal H}\over \partial p_{\alpha_1\cdots \alpha_n}}
\left( {\partial f^i\over \partial q^{\alpha_\mu}} -\lambda^i
{\partial {\cal H}\over \partial q^{\alpha_\mu}}\right) \neq 0.
\]
This condition can be expressed in terms of minors of size $n-p+1$ from the
matrix 
$\left( 
{\partial f^i\over \partial q^{\alpha_\mu}} -\lambda^i
{\partial {\cal H}\over \partial q^{\alpha_\mu}}\right)_{i,\alpha_\mu}$.
For that purpose let us denote by\\

\noindent
$\displaystyle \{\{ {\cal H},f^1,\cdots ,f^{n-p+1}\}\} :=
\sum_{1\leq \alpha_1<\cdots <\alpha_n\leq n+k}\;\sum_{1\leq \mu_1<\cdots <\mu_{n-p+1}\leq n}$
$$\left\langle {\partial \over \partial p_{\alpha_1\cdots \alpha_n}}
\wedge {\partial \over \partial q^{\alpha_{\mu_1}}}\wedge \cdots 
\wedge {\partial \over \partial q^{\alpha_{\mu_{n-p+1}}}},
d{\cal H}\wedge df^1\wedge \cdots \wedge df^{n-p+1}\right\rangle
dp_{\alpha_1\cdots \alpha_n}\wedge dq^{\alpha_{\mu_1}}\wedge \cdots  \wedge
dq^{\alpha_{\mu_{n-p+1}}}.$$
We deduce the following.
\begin{prop}
Let ${\cal M}$ be an open subset of $\Lambda^nT^{\star}{\cal N}$ endowed with its
standard multisymplectic form $\Omega$, let ${\cal H}:{\cal M}\longrightarrow \Bbb{R}$
be a smooth Hamiltonian function and let $f:{\cal M}\longrightarrow \Bbb{R}^{n-p+1}$ be
a smooth function. Let $m\in {\cal M}$ and assume that
$d_p{\cal H}\neq 0$ and $[X]^{\cal H}\neq \emptyset$ everywhere. Then
$X\in [X]^{\cal H}$ $\Longrightarrow$
$df_{m|X}$ is of rank $n-p+1$ if and only if
\begin{itemize}
\item
$\exists \lambda^1(m),\cdots ,\lambda^{n-p+1}(m)\in \Bbb{R}$,
$\forall 1\leq i\leq n-p+1$, $d_pf^i_m=\lambda^i(m)d_p{\cal H}_m$.
\item $\{\{ {\cal H},f^1,\cdots ,f^{n-p+1}\}\} (m)\neq 0$.
\end{itemize}
\end{prop}
And we deduce the global result:
\begin{theo}\label{3.1.theo}
Let ${\cal M}$ be an open subset of $\Lambda^nT^{\star}{\cal N}$ endowed with its
standard multisymplectic form $\Omega$, let ${\cal H}:{\cal M}\longrightarrow \Bbb{R}$
be a smooth Hamiltonian function and let $f:{\cal M}\longrightarrow \Bbb{R}^{n-p+1}$ be
a smooth function. Assume that $d_p{\cal H}\neq 0$ and $[X]^{\cal H}\neq \emptyset$
everywhere. Then all level sets of $f$ are slices if and only if 
$\exists (q,h)\in {\cal N}\times \Bbb{R}$ such that $f(q,p)=\widehat{f}(q,{\cal H}(q,p))$
and $\forall m\in {\cal M}$, $\{\{ {\cal H},f^1,\cdots ,f^{n-p+1}\}\} (m)\neq 0$.
\end{theo}

\subsection{Brackets}
We now consider observable $(p-1)$-forms for $1\leq p\leq n$ and discuss
how to define a Poisson bracket between these observable forms,
which could be relevant for quantization. This is slightly more delicate
than for forms of degree $n-1$ and the definitions proposed here
are based on  empirical observations. We first assume a further hypothesis
on the copolarisation (which is satisfied on $\Lambda^nT^*{\cal N}$ or which could also
be a consequence of Conjecture 2 in Section 5.2).\\

\noindent
{\bf Hypothesis on $P^1T^*_m{\cal M}$} ---
{\em For all $m\in {\cal M}$, every 1-form $a\in T^*_m{\cal M}$
which is proper on ${\cal O}_m$ (see Definition \ref{5.3/2.defi1})
is in $P^1T^*_m{\cal M}$.}\\

\noindent 
Let $1\leq p,q\leq n$ and $F\in \mathfrak{P}^{p-1}{\cal M}$ and
$G\in \mathfrak{P}^{q-1}{\cal M}$ and let us analyze what condition should
satisfy the bracket $\{F,G\}$. We will consider smooth functions
$f^1,\cdots ,f^{n-p}$, $g^1,\cdots ,g^{n-q}$ and $t$ on ${\cal M}$ such that
the level sets of $t$ are slices (we may think $t$ as a time coordinate) and
$\forall m\in {\cal M}$, $df^j_m$ and $dg^j_m$ are proper on ${\cal O}_m$
and $df^1_m\wedge \cdots \wedge df^{n-p}_m\wedge dg^1_m\wedge \cdots \wedge dg^{n-q}_m\neq 0$.
Then, because of the hypothesis on $P^1T^*_m{\cal M}$,
\[
\widetilde{F}:= df^1\wedge \cdots \wedge df^{n-p}\wedge F\quad \hbox{and}\quad
\widetilde{G}:= dg^1\wedge \cdots \wedge dg^{n-q}\wedge G
\]
are in $\mathfrak{P}^{n-1}{\cal M}$. Let us moreover assume\footnote{see the ``Hypothesis on
$\mathfrak{P}^{n-1}{\cal M}$'' below about this assumption} that
\begin{equation}\label{6.4.hypotheseprovisoire}
\xi_{\widetilde{F}}\iN df^\mu = 0,\ \forall 1\leq \mu\leq n-p,\quad \hbox{and}\quad
\xi_{\widetilde{G}}\iN dg^\mu = 0,\ \forall 1\leq \mu\leq n-q.
\end{equation}
Lastly let $\Gamma$ be a Hamiltonian $n$-curve and $\Sigma$ be a level set of $t$. Then
\begin{equation}\label{6.4.bracket.n-1n-1}
\left\{
\int_{\Sigma}\widetilde{F},\int_{\Sigma}\widetilde{G}
\right\}(\Gamma) = \left(
\int_{\Sigma}\left\{ \widetilde{F},\widetilde{G}\right\}
\right)(\Gamma) = 
\int_{\Sigma\cap \Gamma}\left\{ \widetilde{F},\widetilde{G}\right\}.
\end{equation}
We now suppose that the functions $f:=(f^1,\cdots ,f^{n-p})$ and
$g:=(g^1,\cdots ,g^{n-q})$ concentrate around submanifolds denoted
respectively by $\widehat{\gamma}_f$ and $\widehat{\gamma}_g$ of codimension
$n-p$ and $n-q$ respectively. More precisely we suppose that the image of $f$
(reps. $g$) covers the unit cube in $\Bbb{R}^{n-p}$ (resp. in $\Bbb{R}^{n-q}$),
that $df^1\wedge \cdots \wedge df^{n-p}$ (resp. $dg^1\wedge \cdots \wedge dg^{n-q}$)
is zero outside a tubular neighborhood of $\widehat{\gamma}_f$ (resp. of
$\widehat{\gamma}_g$) of width $\varepsilon$ and that the integral of $df^1\wedge \cdots \wedge df^{n-p}$
(resp. $dg^1\wedge \cdots \wedge dg^{n-q}$) on a disc submanifold of dimension $n-p$ (resp. $n-q$)
which cuts transversaly $\widehat{\gamma}_f$ (resp. $\widehat{\gamma}_g$)
is equal to 1. Moreover we suppose that $\widehat{\gamma}_f$ and $\widehat{\gamma}_g$
cut transversaly $\Sigma\cap \Gamma$ along submanifolds denoted by $\gamma_f$
and $\gamma_g$ respectively. Then, as $\varepsilon\rightarrow 0$, we have
\[
\int_{\Sigma\cap \Gamma}df^1\wedge \cdots \wedge df^{n-p}\wedge F \rightarrow
\int_{\Sigma\cap \widehat{\gamma}_f\cap \Gamma}F, \quad 
\int_{\Sigma\cap \Gamma}dg^1\wedge \cdots \wedge dg^{n-q}\wedge G \rightarrow
\int_{\Sigma\cap \widehat{\gamma}_g\cap \Gamma}G.
\]
This tells us that the left hand side of (\ref{6.4.bracket.n-1n-1}) is an approximation
for
\[
\left\{
\int_{\Sigma\cap \widehat{\gamma}_f}F,
\int_{\Sigma\cap \widehat{\gamma}_g}G
\right\}(\Gamma).
\]
\begin{figure}[h]
\begin{center}
\includegraphics[scale=1]{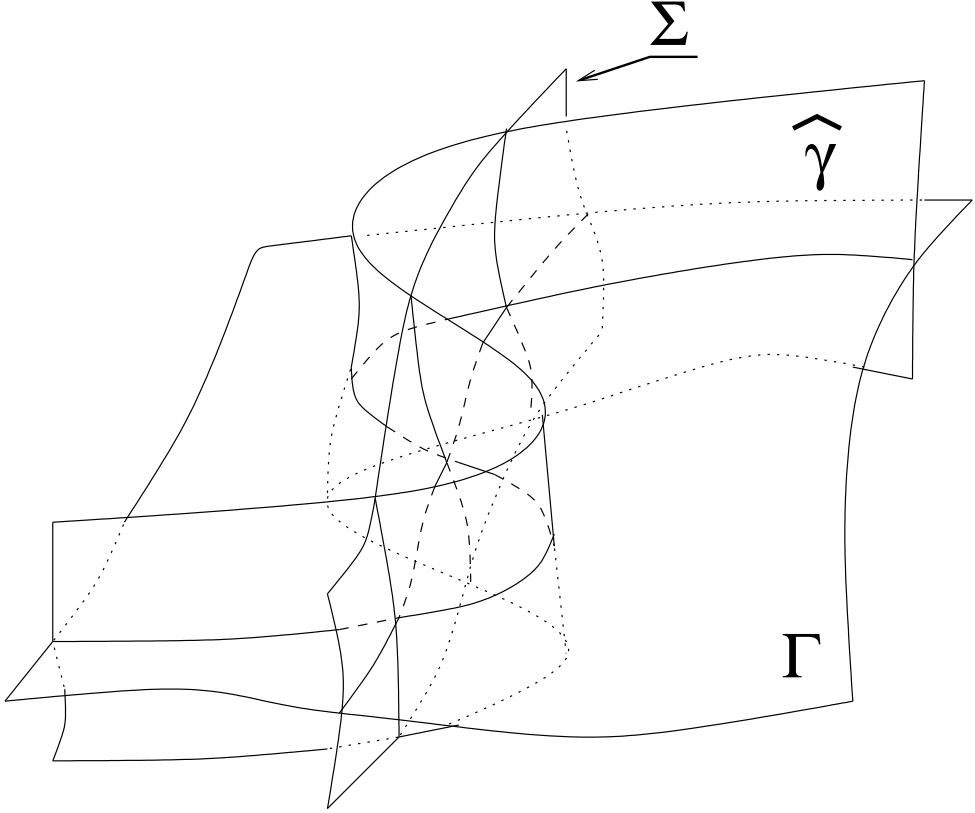}
\caption{Intersection of $\Gamma$, $\widehat{\gamma}$ and $\Sigma$}
\end{center}
\end{figure}
We now want to compute what is the limit of the right hand side of (\ref{6.4.bracket.n-1n-1}).
Using the condition (\ref{6.4.hypotheseprovisoire}) we have
\[
\begin{array}{ccl}
\{\widetilde{F},\widetilde{G}\} & = & \xi_{\widetilde{F}}\iN d\left( 
dg^1\wedge \cdots \wedge dg^{n-q}\wedge G\right) \\
& = & (-1)^{n-q}\xi_{\widetilde{F}}\iN dg^1\wedge \cdots \wedge dg^{n-q}\wedge dG\\
& = & dg^1\wedge \cdots \wedge dg^{n-q}\wedge \left( \xi_{\widetilde{F}}\iN dG\right).
\end{array}
\]
But we have similarly:
\[
\{\widetilde{F},\widetilde{G}\} = - df^1\wedge \cdots \wedge df^{n-p}\wedge 
\left( \xi_{\widetilde{G}}\iN dF\right).
\]
We now use the following result.
\begin{lemm}\label{6.4.lemmetechnique}
Let $\phi\in \Lambda^{n-1}T^*_m{\cal M}$ with $\phi\neq 0$ and $1\leq p,q\leq n$ such that $p+q\geq n+1$.
Suppose that there exists $2n-p-q$ linearly independant 1-forms
$a^1,\cdots ,a^{n-p},b^1,\cdots ,b^{n-q}\in T^*_m{\cal M}$,
$\alpha\in \Lambda^{p-1}T^*_m{\cal M}$ and $\beta\in \Lambda^{q-1}T^*_m{\cal M}$
such that $\phi = a^1\wedge \cdots \wedge a^{n-p}\wedge \alpha$ and $\phi =
b^1\wedge \cdots \wedge b^{n-q}\wedge \beta$. Then there exists 
$\chi \in \Lambda^{p+q-n-1}T^*_m{\cal M}$ such that
$\phi = a^1\wedge \cdots \wedge a^{n-p}\wedge b^1\wedge \cdots \wedge b^{n-q}\wedge 
\chi$. This $\chi$ is not unique in general and is defined modulo forms in the ideal
in $\Lambda^*T^*_m{\cal M}$ spanned by the $a_j$'s and the $b_j$'s. However
it is a unique real scalar if $p+q = n+1$.
\end{lemm}
{\em Proof} --- This is a consequence of Proposition 1.4 in \cite{bcggg}. 
The idea is based on the observation that $a^1,\cdots ,a^{n-p},b^1,\cdots ,b^{n-q}$
are in $\{a\in T^*_m{\cal M}/a\wedge \phi=0\}$.\bbox
\noindent
We deduce from Lemma \ref{6.4.lemmetechnique} that there exist a form
$\chi \in \Lambda^{p+q-n-1}T^*_m{\cal M}$ (not unique a priori) such that
$\{\widetilde{F},\widetilde{G}\} = df^1\wedge \cdots \wedge df^{n-p}\wedge
dg^1\wedge \cdots \wedge dg^{n-q}\wedge \chi$. We thus require that
\begin{equation}\label{6.4.implicite}
\{\widetilde{F},\widetilde{G}\} = df^1\wedge \cdots \wedge df^{n-p}\wedge
dg^1\wedge \cdots \wedge dg^{n-q}\wedge \{F,G\}.
\end{equation}
This does not characterize completely $\{F,G\}$,
unless $p+q = n+1$, the case $\{F,G\}$ where is a scalar. We can now write the
right hand side of (\ref{6.4.bracket.n-1n-1}) as
\[
\int_{\Sigma\cap \Gamma} df^1\wedge \cdots \wedge df^{n-p}\wedge
dg^1\wedge \cdots \wedge dg^{n-q}\wedge \{F,G\}.
\]
Letting $\varepsilon \rightarrow 0$, and assuming that $\widehat{\gamma}_f$
and $\widehat{\gamma}_g$ cross transversaly this integral converge to
\[
\int_{\Sigma\cap \widehat{\gamma}_f\cap \widehat{\gamma}_g\cap \Gamma}
\{F,G\},
\]
so that we have
\[
\left\{
\int_{\Sigma\cap \widehat{\gamma}_f}F,
\int_{\Sigma\cap \widehat{\gamma}_g}G
\right\}(\Gamma) =
\int_{\Sigma\cap \widehat{\gamma}_f\cap \widehat{\gamma}_g\cap \Gamma}
\{F,G\}.
\]
Here the intersection $\Sigma\cap \widehat{\gamma}_f\cap \widehat{\gamma}_g\cap \Gamma$
is oriented by assuming that $X\Ni dt \wedge df\wedge dg$ is oriented
positively, if $X\in [X]^{\cal H}$ orients positively $T_m\Gamma$. Hence if we had started
with 
$\int_{\Sigma\cap \Gamma}\{\widetilde{F},\widetilde{G}\} =
- \int_{\Sigma\cap \Gamma}\{\widetilde{G},\widetilde{F}\}$ we would
have obtained $-\int_{\Sigma\cap \widehat{\gamma}_g\cap \widehat{\gamma}_f\cap \Gamma}
\{G,F\} = -(-1)^{(n-p)(n-q)}
\int_{\Sigma\cap \widehat{\gamma}_f\cap \widehat{\gamma}_g\cap \Gamma}\{G,F\}$. Since the
resulting brackets should coincide we deduce that
\[
\{F,G\} + (-1)^{(n-p)(n-q)}\{G,F\} = 0.
\]

\noindent
Let us now discuss if we can guess a more direct definition of
$\{F,G\}$. A first case is when one of the two forms
$F$ or $G$ is in $\mathfrak{P}^{n-1}{\cal M}$, let us
say $F\in \mathfrak{P}^{p-1}{\cal M}$ and $G\in \mathfrak{P}_0^{n-1}{\cal M}$, then
we let
\begin{equation}\label{5.5.1.p-1n-41}
\{F,G\} := -\xi_G\iN dF.
\end{equation}
This is the idea of external bracket as in Paragraph 3.3.2.
We remark that if $f^1,\cdots ,f^{n-p}$ are in $\mathfrak{P}^0{\cal M}$ and are such that
$\xi_G\iN df^1\wedge \cdots \wedge df^{n-p} = 0$, then
$df^1\wedge \cdots \wedge df^{n-p}\wedge F\in \mathfrak{P}^{n-1}{\cal M}$ and
\[
\{df^1\wedge \cdots \wedge df^{n-p}\wedge F, G\} =
df^1\wedge \cdots \wedge df^{n-p}\wedge \{F,G\}
\]
so that the requirement (\ref{6.4.implicite}) is satisfied.\\

\noindent
Now if $F\in \mathfrak{P}^{p-1}{\cal M}$, $G\in \mathfrak{P}^{q-1}{\cal M}$
and $1<p,q<n$, we do not see an analogue of (\ref {5.5.1.p-1n-41}) for defining
$\{F,G\}$. However we can observe empirically that in most examples the following is true.\\

\noindent
{\bf Hypothesis on $\mathfrak{P}^{n-1}{\cal M}$} ---
{\em For any $F\in \mathfrak{P}^{n-1}{\cal M}$ such that there exists $1<p<n$
such that $dF\in \left( P^{n-p}T^*{\cal M}\right) \wedge \left( P^pT^*{\cal M}\right)$
and for any $f\in \mathfrak{P}^1{\cal M}$ such that $df$ is a point-slice, we have
$\{F,f\}=0$.}\\

\noindent The nice point is then that this hypothesis implies (\ref{6.4.hypotheseprovisoire}).
So using this hypothesis we can define $\{F,G\}$ at least for all cases $p+q = n$. We believe
that this hypothesis (together with the hypothesis on $P^1T^*_m{\cal M}$)
has a physical content, since according to the examples we know,
$(n-1)$-forms $F\in \mathfrak{P}^{n-1}{\cal M}$ such that $dF$ cannot be decomposed
in $\left( P^{n-p}T^*{\cal M}\right) \wedge \left( P^pT^*{\cal M}\right)$ for $1<p<n$
are ``pure momentum'' observable forms and are canonically conjugate to functions
whose differential is proper on ${\cal O}_m$, i.e.\,``position'' 0-forms.\\

\noindent {\bf Example 16} --- {\em Sigma models ---
Let ${\cal M}:= \Lambda^nT^{\star}({\cal X}\times {\cal Y})$ as in Section 2. For simplicity
we restrict ourself to the de Donder--Weyl submanifold ${\cal M}^{dDW}$
(see Section 2.3), so that the Poincar\'e-Cartan
form is $\theta = e\omega + p^\mu_idy^i\wedge \omega_\mu$ and the multisymplectic
form is $\Omega = d\theta$. Let $\phi$ be a function on ${\cal X}$ and 
consider the observable 0-form $y^i$ (for $1\leq i\leq k$) and the observable
$(n-1)$-form $P_{j,\phi}:= \phi(x)\partial / \partial y^j\iN \theta.$
Then $\xi_{P_{j,\phi}} = \phi\partial / \partial y^j -
p^\mu_j\left(\partial \phi/ \partial x^\mu\right)
\partial / \partial e$ and thus $\{P_{j,\phi},y^i\}= \xi_{P_{j,\phi}}\iN dy^i
= \delta^i_j \phi$. It gives the following bracket for observable functionals
\[
\left\{ \int_{\Sigma}P_{j,\phi},\int_{\Sigma\cap \widehat{\gamma}}y^i\right\}(\Gamma)
= \int_{\Sigma\cap \widehat{\gamma}\cap \Gamma}\delta^i_j \phi(x)
= \delta^i_j \sum_{m\in \Sigma\cap \widehat{\gamma}\cap \Gamma}
\hbox{sign}(m)\phi(m),
\]
where $\Sigma$, $\widehat{\gamma}$ and $\Gamma$ are supposed to cross transversaly
and $\hbox{sign}(m)$ accounts for the orientation of their intersection points.}\\

\noindent
{\bf Example 13''} --- {\em Maxwell equations --- In this case we find that,
for all functions $f,g_1,g_2:{\cal M}^{Max}\longrightarrow \Bbb{R}$ whose differentials
are proper on ${\cal O}_m$, $\{df\wedge \pi,dg_1\wedge dg_2\wedge a\} =
df\wedge dg_1\wedge dg_2$. We hence deduce that $\{\pi,a\} =1$: these forms are
canonically conjugate. We deduce the following bracket for observable functionals
\[
\left\{ \int_{\Sigma\cap \widehat{\gamma}_f}\pi,\int_{\Sigma\cap \widehat{\gamma}_g}a\right\}(\Gamma)
=  \sum_{m\in \Sigma\cap \widehat{\gamma}_f\cap \widehat{\gamma}_g\cap \Gamma}
\hbox{sign}(m),
\]
where $\Sigma\cap \widehat{\gamma}_f\cap \Gamma$ is a surface
and $\Sigma\cap \widehat{\gamma}_g\cap \Gamma$ is a curve in the three-dimensional
space $\Sigma\cap \Gamma$. Note that this conclusion was achieved by I.\,Kanatchikov with
its definition of bracket $\{\pi,a\}_{Kana}:=\xi_\pi\iN da$, where $\xi_\pi \in
\Lambda^2T^*_m{\cal M}^{Max}$ is such that $\xi_\pi\iN \Omega = d\pi$. But there
by choosing $\xi_\pi=(1/2)\sum_\mu (\partial /\partial a_\mu)\wedge (\partial /\partial x^\mu)$
one finds (in our convention) that $\{\pi,a\}_{Kana} = n/2$ ($= 2$, if $n=4$).
So the two brackets
differ (the new bracket in this paper differs also from the one that we proposed
in \cite{HeleinKouneiher}).}

\section{Dynamical observable forms and functionals}
We have seen in Paragraph 3.3.4 that integration of algebraic observable $(n-1)$-forms
over the intersection
of a slice of codimension 1 with a Hamiltonian $n$-curve defines observable functionals
on ${\cal E}^{\cal H}$. These observable functionals are of the same type that the ones built
using the standard
canonical formalism for field theory (based on choosing a time variable on
the space-time ${\cal X}$: then the slices correspond to constant time hypersurfaces).
Moreover the bracket between observable functionals defined in Paragraph 3.3.4 
coincides with the Poisson bracket obtained by means of the standard canonical formalism
(based on the well-known symplectic structure on infinite dimensional
manifolds), see \cite{HeleinKouneiher}, \cite{Kijowski1}, \cite{Kanatchikov1}.
But there is one problem left: 
to make sense of the Poisson bracket of two observable functionals supported on different slices.
This is essential in an Einstein picture (classical analogue of the Heisenberg picture)
which seems inavoidable in a completely covariant theory. It will lead to the
notion of {\em dynamical} observable forms (in contrast with kinematic observable functionals)
which corresponds more or less to the notion used by J. Kijowski in \cite{Kijowski1}.
This is the subject of this Section.

\subsection{Dynamical observable $(n-1)$-forms}

We come back here to the Poisson bracket between two observable functionals
of the form $\int_{\Sigma}F$ and $\int_{\Sigma}G$, i.e.\,$\int_{\Sigma}\{F,G\}$.
We see a difficulty: if $\Sigma$ and $\Sigma'$ are two different slices of
codimension 1, there is no way a priori to define the Poisson bracket between
$\int_{\Sigma}F$ and $\int_{\Sigma'}G$. At this point we can choose between two
options: either we accept that and try to construct a quantum field theory using
a kind of Schr\"odinger picture --- but the loose the covariance of the theory ---,
or we wish to use the Heisenberg picture and we
try to extend the above concepts
in order to make sense of the Poisson bracket between $\int_{\Sigma}F$ and $\int_{\Sigma'}G$.
Let us explore this strategy.\\

\noindent 
One way to define the Poisson bracket between $\int_{\Sigma}F$ and $\int_{\Sigma'}G$
is to express one of the two observable functionals, say $\int_{\Sigma'}G$, as
an integral over $\Sigma$. This motivates the search for observable $(n-1)$-form $G$ such that
\[
\int_{\Sigma'}G = \int_{\Sigma}G\quad \hbox{on}\quad {\cal E}^{\cal H}.
\]
This can be achieved for all slices $\Sigma$ and $\Sigma'$ which are  
cobordism equivalent, i.e.\,such that there exists a smooth
domain ${\cal D}$ in ${\cal M}$ with $\partial {\cal D} = \Sigma' - \Sigma$, if
$dG_{|\Gamma} = 0$, $\forall \Gamma \in {\cal E}^{\cal H}$.
Then indded
\begin{equation}\label{3.1.cobord}
\int_{\Sigma\cap \Gamma}G - \int_{\Sigma'\cap \Gamma}G =
\int_{\partial {\cal D}\cap \Gamma}G
=\int_{{\cal D}\cap \Gamma}dG = 0.
\end{equation}
Thus we are led to the following.

\begin{defi}
A {\em dynamical observable $(n-1)$-form} is an observable form
$G\in \mathfrak{P}^{n-1}{\cal M}$ such that
$$ \{{\cal H}, G\} = 0$$
\end{defi}
Now Corollary \ref{3.2.2.corodyn} implies immediately that if $G$ is a dynamical
observable $(n-1)$-form then $dG_{|\Gamma} = 0$ and hence
(\ref{3.1.cobord}) holds. As a consequence if
$F$ is any observable $(n-1)$-form and $G$ is a dynamical
observable $(n-1)$-form (and if one of both is an algebraic one), then we can state
$$\left\{ \int_{\Sigma}F, \int_{\Sigma'}G\right\} :=
\left\{ \int_{\Sigma}F, \int_{\Sigma}G\right\} .$$
The concept of dynamical observable form is actually more or less the one used by J. Kijowski
in \cite{Kijowski1}, since his theory corresponds to working on the
restriction of $({\cal M}, \Omega)$ on the hypersurface ${\cal H} = 0$.\\

\noindent
Hence we are led to the question of characterizing dynamical observable $(n-1)$-forms.
(We shall consider mostly algebraic observable forms.)
This question was already investigated for some particular case in \cite{Kijowski1}
(and discussed in \cite{GoldschmidtSternberg})
and the answer was a (surprising) deception: as long as the variational problem
is linear (i.e.\,the Lagrangian is a quadratic function of all variables)
there are many observable functionals (basically all smeared integrals of fields using test functions
which satisfy the Euler-Lagrange equation), but as soon as
the problem is non linear the choice of dynamical observable forms is dramatically reduced
and only global dynamical observable exists. For instance for a non nonlinear
scalar field theory with $L(u, du) = {1\over 2}(\partial _tu)^2
-|\nabla u|^2 + {m^2\over 2}u^2 + {\lambda \over 3}u^3$, the only dynamical
observable forms $G$ are those for which $\xi_G$ is a generator of the Poincar\'e
group. One can also note that in general dynamical observable forms correspond
to momentum or energy-momentum observable functionals.\\

\noindent
Several possibilities may be considered to go around this difficulty. If the variational
problem can be seen as a deformation of a linear one (i.e.\,of a free field theory)
then it could be
possible to construct a perturbation theory, leading to Feynman type expansions for classical
fields. For an example of such a theory, see \cite{perturb}. Another interesting direction would be to
explore completely integrable systems. We present here a third alternative, which relies
on symmetries and we will see on a simple example how the purpose of constructing dynamical observable forms leads
naturally to gauge theories.

\subsection{An example: complex scalar fields}
We consider on the set of maps $\varphi:\Bbb{R}^n\longrightarrow \Bbb{C}$
the variational problem with Lagrangian
\[
L_0(\varphi,d\varphi) = {1\over 2}\eta^{\mu\nu}
{\partial \overline{\varphi}\over \partial x^\mu}
{\partial \varphi\over \partial x^\nu} + V\left( {|\varphi|^2\over 2}\right) =
{1\over 2}\eta^{\mu\nu}\left(\
{\partial \varphi^1\over \partial x^\mu}{\partial \varphi^1\over \partial x^\nu} +
{\partial \varphi^2\over \partial x^\mu}{\partial \varphi^2\over \partial x^\nu}\right)
+ V\left( {|\varphi|^2\over 2}\right) .
\]
Here $\varphi = \varphi^1+i\varphi^2$. We consider the multisymplectic manifold ${\cal M}_0$, with
coordinates $x^\mu$, $\phi^1$, $\phi^2$, $e$, $p^\mu_1$ and $p^\mu_2$ and the multisymplectic
form $\Omega_0= de\wedge \omega + dp^\mu_a\wedge d\phi^a\wedge \omega_\mu$ (which is the differential
of the Poincar\'e-Cartan form $\theta_0:= e\omega + p^\mu_ad\phi^a\wedge \omega_\mu$).
Then the Hamiltonian is
\[
{\cal H}_0(x,\phi,e,p) = e + {1\over 2}\eta_{\mu\nu}(p^\mu_1p^\nu_1
+p^\mu_2p^\nu_2) - V\left( {|\phi|^2\over 2}\right) .
\]
We look for $(n-1)$-forms $F_0$ on ${\cal M}_0$ such that
\begin{equation}\label{7.2.1}
dF_0 + \xi_{F_0}\iN \Omega_0 = 0,\quad \hbox{for some vector field}\quad
\xi_{F_0},
\end{equation}
\begin{equation}\label{7.2.2}
d{\cal H}_0(\xi_{F_0}) = 0.
\end{equation}
The analysis of this problem can be dealt by looking for all vector fields
\[
\xi_0 = X^\mu(x,\phi,e,p){\partial \over \partial x^\mu} + \Phi^a(x,\phi,e,p)
{\partial \over \partial \phi^a}
+ E(x,\phi,e,p) {\partial \over \partial e} + P^\mu_a(x,\phi,e,p)
{\partial \over \partial p^\mu_a}.
\]
satisfying (\ref{7.2.1}) and (\ref{7.2.2}).
For simplicity we will assume that $X^\mu = 0$ (this will exclude stress-energy tensor observable
forms $X^\mu{\partial \over \partial x^\mu}\iN \theta_0$, for $X^\mu$ constant). Then we find two cases:\\

\noindent
If $V(|\phi|^2/2)$ is quadratic in $\phi$, i.e.\,if $V(|\phi|^2/2) = m^2|\phi|^2/2$,
then Equations (\ref{7.2.1}) and (\ref{7.2.1}) have the solutions
\[
\xi_0 = \lambda \vec{j}_0
+ U^a(x){\partial \over \partial \phi^a}
- \left(p^\mu_a{\partial U^a\over \partial x^\mu}(x) + \delta_{ab}LU^a(x)\phi^b\right){\partial \over \partial e}
+ \eta^{\mu\nu}\delta_{ab}{\partial U^a\over \partial x^\mu}(x){\partial \over \partial p^\mu_b},
\]
where $\lambda$ is a real constant,
\[
\vec{j}_0 := \left( \phi^2{\partial \over \partial \phi^1} - \phi^1{\partial \over \partial \phi^2}\right)
+ \left(p^\mu_2{\partial \over \partial p^\mu_1} - p^\mu_1{\partial \over \partial p^\mu_2}\right),
\]
$L:= - \eta^{\mu\nu}{\partial ^2\over \partial x^{\mu}\partial x^{\nu}}$
and $U^1$ and $U^2$ are arbitrary solutions of the linear equation $LU + m^2U = 0$.
Then $F_0=U^ap^\mu_a\omega_\mu - \eta^{\mu\nu}\left( 
{\partial U^1\over \partial x^\nu}\phi^1 + {\partial U^2\over \partial x^\nu}\phi^2
\right) \omega_\mu + \lambda \left( p^\mu_1\phi^2 - p^\mu_2\phi^1\right) \omega_\mu$.\\

\noindent
However if $V'$ is not a constant, then system (\ref{7.2.1}) and (\ref{7.2.1}) has only the solutions
$\xi_0 = \lambda \vec{j}_0$ and the resulting dynamical observable $(n-1)$-form is
$F_0 = \lambda(p^\mu_1\phi^2 - p^\mu_2\phi^1)\omega_\mu$, which
corresponds to the global charge due to the $U(1)$ invariance of the Lagrangian.\\

\noindent
For instance we would like to replace $\lambda$ by a smooth function $\psi$ of $x$,
i.e.\,to look at $F = \psi(x)(p^\mu_1\phi^2 - p^\mu_2\phi^1)\omega_\mu$. These are
non dynamical algebraic observable $(n-1)$-forms since we have 
$dF_1 + \widetilde{\xi}\iN \Omega_0 = 0$, where
$\widetilde{\xi}:= \psi\vec{j}_0 - (p^\mu_1\phi^2 - p^\mu_2\phi^1){\partial \psi\over \partial x^\mu}
{\partial \over \partial e}$,
but $d{\cal H}_0(\widetilde{\xi}) = - (p^\mu_1\phi^2 - p^\mu_2\phi^2){\partial \psi\over \partial x^\mu}\neq 0$.\\

\noindent
In order to enlarge the set of dynamical observable forms, we further incorporate the
gauge potential field $A:= A_\mu dx^\mu$ and consider the Lagrangian
\[
L_1(\varphi,A,d\varphi):= {1\over 2}\eta^{\mu\nu}
\left( \overline{ {\partial \varphi\over \partial x^\mu} + iA_\mu \varphi}\right)
\left( {\partial \varphi\over \partial x^\nu} + iA_\nu \varphi\right) - {1\over 4}\eta^{\mu\lambda}\eta^{\nu\sigma}
F_{\mu\nu}F_{\lambda\sigma} + V\left( {|\varphi|^2\over 2}\right),
\]
where $F_{\mu\nu}:={\partial A_\nu\over \partial x^\mu} - {\partial A_\mu\over \partial x^\nu}$.
It is invariant under gauge transformations $\varphi\longmapsto e^{i\theta}\varphi$,
$A\longmapsto A - d\theta$. Note that we did incorporate an energy
for the gauge potential $A$. We now consider the multisymplectic manifold
${\cal M}_1$ with coordinates $x^\mu$, $\phi^1$, $\phi^2$, $e$,
$p^\mu_1$, $p^\mu_2$, $a_\mu$ and $p^{\mu\nu}$.
The multisymplectic form is: $\Omega_1= de\wedge \omega + dp^\mu_a\wedge d\phi^a\wedge \omega_\mu
- (da_\lambda\wedge dx^\lambda )\wedge ({1\over 2}dp^{\mu\nu}\wedge \omega_{\mu\nu})$.
The Hamiltonian is then
\[
{\cal H}_1(x,\phi,a,e,p) = e + {1\over 2}\eta_{\mu\nu}(p^\mu_1p^\nu_1 + p^\mu_2p^\nu_2)
+ (p^\mu_1\phi^2 - p^\mu_2\phi^1)a_\mu
- {1\over 4}\eta_{\mu\lambda}\eta_{\nu\sigma}p^{\mu\nu}p^{\lambda\sigma}
- V\left( {|\phi|^2\over 2}\right) .
\]
The gain is that we may now consider the algebraic observable $(n-1)$-form
\[
F_1:= \psi(x) (p^\mu_1\phi^2 - p^\mu_2\phi^1)\omega_\mu
-{1\over 2}p^{\mu\nu}d\psi \wedge \omega_{\mu\nu}.
\]
where $\psi$ is any smooth function of $x$. We indeed still have on the one hand
$dF_1 = - \xi_1 \iN \Omega_1$,
where 
\[
\xi_1 := \psi\vec{j}_0 - (p^\mu_1\phi^2 - p^\mu_2\phi^1){\partial \psi\over \partial x^\mu}
{\partial \over \partial e} 
+ {\partial \psi\over \partial x^\mu}{\partial \over \partial a_\mu}.
\]
Then $d{\cal H}_1(\xi_{F_1}) = 0$.
Thus $F_1$ is a dynamical observable $(n-1)$-form.

\end{document}